\documentclass[12pt,notitlepage,a4paper]{article}

\pdfoutput=1

\usepackage{a4wide}
\usepackage{epsfig}
\usepackage{color,graphicx}
\usepackage{cite}
\usepackage{amssymb,amsmath}

\newcommand{\be}{\begin{equation}}
\newcommand{\ee}{\end{equation}}
\newcommand{\bea}{\begin{eqnarray}}
\newcommand{\eea}{\end{eqnarray}}

\usepackage[dvipsnames]{xcolor}

\usepackage{graphics,graphicx, color,tikz,mathtools, colortbl, xcolor,enumerate, tikz-3dplot, slashed,bm}

\usetikzlibrary{positioning}
\usetikzlibrary{chains}
\usetikzlibrary{arrows, arrows.meta ,fit,decorations.pathreplacing}
\tikzstyle{every picture}+=[remember picture]
\tikzstyle{na} = [baseline]

\usetikzlibrary{arrows, decorations.markings, calc, fadings, decorations.pathreplacing, patterns, decorations.pathmorphing, positioning}
\tikzset{>={Latex[width=1.5mm,length=1.5mm]}}
\usetikzlibrary{graphs,graphs.standard,quotes}

\tikzset{->-/.style={decoration={
  markings,
  mark=at position #1 with {\arrow{>}}},postaction={decorate}}}    
\tikzset{-<-/.style={decoration={
  markings,
  mark=at position #1 with {\arrow{<}}},postaction={decorate}}}  
  
\newcommand{\nn}{\nonumber}

\begin{document}

\begin{center}  

\vskip 2cm 

\centerline{\Large {\bf Compactifying 5d superconformal field theories to 3d}}

\vskip 1cm

\renewcommand{\thefootnote}{\fnsymbol{footnote}}

   \centerline{
    {\large \bf Matteo Sacchi${}^{a}$} \footnote{m.sacchi13@campus.unimib.it}{\large \bf , Orr Sela${}^{b}$} \footnote{sorrsela@campus.technion.ac.il} {\large \bf and Gabi Zafrir${}^{a}$} \footnote{gabi.zafrir@unimib.it}}

\vspace{1cm}
\centerline{{\it ${}^a$ Dipartimento di Fisica, Universit\`a di Milano-Bicocca \& INFN, Sezione di Milano-Bicocca,}}
\centerline{{\it I-20126 Milano, Italy}}
\centerline{{\it ${}^b$ Department of Physics, Technion, Haifa, 32000, Israel}}
\vspace{1cm}

\end{center}

\vskip 0.3 cm

\setcounter{footnote}{0}
\renewcommand{\thefootnote}{\arabic{footnote}}   
   
\begin{abstract}

Building on recent progress in the study of compactifications of $6d$ $(1,0)$ superconformal field theories (SCFTs) on Riemann surfaces to $4d$ $\mathcal{N}=1$ theories, we initiate a systematic study of compactifications of $5d$ $\mathcal{N}=1$ SCFTs on Riemann surfaces to $3d$ $\mathcal{N}=2$ theories. Specifically, we consider the compactification of the so-called rank 1 Seiberg $E_{N_f+1}$ SCFTs on tori and tubes with flux in their global symmetry, and put the resulting $3d$ theories to various consistency checks. These include matching the (usually enhanced) IR symmetry of the $3d$ theories with the one expected from the compactification, given by the commutant of the flux in the global symmetry of the corresponding $5d$ SCFT, and identifying the spectrum of operators and conformal manifolds predicted by the $5d$ picture. As the models we examine are in three dimensions, we encounter novel elements that are not present in compactifications to four dimensions, notably Chern-Simons terms and monopole superpotentials, that play an important role in our construction. The methods used in this paper can also be used for the compactification of any other $5d$ SCFT that has a deformation leading to a $5d$ gauge theory.

\end{abstract}
 
 \newpage
 
\tableofcontents

\section{Introduction}  

The study of the dynamics of supersymmetric quantum field theory in three spacetime dimensions is a very rich research topic. One reason for this is that it is believed that any supersymmetric gauge theory with sufficient matter flows to an interacting superconformal field theory (SCFT), leading to a wealth of models. Furthermore, in many cases these models exhibit interesting non-perturbative dynamical effects. An example of this is given by the phenomenon of symmetry enhancement, where the global symmetry of the IR theory is larger than that expected from the UV description. A further example is given by the phenomenon of duality, where two different UV theories flow to the same IR SCFT. 

For theories with at least four supercharges, the advent of supersymmetric localization has brought about many tools that can be used to uncover cases of such phenomena facilitating the discovery of many examples of these. However, this discovery raises the need for an organizing principle, allowing for the organization, motivation and prediction of such models. 

In the case of four dimensional quantum field theory, where similar phenomena also occur, one such organizing principle is given by the compactification of $6d$ SCFTs on Riemann surfaces to $4d$, potentially with flux in the global symmetry of the $6d$ SCFT. The way this method works is that we can break any complicated Riemann surface to a collection of three-punctured spheres, called trinions, that are glued together along their punctures. This is usually refereed to as a pair of pants decomposition, and has interesting implications for $4d$ physics. Specifically, to each trinion one associates a $4d$ theory, given by the compactification of the $6d$ SCFT on this surface. The $4d$ theory resulting from the compactification on the full surface can be obtained by gluing together the $4d$ theories associated with each trinion, usually by gauging mutual global symmetries, in the same manner as the pair of pants decomposition constructs the full Riemann surface. However, the pair of pants decomposition is not unique. This naturally leads to dualities between different $4d$ theories, corresponding to different pair of pants decompositions of the same surface, and to symmetry enhancement, where the surface is built from trinions that individually preserve less symmetry than that expected from the full surface.    

Given the success of this approach in understanding the behavior of $4d$ theories, it is thus tempting to seek similar such organizing principles also for three dimensional theories. One option is to consider the compactification of $6d$ SCFTs on $3$-surfaces to $3d$, which is usually referred to as the $3d$-$3d$ correspondence\cite{Dimofte:2011ju,Gang:2018wek}. In this approach we keep the UV theory to be a $6d$ SCFT, but change the dimension of the compactification surface to three. This means that in this approach we need to deal with $3$-surfaces whose geometry is significantly more complicated than the geometry of $2$-surfaces.

Another option is to keep the dimension of the compactification surface, but change the UV theory to be a $5d$ SCFT. This allows us to keep the relatively simple geometry of two dimensional surfaces. Despite this appealing feature, this approach has remained relatively unexplored. Here we aim to begin the study of this method of realizing $3d$ theories. Specifically, we shall rely on the advancement made in the study of the compactification of $6d$ $(1,0)$ SCFTs on Riemann surfaces to $4d$ $\mathcal{N}=1$ theories, and apply the methods developed there also to the study of the compactification of $5d$ $\mathcal{N}=1$ SCFTs on Riemann surfaces to $3d$ $\mathcal{N}=2$ theories.

To study the compactification of $5d$ SCFTs, we first need to make a choice of $5d$ SCFT. These can be realized by a variety of methods. Many of them can be realized in field theory as UV completions of $5d$ gauge theories\cite{SEI,SM,SMI}. However, the most prolific method to realize them is using string theory, either through brane systems\cite{AH,AHK,KB,Bergman:2015dpa,Zafrir:2015ftn,Hayashi:2015vhy} or by the compactification of M-theory on Calabi-Yau threefolds\cite{Douglas:1996xp,Jefferson:2018irk,Bhardwaj:2018vuu,Bhardwaj:2018yhy,Closset:2018bjz,Apruzzi:2018nre,Apruzzi:2019opn,Apruzzi:2019enx,Bhardwaj:2019jtr,Bhardwaj:2019fzv,Saxena:2019wuy,Apruzzi:2019kgb}. While a complete classification of $5d$ SCFTs is still lacking, there are by now many known examples of them.   

Here we shall make a choice of a specific family of $5d$ SCFTs. This family is the so-called Seiberg $E_{N_f+1}$ SCFTs, first discussed in \cite{SEI}. These correspond to one of the simplest families of $5d$ SCFTs, that still possess a rich flavor symmetry, which is the reason we shall choose to concentrate on this case. It can be conveniently described in field theory as the SCFT UV completion of the $5d$ gauge theory made from an $SU(2)$ vector multiplet and $N_f$ fundamental hypermultiplets. We note that the methods used here can also be used to analyze the compactifications of other $5d$ SCFTs, as long as they have a deformation leading to a $5d$ gauge theory.

We shall also concentrate on the cases where the surfaces are tori and tubes, that is spheres with two punctures, with flux in their global symmetry supported on the surface. This choice is motivated by the strategy used to study the compactification of $6d$ SCFTs to $4d$ on Riemann surfaces. There, the starting point is to first consider the compactification on these surfaces, which can be handled by reducing along the circle to $5d$. Once determined for a family of theories, it is possible to use this to also study theories resulting from the compactification on trinions, that is spheres with three punctures, and from there to general surfaces.

While many steps in the computation follow through the application of the methods used in the study of compactifications to four dimensions, there are several novel elements that make an appearance due to the fact that we now get $3d$ theories. Particularly, in $3d$ gauge theories we can add Chern-Simons terms to the Lagrangian. Additionally, $3d$ gauge theories have a non-perturbative sector of operators, the so called monopole operators, which can be used to introduce interactions through monopole superpotentials. As we shall see, both of these new aspects appear and play an important role in the construction.

Once we have motivated a potential $3d$ theory, we shall test it using various methods. This will be done by considering the theories associated with tori. The $5d$ construction suggests that these should have a specific symmetry, the commutant of the flux in the global symmetry of the $5d$ SCFT. However, the $3d$ Lagrangian models generally manifest less symmetry. An important test then is to show that the symmetry in the IR is consistent with that expected from the geometric picture. This is done by studying supersymmetric partition functions, specifically the superconformal index and the $\mathbb{S}^3$ partition function. The former is used to check that the BPS operators indeed form characters of the expected symmetry, up to possible merging of operators. The latter is used to evaluate central charges and check that these are consistent with the enhancement. 

Moreover, the $5d$ picture has several implications on the operator spectrum that can be checked using the superconformal index. For instance, the theory is expected to inherit a conformal manifold from the compactification, and we can use the superconformal index to check for its existence. Additionally, there are operators associated with the currents broken by the flux, and these lead to special relevant and irrelevant operators. An extra consistency check then is to verify that we indeed see these extra states in the representations that would correctly enhance the expected global symmetry to that of the $5d$ SCFT. 

We do note that most of our consistency checks are preformed on theories associated to tori compactification, and not to the tubes themselves. The main consistency check for the tubes is rather indirect, that the theories associated to tori which we build from them pass the necessary consistency checks for many different models. While this gives a strong indication that these theories indeed correspond to the claimed tubes, it still leaves open the possibility that the two theories differ by some modification that cancels out once the theories are glued to a closed surface. Additionally, the checks mainly involve predictions for symmetry enhancement, dualities, and the existence of a certain special subsector of operators. This leaves open the possibility of modifying this theory by something that does not affect these, like the addition of singlet fields. As such, the best we can claim is that these theories are closely related to the theories engineered through the compactification of the rank $1$ $E_{N_f+1}$ $5d$ SCFTs on a torus with flux, at worst differing from these by minor modifications.

The structure of this paper is as follows. We begin in section \ref{compgen} by studying the compactification of the $5d$ Seiberg $E_{N_f+1}$ SCFTs on tubes with fluxes for the global symmetry, and construct the $3d$ $\mathcal{N}=2$ theories expected to be the result of this procedure. Then, after discussing the consistency checks one can perform to test such compactifications, we turn to sections \ref{2t}, \ref{4t} and \ref{addc} where we consider tori with different fluxes as the compactifying surfaces. These cases are obtained by gluing tubes together, and reveal the role played by Chern-Simons terms and monopole superpotentials. In section \ref{2t} we consider tori obtained by gluing two copies of the basic tube together, while in section \ref{4t} the tori are constructed from four such tubes corresponding to a higher value of flux. Section \ref{addc} contains several additional cases of interest, including tori obtained by gluing more than four copies of the basic tube together or by gluing just a single such tube to itself. Then, in section \ref{gr} we discuss gluing rules in general and refer to the novel elements we encountered that are not present in compactifications to $4d$, and later conclude in section \ref{conc}. Several appendices include additional information and notations.

\section{The compactification of $5d$ SCFTs to $3d$ $\mathcal{N}=2$ theories}
\label{compgen}

We shall begin by motivating the $3d$ $\mathcal{N}=2$ theories that we expect result from the compactification of the $5d$ $E_{N_f+1}$ theories on tubes, that is spheres with two punctures. While we shall try to make the discussion self-contained, the analysis is heavily motivated by similar methods used in the study of the compactification of $6d$ $(1,0)$ SCFTs to $4d$. Specifically, we can consider the compactification of $6d$ $(1,0)$ SCFTs on a Riemann surfaces, which should lead to a $4d$ theory at low-energies. In order to preserve some supersymmetry, we couple the R-symmetry bundle to a background connection equal to the spin connection on the Riemann surfaces so that the two will cancel for some of the supercharges. This is usually refereed to as twisting, and with it we can generically preserve four supercharges, leading to $\mathcal{N}=1$ supersymmetric theories in $4d$. It is also possible to turn on fluxes in the global symmetry of the $6d$ SCFT without breaking additional supersymmetry\cite{Chan:2000qc,RVZ}.

These types of relations were studied extensively in the context of $6d$ $(2,0)$ SCFTs\cite{Gai,BTW,BBBW}, and recently much progress has been made in the study of similar relations also for purely $6d$ $(1,0)$ SCFTs\cite{Gaiotto:2015usa,RVZ,Bah:2017gph,KRVZ,KRVZ1,KRVZ2,Razamat:2018gro,Zafrir:2018hkr,Chen:2019njf,Razamat:2019mdt,Pasquetti:2019hxf,Razamat:2019ukg,Razamat:2020bix,Sabag:2020elc}. The culmination of this work is a general procedure to study and determine the resulting $4d$ theories. We shall rely on this procedure, and mimic it in an attempt to make progress on the study of compactifications of $5d$ SCFTs to $3d$ on Riemann surfaces. For more information on the many steps of the analysis as they apply to the study of the compactification of $6d$ SCFTs, we refer the interested readers to the original papers\cite{KRVZ,KRVZ1,KRVZ2}.     

Similarly to the study of the compactification of $6d$ SCFTs, we wish to study the compactification of $5d$ SCFTs on Riemann surfaces to $3d$, potentially with flux in their global symmetry supported on the Riemann surface. Like in the case of the compactification of $6d$ SCFTs, to preserve SUSY we need to perform a twist when compactifying on a curved Riemann surface. While we shall mostly concentrate on compactifications on flat surfaces, we still comment that to preserve SUSY on curved surfaces we need to couple the Cartan of the $SU(2)$ R-symmetry to a background connection equal to the spin connection. This allows us to preserve four supercharges in $3d$ leading to $\mathcal{N}=2$ theories. We can also add flux in the global symmetries of the $5d$ SCFT, without breaking the $3d$ $\mathcal{N}=2$ supersymmetry.

So far the discussion has been general. However, in the proceeding discussion we shall concentrate on a specific choice of both $5d$ SCFTs and compactification surfaces. For the choice of $5d$ SCFTs, we shall take the rank $1$ Seiberg $E_{N_f+1}$ theories introduced in \cite{SEI}. This choice is motivated as these are one of the simpler and most well studied $5d$ SCFTs, that furthermore also support a rich global symmetry. For the choice of compactification surface, we shall take tori and tubes, that is a spheres with two punctures, with flux in the global symmetry of the $5d$ SCFT supported on the surface. The choice of surfaces is motivated by the study of compactifications of $6d$ SCFTs. There the basic starting point is determining the theories corresponding to tubes. From them we can form theories associated with tori with flux by gluing tubes together to form a closed surface. There are then methods to determine theories associated to trinions, that is three-punctured spheres, once the tube theories have been determined for a family of theories, see \cite{Razamat:2019mdt,Razamat:2019ukg,Razamat:2020bix,Sabag:2020elc}. Knowledge of the trinion theories then allows us to determine the theories associated with compactifications on arbitrary surfaces. Here we shall take the first step by considering only the cases of tubes and tori, postponing thought on other surfaces for future work.   

We shall begin by discussing some of the properties of the Seiberg $E_{N_f+1}$ SCFTs. Following that, we shall begin considering the compactification of the $5d$ $E_{N_f+1}$ SCFTs to $3d$ on tubes with flux in their global symmetry.

\subsection{Properties of the $5d$ rank $1$ $E_{N_f+1}$ SCFTs}

Here we shall concentrate on the rank $1$ Seiberg $E_{N_f+1}$ theories. These are a family of $5d$ SCFTs, defined for $N_f=-1,0,1,...,7$. One of their defining properties is that they have an $E_{N_f+1}$ global symmetry\footnote{For $N_f=5,6$ and $7$ these are just the exceptional groups $E_6$, $E_7$ and $E_8$. For smaller values of $N_f$ these are identified with some combination of classical groups, specifically: $E_5=SO(10)$, $E_4=SU(5)$, $E_3=SU(2)\times SU(3)$, $E_2=U(1)\times SU(2)$, $E_1=SU(2)$.}. For $N_f \geq 0$ they are conveniently defined as the SCFT UV completions of the $\mathcal{N}=1$ supersymmetric gauge theories $SU(2)+N_f F$, where this shortened notation stands for an $SU(2)$ vector multiplet with $N_f$ fundamental hypermultiplets. Specifically, the $E_{N_f+1}$ SCFTs can be deformed by a mass deformation, breaking $E_{N_f+1}\rightarrow U(1)\times SO(2N_f)$. This initiates an RG flow leading at low-energies to the $SU(2)+N_f F$ $5d$ gauge theories. Here the $SO(2N_f)$ factor of the preserved symmetry is identified with the symmetry rotating the $N_f$ fundamentals, while the $U(1)$ is identified with the topological instanton symmetry. The gauge coupling of the low-energy theory is then associated with the mass deformation.

Besides the mass deformation sending the $5d$ SCFTs to the $SU(2)$ gauge theories, the $E_{N_f+1}$ SCFTs are also connected to one another via mass deformations. Specifically, there are mass deformations that initiate a flow from an $E_{N_f+1}$ SCFT to an $E_{N_f}$ SCFT. These are manifested in the gauge theories by the integrating out of fundamental matter. These flows proceed straightforwardly until we reach the $E_2$ SCFT. The curious thing at this stage is that there are two $5d$ SCFTs, dubbed the $E_1$ and $\tilde{E}_1$ theories, that one can go to. True to its name, the $E_1$ SCFT has an $E_1=SU(2)$ global symmetry, while the $\tilde{E}_1$ SCFT has only a $U(1)$ global symmetry. Both SCFTs UV complete the $5d$ gauge theory consisting of a pure $SU(2)$ $\mathcal{N}=1$ vector multiplet, but differ by the $\theta$ angle, which is a $\mathbb{Z}_2$ valued parameter available for pure $USp$ type $5d$ gauge theories\cite{SM}. Specifically, the $E_1$ SCFT UV completes the case with $\theta$ angle $0$, while the $\tilde{E}_1$ SCFT UV completes the case with $\theta$ angle $\pi$.

The mass deformations to the gauge theories of the $E_1$ and $\tilde{E}_1$ SCFTs are associated to the single Cartan factor of their global symmetry. Mass deformations in $5d$ $\mathcal{N}=1$ theories are real, and as such can be both positive and negative. For the case of the $E_1$ SCFT, as it has an $SU(2)$ global symmetry, both the positive and negative mass deformations give the gauge theory. This property, that both the positive and negative mass deformations lead to the same theory, also holds for the mass deformations leading to the $SU(2)$ gauge theories for the $E_{N_f+1}$ SCFTs with $N_f>0$\cite{Mitev:2014jza}, again due to the enhanced symmetry. This will be important to us later on. However, this is not the case for the $\tilde{E}_1$ SCFT. Here one sign gives the gauge theory, but the other leads to a new $5d$ SCFT. This $5d$ SCFT is dubbed the $E_0$ SCFT, has no global symmetry, and as such no SUSY preserving mass deformations. 

Here we shall mostly concentrate on the $E_{N_f+1}$ theories for $N_f\geq 0$. This follows as we shall rely on their gauge theory deformations to motivate the resulting $3d$ theories. Specifically, we shall rely on the expectation that these theories can be made to flow to the corresponding $4d$ theories by compactifying them with a suitable holonomy. To understand this we first consider what happens when we reduce theories with eight supercharges on a circle with an holonomy in a flavor symmetry. In these cases, the holonomy sits in the component along the circle reducing from $5d$ to $4d$, of a background vector field belonging to a $5d$ vector multiplet. Such $5d$ vector multiplets also contain a scalar field, whose vevs leads to mass deformations. 

In that case, we can take the holonomy to be in the vector field belonging to the same $5d$ vector multiplet in which resides also the scalar whose vev leads to the mass deformation sending the $E_{N_f+1}$ SCFTs to the $5d$ gauge theories $SU(2)+N_f F$. In such compactified theories, the component of the vector field along the compact direction and the real scalar join to form the complex scalar parameterizing mass deformations in $4d$ $\mathcal{N}=2$ theories. As such this deformation in $5d$ should be equivalent to the purely $5d$ mass deformation, and so we expect that the $5d$ $E_{N_f+1}$ SCFTs will flow to the $4d$ $SU(2)+N_f F$ gauge theories when compactified in the presence of this holonomy. We shall make use of this next when we discuss the compactification of the $E_{N_f+1}$ SCFTs to $3d$ on tubes.

\subsection{$4d$ compactification}

Our first step in analyzing the compactification of the $5d$ $E_{N_f+1}$ SCFTs on tubes is to reduce on the circle, leading to a $4d$ theory compactified on the interval. At the boundaries of the interval, corresponding to the boundaries of the tube, we will need to put boundary conditions, which we shall describe momentarily. One issue here is what happens to the flux in the global symmetry of the $5d$ SCFT supported on the surfaces when we perform this reduction. 

The central idea here, that we shall next describe, is that this flux can be implemented using a variable holonomy. Specifically, let us call the $4d$ compact direction $x_4$, and consider the boundaries to be at $x_4 = \pm a$. Then we consider an holonomy inside a $U(1)$ subgroup of the global symmetry such that its value at $x_4 = a$ is say $m$, but its value at $x_4 = -a$ is $-m$. Furthermore, we shall take it to be constant everywhere except at $x_4=0$ where it jumps, and as such we have flux located at that point, leading to flux supported on the surface. This is a convenient choice to take as it allows us to relate the flux to domain walls. This approach has been extensively used in the study of compactifications of $6d$ SCFTs, and its success there suggests that the actual profile of the flux has an either irreverent or at best marginal effect on the resulting $4d$ theory, so we do not lose anything in taking this specific profile. We shall assume here that this continues to hold also for the case of compactifications of $5d$ SCFTs.

The presence of the holonomy leads to two $4d$ theories. One at $x_4>0$ and the other at $x_4<0$, while at $x_4=0$ we have a domain wall separating the two. The type of $4d$ theories we have at the two segments depends on the choice of holonomy, and thus the choice of flux. Here we shall make a special choice and choose this holonomy to be the one that is expected to give the $4d$ gauge theory $SU(2)+N_f F$, as explained previously. As such, in the $4d$ picture, we expect the $4d$ theory at both $x_4>0$ and $x_4<0$ to be the gauge theory $SU(2)+N_f F$. 

The nice property of this choice is that it leads to Lagrangian gauge theories at almost all points of the $4d$ spacetime. This facilitates the study of the boundary conditions and the determination of the resulting $3d$ theories. For $N_f>4$ the $4d$ theories away from the domain wall are IR free and so expected to be weakly coupled. However, for $N_f=4$ the gauge coupling is marginal, and becomes relevant for $N_f<4$ leading to potentially strongly coupled $4d$ interactions. This seems to lead to a difficulty in the study of the resulting $3d$ theories for $N_f\leq 4$, on which we shall elaborate in the later sections.

This specific choice of holonomy leads to a special choice of flux. To specify the flux, we need to choose a Cartan basis of the global symmetry and specify the flux in each Cartan element. For our considerations here, it is convenient to choose a basis manifesting the $U(1)\times SO(2N_f)$ subgroup of $E_{N_f+1}$, which is the one preserved by the gauge theory deformation that we are frequently using. Additionally, we shall span the $SO(2N_f)$ part by its $SO(2)^{N_f}$ subgroup. As such, we shall denote the flux by $(F_{U(1)};F_{SO(2)_1},F_{SO(2)_2},...,F_{SO(2)_{N_f}})$, where we use $F_{U(1)}$ for the flux in the $U(1)$ part of $U(1)\times SO(2N_f)$, and $F_{SO(2)_i}$ for the flux in the $N_f$ independent $SO(2)$ subgroups of $SO(2N_f)$. In this basis, as we shall see in the later sections, the flux associated with this specific choice of holonomy appears to be $(\frac{\sqrt{8-N_f}}{4};\underbrace{\frac{1}{4},\frac{1}{4},...,\frac{1}{4}}_{N_f})$. We note here that the square root in the flux comes from our chosen normalization of the roots of $E_{N_f+1}$. Specifically, we normalize one of the $U(1)$ Cartans such that the minimal charge under it is proportional to the square root. This flux is then one of the roots of $E_{N_f+1}$ and more specifically, is one of the roots enhancing $U(1)\times SO(2N_f)$ to $E_{N_f+1}$. As such we could also choose to represent it by any root of $E_{N_f+1}$, that is equivalent to it by the action of the Weyl group. Here, we have chosen this specific representation for reasons that will become apparent later. We refer the reader to appendix \ref{App:flux} for more information on the fluxes and our chosen conventions for them.

With this choice we have two $\mathcal{N}=2$ $SU(2)+N_f F$ gauge theories separated by a domain wall at $x_4=0$. Next we need to consider the reduction to $3d$, but before that we need to discuss the boundary conditions.  

\subsubsection{Boundary conditions}

Next we discuss the boundary conditions that we can enforce at the two boundaries. As we are interested in having $\mathcal{N}=2$ theories in $3d$, we shall concentrate on boundary conditions preserving 4 supercharges. For this we consider the $4d$ fields, that we have in the bulk, close to the boundary. Specifically, we have $4d$ $\mathcal{N}=2$ vector multiplets and hypermultiplets. Close to the boundary, the former decomposes to the $3d$ $\mathcal{N}=2$ vector and adjoint chiral multiplets, while the latter decomposes to two chiral fields in conjugate representations. Boundary conditions, preserving $3d$ $\mathcal{N}=2$ SUSY, are then given by assigning Neumann and Dirichlet boundary conditions to these.

In general there are many possible choices of such boundary conditions. Here we shall not concentrate on classifying all types of boundary conditions. Instead, we consider a specific type of boundary conditions. These are the analogues of the boundary conditions that correspond to maximal punctures in compactifications of $6d$ SCFTs to $4d$, see the discussion in \cite{KRVZ,KRVZ1,KRVZ2}. In terms of the $3d$ fields, it is given by assigning Dirichlet boundary conditions to the vector in the vector multiplet and Neumann boundary conditions to the adjoint chiral. For the hypermultiplets, we give one of the chiral fields Dirichlet boundary conditions and Neumann boundary conditions to the other. Here we have a choice of which field to give which boundary condition, which translate to a discrete label associated with the puncture, which is usually refereed to as the sign or color of the puncture. 

\subsubsection{Conjecturing the $3d$ theories}

Having discussed the boundary conditions, we next move to discuss how we can conjecture the $3d$ theories. For this we first consider the fields coming from the bulk. There we have a Lagragian gauge theory which we can analyze. For $N_f>4$ the gauge theory is IR free, and we need not consider strong coupling effects. Other values of $N_f$ might involve such effects. Nevertheless, we shall here only analyze the Lagrangian gauge theory. The Dirichlet boundary conditions for the vector field at the two boundaries mean that it becomes non-dynamical. As such we expect two $SU(2)$ global symmetry groups, coming from the bulk $SU(2)$ gauge symmetry, associated with the two punctures.

The Dirichlet boundary conditions are also assigned to one of the chiral fields in the hypermultiplet and so it is not expected to contribute to the $3d$ theory. The other chiral field in the hypermultiplet, that receives Neumann boundary conditions, is expected to contribute. These should give chiral fields in the bifundamental representation of $SU(2)\times SU(N_f)$, where $SU(2)$ is the global symmetry associated with the puncture they are connected to, while the $SU(N_f)$ is part of the global symmetry inherited from the $5d$ theory. We expect additional fields coming from the domain wall, and so, properly understanding the resulting $3d$ theory necessitates understanding the fields and boundary conditions present at the domain wall.

To make progress, we shall assume that the domain walls for the theories considered here behave similarly to the domain walls that appear in the study of compactifications of $6d$ SCFTs to $4d$. Specifically, we consider the case of compactifications of the E-string $6d$ SCFT, which is the closest analogue to the theories considered here. The domain walls involved in that case were studied in \cite{KRVZ}. In there, it was found that the fields living on the domain walls are an $SU(2)\times SU(2)$ bifundamental, interpolating between the two $SU(2)$ groups associated with the two punctures, and a singlet chiral field. These interact with the other fields and with themselves as follows. There is a cubic superpotential involving the $SU(2)\times SU(2)$ bifundamental coming from the domain wall, and the two $SU(2)\times SU(N_f)$ bifundamentals that survive at the two boundaries. Additionally, there is a cubic superpotential connecting the singlet field and the quadratic $SU(2)\times SU(2)$ invariant made from the bifundamental. Finally, the domain wall assigns Dirichlet boundary conditions to the adjoint chiral field in the vector multiplet and so it is not expected to survive in the low-energy theory.

\begin{figure}
\center
\includegraphics[width=0.35\textwidth]{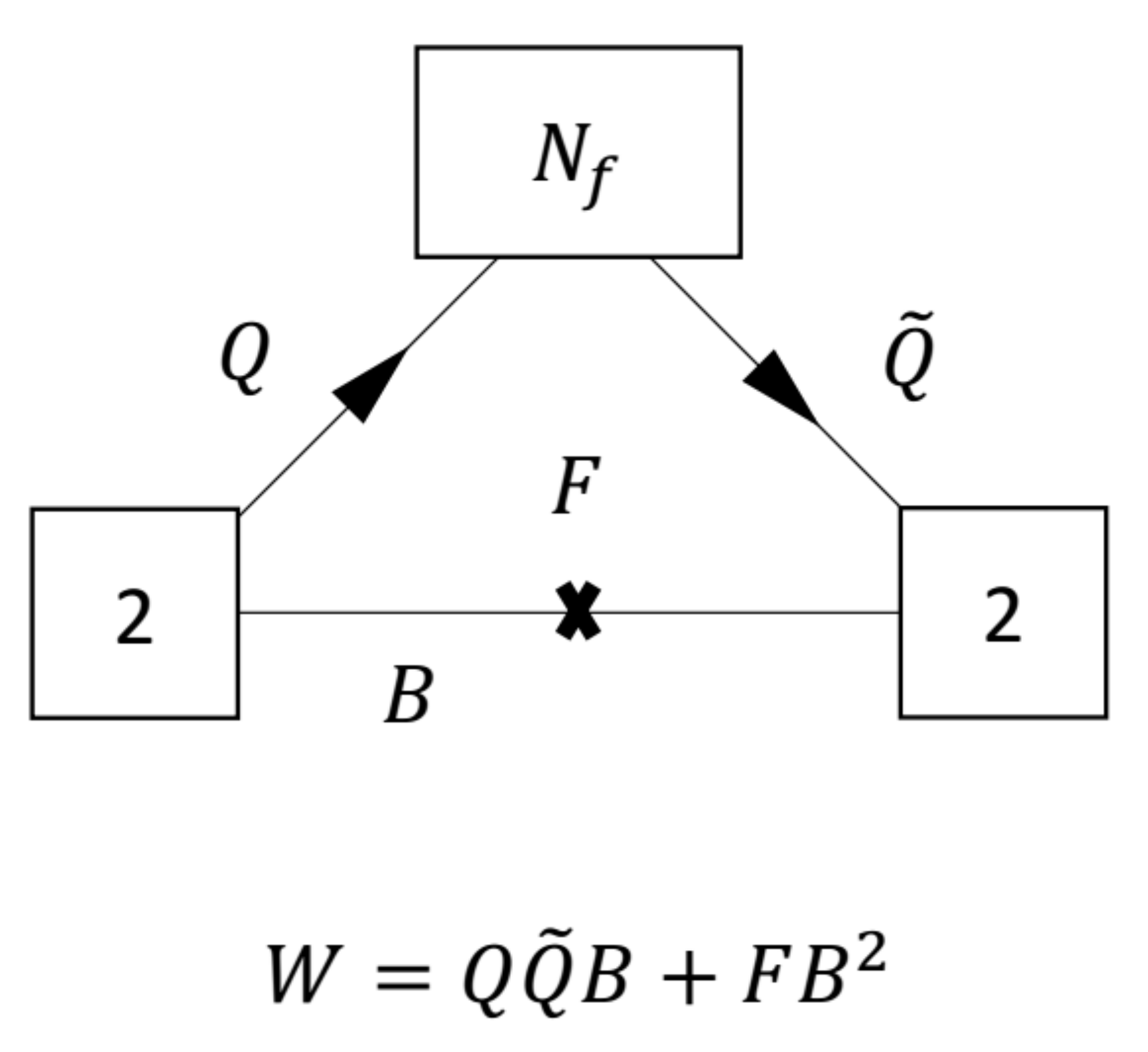} 
\caption{The conjectured $3d$ theory associated with the compactification of the rank $1$ $5d$ $E_{N_f+1}$ SCFT on a tube with two maximal punctures and flux $(\frac{\sqrt{8-N_f}}{4};\frac{1}{4},\frac{1}{4},...,\frac{1}{4})$. As customary, squares denote $SU$ type global symmetries, and lines connecting them denote bifundamental chiral fields under the connected groups. We also use crosses over fields to denote flipping, that is that there is an additional chiral field linearly coupled to the invariant made from the field being flipped under the non-abelian global symmetries. Finally, $W$ denotes the superpotentials, with the last term being the flipping one.}
\label{BTube}
\end{figure}

Assuming the domain walls considered here behave similarly to those studied in \cite{KRVZ}, we are lead to conjecture the $3d$ theory shown in figure \ref{BTube}. These are expected to be the result of the compactification of the rank $1$ $5d$ $E_{N_f+1}$ SCFT on a tube with two maximal punctures and flux. As we shall motivate later the specific value of flux is $(\frac{\sqrt{8-N_f}}{4};\underbrace{\frac{1}{4},\frac{1}{4},...,\frac{1}{4}}_{N_f})$.    

We could have equally chosen any other value consistent with the Weyl group. Specifically, the last $N_f$ numbers represent the flux under the $SO(2N_f)$ subgroup of the global symmetry and we are free to act on these via the Weyl group of $SO(2N_f)$, which is given by the permutation of these $N_f$ values and the multiplication by $-1$ of any even number of them. The reason to consider this specific subgroup of the full Weyl group is that the $SO(2N_f)$ subgroup of the global symmetry is manifest in the gauge theory living in the bulk, and so its action on the flux also has an effect on the gauge theory that will be of use to us later on.

Specifically, the permutation symmetry acts by permuting the $N_f$ doublets in the bulk $SU(2)+N_f F$ gauge theory, while the operation of multiplying an element by $-1$ acts by exchanging the two chiral fields in the hypermultiplets. This latter operation is of interest to us. The reason for this is that the boundary conditions generally give Dirichlet boundary conditions to one of them and Neumann boundary conditions to the other. Thus, tubes with flux of $(\frac{\sqrt{8-N_f}}{4};\underbrace{\pm \frac{1}{4},\pm \frac{1}{4},...,\pm \frac{1}{4}}_{N_f})$, with even number of minus signs, are all equivalent, but with different component of the $N_f$ bulk doublet hypermultiplets receiving the Neumann and Dirichlet boundary conditions. This is true both at the domain wall, and at the two punctures, which as such have differing signs. 

Next, we can take two such tubes and glue them together to form the theory corresponding to the compactification on a torus with flux $(\frac{\sqrt{8-N_f}}{2};\underbrace{\frac{1}{2},\frac{1}{2},...,\frac{1}{2}}_{N_f})$. The gluing is done along two punctures in the glued tube, and involves gauging the symmetries of the punctures. Additionally, we need to also introduce a bifundamental in the gauge $SU(2)$ and $SU(N_f)$ global symmetry, $\phi$, and couple it to the fields $Q$ or $\tilde{Q}$ of the punctures through the superpotential: $\phi (Q_1-Q_2)$, with the sub-index indicating the tube the field is associated with. These rules stem from the fact that when we glue we essentially eliminate the boundary and so must reintroduce the fields that were eliminated by the boundary conditions. Specifically, here these are the $SU(2)$ gauge field and the other chiral in the $N_f$ doublet hypermultiplets. Additionally the fields $Q_i$, associated with the other component of the doublet hypers, must be identified which is achieved by the superpotential.

When gluing we are also faced with several options that do not exist in $4d$. Specifically, besides the Yang-Mills term, we can also introduce a Chern-Simons term for the gauge fields. Additionally, we can introduce superpotentials involving monopole operators. As we shall see when we explicitly study the model, it appears that we need to introduce Chern-Simons terms with value of $\frac{6-N_f}{2}$ and alternating signs, and introduce superpotentials corresponding to monopole operators with unit charges in all two adjacent $SU(2)$ gauge groups. This will be motivated in the next section when we explicitly study the $3d$ theories associated with the compactified theory. We shall return to the question of the exact gluing rules later on.

We can also consider gluing more tubes to generate torus compactifications corresponding to higher values of flux preserving the same symmetry. Next, we will be interested in using the basic tube to generating tubes associated with other values of flux. This follows the strategy used in \cite{KRVZ,KRVZ1,KRVZ2} to generate more general tubes by a special gluing of the basic tube. As done so far, we shall adopt the strategy of assuming this model works similarly to the models studied previously to derive potential tube theories, and then test them for consistency.  

In the study of $6d$ compactifications, particularly that of the rank $1$ E-string SCFT in \cite{KRVZ}, it was noted that given the basic tube one can construct more general tubes as follows. First, we can glue two tubes together to generate a new tube with twice the value of flux of the original tube. As previously covered, to do this we need to gauge the puncture symmetry and reintroduce the fields eliminated by the Dirichlet boundary conditions, which we have collectively denoted as $\phi$. 

\begin{figure}
\center
\includegraphics[width=1.05\textwidth]{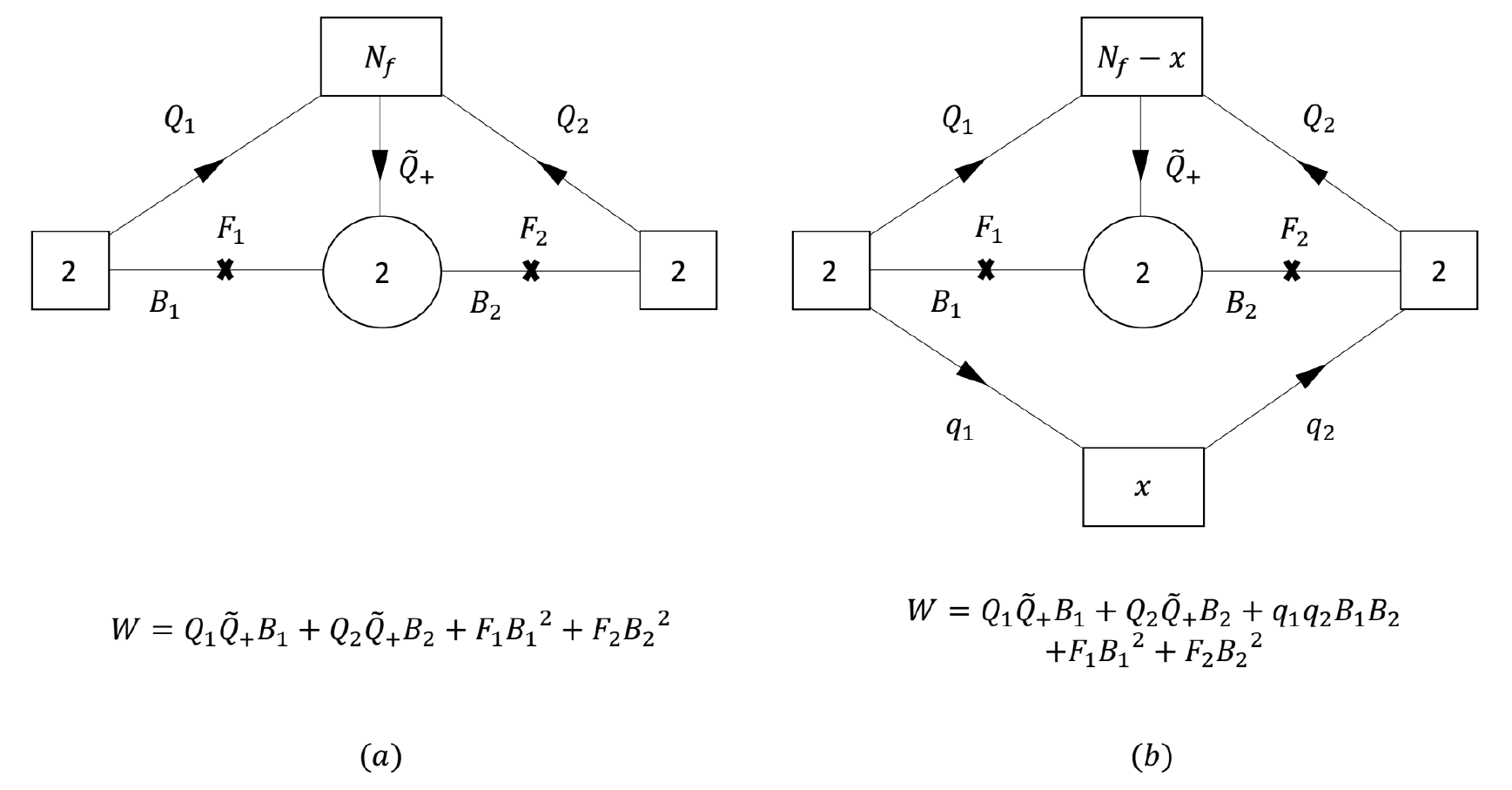} 
\caption{The conjectured $3d$ theory associated with the compactification of the rank $1$ $5d$ $E_{N_f+1}$ SCFT on a tube with two maximal punctures for various values of flux. Here we use the same conventions for denoting quiver theories, with the addition of using circles to denote gauge groups, again as customary. The case in (a) comes from gluing the basic tube in figure \ref{BTube} to itself, while the case in (b) comes from gluing it to an equivalent one made by acting on it with a specific Weyl group element. Like in the previous case, W denotes the superpotentials, with the last terms being the flipping ones.}
\label{DTubes}
\end{figure}

This leads to the theory shown in figure \ref{DTubes} (a). This theory can be associated to the configuration involving two domain walls, of the type we covered previously, with two maximal punctures, one at each end of the tube. The $SU(2)$ gauge symmetry comes from the dynamical gauge field living in the interval between the two domain walls, which receives Neumann boundary conditions at the two domain walls. Similarly, the $N_f$ bifundamentals between this gauge $SU(2)$ and the global $SU(N_f)$ come from the chiral fields in the $SU(2)+N_f F$ gauge theory living on the interval between the two domain walls, which receives Neumann boundary conditions at the two domain walls.

As we pointed out we can get equivalent tubes by acting with the Weyl symmetry, but these differ by the choice of boundary conditions and by the flux representation in our chosen basis. This allows us to generate more general tubes by gluing together two tubes with slightly different boundary conditions. For this, consider the case where we again have two domain walls on the interval, one of the type we considered previously but the other related to it by a Weyl action of $-1$ on $x$ fields\footnote{Here $x$ is even as the Weyl action of $D$ type groups does not contain the action with an odd number of minus signs.}. 

We can next derive the $3d$ theory expected from this configuration from our conjecture about the domain wall. The resulting theory is shown in figure \ref{DTubes} (b). Like the previous case, the $SU(2)$ gauge group and the two $SU(2)$ global symmetry groups come from the gauge field in the bulk. Specifically, as the gauge field is given Neumann boundary conditions on the domain walls, the gauge field between them survives in $3d$. However, at the boundaries the gauge field receives Dirichlet boundary conditions leading to it becoming non-dynamical at the segments connecting to them. This in turn leads to the $4d$ $SU(2)$ gauge symmetries living on these segments becoming non-dynamical, and so global symmetries, in $3d$. 

We also have the contribution of the $SU(2)\times SU(N_f)$ bifundamental hypers. As we mentioned, half of these receive Dirichlet boundary conditions while half receive Neumann boundary conditions. For the $N_f-x$ hypers, associated with the elements on which we did not act with the Weyl group, the fields receiving which boundary condition are the same across the two domain walls. As such these reduce as in the previous case. However, for the $x$ hypers, associated with the elements on which we did act with the Weyl group, the boundary conditions differ between the two domain walls so that they don't survive in the $3d$ limit. Despite this, in the segments ending on the boundaries, these fields receive the same boundary conditions on the edges and on the domain walls leading to these fields contributing in $3d$, although the surviving fields being charged in opposite representations of the $SU(x)$ symmetry rotating them as they come from opposite components in the bifundamental. Additionally, we expect to get $SU(2)\times SU(2)$ bifundamentals, together with their flipping fields, coming from the domain walls. Summing everything up, we end up with the quiver in figure \ref{DTubes} (b).

The fields also interact through various superpotentials. Specifically, we expect cubic superpotentials associated with the two upper triangles, and coupling the two flipping fields to their respective quadratic $SU(2)\times SU(2)$ bifundamental invariant. Finally, from analogy with the cases of $6d$ compactifications, we expect a quartic superpotential associated with the lower triangles. There can also be $3d$ specific additions, like Chern-Simons terms and monopole superpotentials. Specifically, we shall later see when we explicitely study the model that we need to introduce a Chern-Simons term of value $\frac{6-N_f}{2}$ for the $SU(2)$ gauge group. We do not appear to need to introduce superpotentials to the tubes, though as previously mentioned, these are needed once the tubes are glued together.  

Finally, we need to consider the flux associated with these tubes. We expect it to be given by summing up the fluxes associated with the two domain wall theories. We recall that for the first domain wall we associated the flux $(\frac{\sqrt{8-N_f}}{4};\underbrace{\frac{1}{4},\frac{1}{4},...,\frac{1}{4}}_{N_f})$, while for the second one we have associated the flux $(\frac{\sqrt{8-N_f}}{4};\underbrace{\frac{1}{4},\frac{1}{4},...,\frac{1}{4}}_{N_f-x},\underbrace{-\frac{1}{4},-\frac{1}{4},...,-\frac{1}{4}}_{x})$. As such, we expect the flux associated with the tubes in figure \ref{DTubes} (b) to be $(\frac{\sqrt{8-N_f}}{2};\underbrace{\frac{1}{2},\frac{1}{2},...,\frac{1}{2}}_{N_f-x},\underbrace{0,0,...,0}_{x})$.

This completes are conjectural derivation of the various building blocks that we shall use to build our $3d$ models. Next, we shall study the behavior of these models and compare them against our $5d$ expectations, but before that we need to discuss what type of tests we can perform on these models.

\subsection{Tests}

We have so far motivated a derivation of the $3d$ theories expected to be the result of the compactification of the $5d$ $E_{N_f+1}$ SCFTs on tubes with specific fluxes in their global symmetries. However, the derivation has some conjectural aspects, as well as several aspects that were left undetermined. To fully determine the theories, and have confidence in the claims, we need to perform some consistency checks, and we shall next discuss the tests that can be performed. 

As previously mentioned, the derivation of the theories was motivated by similar methods used in the study of the compactification of $6d$ SCFTs. Similarly, the tests we shall perform are also motivated by similar tests used in that context. However, not all the tests used as consistency checks for the study of the compactification of $6d$ SCFTs generalize also to the case of $5d$ SCFTs. We shall next review the tests used in the study of the compactification of $6d$ SCFTs, and discuss their utility and implications for the case studied here.


\subsubsection{Anomalies}

One commonly used test in the study of the compactification of $6d$ SCFTs is to compare 't Hooft anomalies. Specifically, the $6d$ SCFT has 't Hooft anomalies that can be encoded in an anomaly polynomial $8$-form. Similarly, $4d$ theories have 't Hooft anomalies that can be encoded in an anomaly polynomial $6$-form. If the $4d$ and $6d$ theories are related via compactification, then we have that the $4d$ anomaly polynomial $6$-form is given by integrating the $6d$ anomaly polynomial $8$-form on the Riemann surface\cite{BTW}. This allows us to compute the anomalies of the resulting $4d$ theories from those of the $6d$ SCFT, at least for those anomalies involving the symmetries visible in $6d$.

This is an extremely useful consistency check as it allows us to make quantitative checks on the proposed theories. However, in theories of odd dimensions, there are no 't Hooft anomalies for continuous symmetries, so we cannot use this test. Instead, we shall mostly need to rely on other consistency checks. There is, however, an interesting exception that we shall next elaborate on although we shall not make use of it here.

The point is that while there are no 't Hooft anomalies for continuous symmetries, there can be 't Hooft anomalies involving discrete symmetries, including both standard 0-form symmetries and higher form symmetries. It is then possible for these to be related across dimensions. This in principle is true also for the case of compactifications of $6d$ theories, though it has not been studied. This of course requires a study of the discrete 0-form symmetries, higher form symmetries and their anomalies for $5d$ SCFTs, a subject that deserves further study\footnote{Nevertheless, there are some known results on these subjects. See for instance \cite{Zafrir:2016wkk} for discrete 0-form symmetries of $5d$ SCFTs, \cite{Morrison:2020ool,Albertini:2020mdx,Bhardwaj:2020phs} for higher form symmetries of $5d$ SCFTs, and \cite{BenettiGenolini:2020doj} for anomalies involving these symmetries.}.

The 't Hooft anomalies in continuous symmetries in even dimensional theories are related to central charges of the theories, which are numbers appearing in various n-point correlation functions involving the currents. These are also related to various terms appearing in the sphere partition function. While 't Hooft anomalies for continuous symmetries don't exist in odd dimensional theories, the central charges, defined either using correlation functions of the currents or the sphere partition function, do exist. The previous discussion implies that the central charges in even dimensional theories are related in theories connected via compactification. One can then speculate that this might be true also for odd dimensional theories. If so, then it can be used instead of anomaly matching, assuming an efficient way to compute the relation between the two can be found. 

Still, the computation of the central charges will be of use to us to check for symmetry enhancement. Specifically, the $5d$ picture leads to various expectations for symmetry enhancement in the $3d$ theory. If two symmetries are to form a larger symmetry, then the central charges need to be compatible with this merging. We can then use this as an additional test. For examples of this type of consistency checks in other odd dimensional theories, see for instance \cite{Chang:2017cdx,Gang:2018huc}.         

\subsubsection{Symmetries and general behavior}

Another set of consistency checks is given by comparing other physical properties against those expected from the higher dimensional picture. Specifically, we mentioned that different, but equivalent ways of constructing the same surface should give equivalent theories. We also mentioned that the theory is expected to manifest a global symmetry given by the commutant of the flux in the global symmetry of the higher dimensional theory. Therefore, an important consistency check is to see that this is indeed obeyed. Additionally, the higher dimensional origin leads to interesting predictions on the conformal manifold and the presence of special operators that can also be checked.

These tests have been extensively used in the study of compactifications of $6d$ SCFTs, and we expect this to still hold also for the case of compactifications of $5d$ SCFTs. As such, these will be the main tests that we employ as consistency checks to verify that the proposed $3d$ models have the necessary properties, and to fill in some of the open details in the derivation. Specifically, as we are dealing with tubes, we don't have that much freedom in terms of pair of pants decompositions. However, we can build theories corresponding to torus compactifications with many different values of fluxes, and for each of these cases the IR theory is expected to manifest the global symmetry preserved by the flux. Additionally, it is possible to build theories corresponding to fluxes related by a Weyl transformation. These are then expected to give the same IR theory although the $3d$ UV descriptions can be different. An important consistency check is to verify that the resulting theories can indeed be dual. This is a novel way to generate dualities via higher dimensional compactifications that is different from a pair of pants decomposition.

These expectations of symmetry enhancements and dualities can then be checked by calculating various RG invariant quantities. Here we shall mainly use the superconformal index, which is a weighted counting of the BPS operators of the theory. Symmetry enhancement can then be checked as it implies that the index forms characters of the enhanced symmetry. Similarly, dualities can be checked as they imply that the indices of the dual theories are equal. Finally, other types of consistency checks that we shall mention here, involving the presence of specific types of operators, can also be performed as these usually contribute to the index. Besides the index, we shall also use the $\mathbb{S}^3$ partition function, notably the central charges that can be computed from it, to test such expectations. For a review of the superconformal index, $\mathbb{S}^3$ partition function and central charges, we refer the reader to appendix \ref{App:PF}. 

Another important consistency check is to study the structure of the conformal manifold. Specifically, the higher dimensional construction implies the existence of marginal operators, and thus leads to a prediction for the dimension of the conformal manifold. While these were originally derived for the case of $6d$ compactifications on Riemann surfaces, the arguments also hold for the case of $5d$ compactifications on Riemann surfaces. We next briefly review these expectations, referring the interested reader to the literature for more details\cite{BTW,RVZ}.    

For simplicity, we shall concentrate on the case of a torus with no punctures, which will be the main case that we shall use to test our proposal. For these cases marginal operators come generically from two sources. One is the complex structure moduli of the torus, which descend to marginal deformations in some cases, notably, the case of the compactification of the $6d$ $(2,0)$ theory to the $4d$ maximally supersymmetric Yang-Mills theory. The second source are flat connections, notably holonomies around the two cycles of the torus. We shall next discuss each one in turn.

The complex structure moduli is expected to give a one dimensional conformal manifold preserving supersymmetry. While its appearance as a marginal operator is familiar in the case of the compactification of the $6d$ $(2,0)$ theory to the $4d$ maximally supersymmetric Yang-Mills theory, it is in fact not a generic property. For instance, if we compactify the E-string $6d$ SCFT on a torus to $4d$ then we expect to get the Minahan-Nemeshansky $E_8$ SCFT\cite{Ganor:1996pc}, which do not have an $\mathcal{N}=2$ preserving conformal manifold. As such in some cases the complex structure moduli of the torus does not contribute a marginal operator. This appears to also extend for flux compactifications of the E-string SCFT on a torus\cite{KRVZ} and to similar compactifications of related theories\cite{KRVZ1,KRVZ2,Zafrir:2018hkr}. Similarly, we shall see that we do not observe a marginal operator coming from the complex structure moduli of the torus when we study the models we propose. This is not inconsistent with the behavior of related models in the study of compactification of $6d$ SCFTs to $4d$ $\mathcal{N}=1$ theories. In fact having the complex structure moduli contributing a marginal operator in $3d$ might be problematic. This follows as the resulting marginal deformation should preserve all supersymmetry, and as such in the case without flux, where all eight supercharges are preserved, would lead to an $\mathcal{N}=4$ preserving marginal deformation. However, such deformations are known to be in conflict with superconformal representation theory, and so cannot exist\cite{Cordova:2016xhm}.

This brings us to consider holonomies, which first were considered in this context in \cite{BTW}. We can turn on holonomies around the two cycles of the torus. These are not independent, but must obey the homotopy group relation of the surface, which for the torus means they must commute. As such they can be simultaneously diagonalized, leading to $2rank(G)$ real numbers, where $rank(G)$ is the rank of the global symmetry group of the higher dimensional SCFT that commutes with the flux, here denoted as $G$. These preserve only four supercharges, and are observed to lead to marginal operators preserving only $\mathcal{N}=1$ SUSY in the study of compactifications of $6d$ SCFTs on Riemann surfaces to $4d$. This leads to a conformal manifold of dimension $rank(G)$, where only four supercharges are preserved, and the global symmetry on generic points is broken down to its Cartan subalgebra.

There is an alternative viewpoint of this that is useful. Specifically, we can associate with the two real holonomies a single complex marginal operator in the $4d$ $\mathcal{N}=1$ theory. Since it originates in an holonomy, there are actually $dim(G)$ such operators, for $dim(G)$ the dimension of $G$, and these in fact transform in the adjoint representation of $G$. However, while these operators are marginal, they are not necessarily exactly marginal. Specifically, in theories with four supercharges, for a marginal operator to be exactly marginal, they must form a non-trivial Kahler quotient with respect to the global symmetry of the theory\cite{GKSTW}, here expected to be $G$. The dimension of the quotient gives the dimension of the resulting conformal manifold, with the remaining marginal operators becoming marginally irrelevant by recombining with the conserved currents of the global symmetries broken by the marginal deformation. For the case at hand, the Kahler quotient of the adjoint of a symmetry $G$ is indeed of order $rank(G)$, and leads to the previously determined structure.

So far the discussion has been geared toward the much more well studied case of flux compactifications of the $6d$ SCFTs to $4d$ $\mathcal{N}=1$ theories. However, everything said here should also apply for the case of flux compactifications of the $5d$ SCFTs to $3d$ $\mathcal{N}=2$ theories, and indeed we shall employ this as a test for the $3d$ models we previously determined.

There are two important loopholes in the arguments so far that we ought to mention. First, there might be additional marginal operators besides the ones coming from the sources listed above. The special features of the marginal operators we mentioned is that they are generic, and are not dependent on the intricacies of the higher dimensional theory. A second issue that may occur is that the global symmetry manifested in lower dimensions can be larger than $G$, that is there is an accidental enhancement of symmetry. If the marginal operators are charged under these symmetries, then some of them may become marginally irrelevant. As such the higher dimensional expectations on the conformal manifold may fail, as the actual conformal manifold may end up larger, if there are additional marginal operators, or smaller, if there are accidental symmetries. The importance of these expectations is not that they always hold, but rather that they hold generically, that is for generic Riemann surfaces and values of flux. Deviations from these are encountered in cases of small flux, but these usually vanish and the expectation is again observed at larger values of fluxes. We shall see examples of this when we study the various $3d$ models.

Finally, when we discussed the holonomies we considered them only in the global symmetry that commutes with the flux. However, we could also consider holonomies in the global symmetries broken by the flux. These have the interesting property of also appearing in the lower dimensional theory but as special cases of relevant and irrelevant operators. Next we review the expectations for how these operators should look like in the lower dimensional theory, particularly through the superconformal index. For a detailed derivation we refer the reader to \cite{BRZtoapp}.   

The general idea is as follows. Consider a higher dimensional theory, which could be a $6d$ or a $5d$ SCFT, with a global symmetry $G'$. These SCFTs generically contain two types of multiplets. One is the energy-momentum tensor multiplet that any SCFT possesses. The second, are the conserved current multiplets of the group $G'$. These are both BPS multiplets in their respective dimensions. Consider the compactification of the SCFT on a Riemann surface. We can ask what BPS operators do these operators give in the lower dimensional theories. This is interesting as these operators are quite generic, and so understanding their reduction provides us with general predictions of the presence of certain operators in the lower dimensional theory.

The main point of \cite{BRZtoapp} is that instead of looking at the general number of such operators, we should look at the contribution of these operators to the superconformal index. While the individual numbers may be difficult to compute, the contribution to the superconformal index, which only counts operators modulo merging, can be related to topological properties of the compactification surface via the Atiyah-Singer index theorem. We refer the readers to \cite{BRZtoapp} for more details. The operators that we mentioned coming from holonomies can be understood in this frame as coming from the higher-dimensional conserved current multiplets, while those associated with complex structure moduli can be understood in this frame as coming from the higher-dimensional energy-momentum tensor multiplet.   

We next describe the results. Say we compactify said theory on a Riemann surface of genus $g$ with flux $F$ in a $U(1)$ subgroup of $G'$ \footnote{For simplicity we assume that there is flux only in one $U(1)$. The generalization to the case of flux in multiple $U(1)$ groups is straightforward.}, which we shall denote as $U(1)_{\alpha}$. Additionally, we shall assume that there are no punctures on the surface. The presence of the flux breaks $G'$ to $U(1)_{\alpha} \times \tilde{G}$. We can then break the adjoint character of $G'$ as follows:

\be \label{Gdecomp}
\chi_{adj} (G') = \sum_i \alpha^{q_i} \chi_{R_i} (\tilde{G}) ,
\ee
where $q_i$ is the $U(1)_{\alpha}$ charge of the $\tilde{G}$ representation $R_i$ appearing in the decomposition of the adjoint representation of $G'$. Here we have used $\alpha$ as the fugacity of $U(1)_{\alpha}$. While the representations $R_i$ depend on the choice of $U(1)_{\alpha}$, they always contain the adjoint of $\tilde{G}$ and a singlet corresponding to the adjoint of $U(1)_{\alpha}$. 

The statement that is of use to us here is that for a generic punctureless Riemann surface and flux $F$ the index of the lower dimensional theories has a special form when written using the $U(1)_R$ symmetry that is the Cartan of the $SU(2)_R$ symmetry, which is the R-symmetry for both $6d$ $(1,0)$ and $5d$ SCFTs. For the case of $3d$ $\mathcal{N}=2$ theories, this form is:

\bea \label{Indexexp}
I = & 1 & + (\sum_{i|q_i>0} \alpha^{q_i} \chi_{R_i} (\tilde{G}) (g-1+q_i F))x^2 + \left(3g-3 + (1+\chi_{adj} (\tilde{G}))(g-1)\right)x^2 \\ \nonumber & + & (\sum_{i|q_i<0} \alpha^{q_i} \chi_{R_i} (\tilde{G}) (g-1+q_i F))x^2 + ...
\eea 

Here the first and last terms are associated with holonomies in the symmetries broken by the flux, while the middle term gives the contributions of the holonomies in the symmetry that commutes with the flux, as well as the complex structure moduli. More correctly, the first and last terms are associated with the contribution of BPS operators originating from the components of the $5d$ conserved current multiplet that are charged under $U(1)_{\alpha}$. The second term in the middle term gives the contribution of BPS operators originating from the components of the $5d$ conserved current multiplet that are uncharged under $U(1)_{\alpha}$, while the first term in the middle term gives the contribution of BPS operators originating from the $5d$ energy-momentum tensor multiplet. Note in particular that $3g-3$ coincides with the complex structure moduli if $g>1$. For the case of $g=1$, which is the case of interest to us here, we have that there are no marginal operators, in accordance with the many observations we mentioned previously. 

All of them contribute as marginal operators under the R-symmetry inherited from the higher dimensional theory. However, this symmetry is not the actual superconformal R-symmetry, which involves mixing with $U(1)_{\alpha}$. This causes these operators to behave differently under it. Specifically, the first group of operators in (\ref{Indexexp}), that is the ones with positive $U(1)_{\alpha}$ charge, are expected to have R-charge less than $2$, and so give relevant operators. As such, the last group of operators in (\ref{Indexexp}), that is the ones with negative $U(1)_{\alpha}$ charge, are expected to have R-charge greater than $2$, and so give irrelevant operators. Finally, the middle group of operators in (\ref{Indexexp}), which are the ones uncharged under $U(1)_{\alpha}$, are expected to have R-charge $2$, and so give the marginal operators under the superconformal R-symmetry.

We can connect the discussion here to our previous discussion on the conformal manifold by noting several properties of the $3d$ $\mathcal{N}=2$ index noted in \cite{Razamat:2016gzx}, with similar observations for the case of the $4d$ $\mathcal{N}=1$ index\cite{GB}. Specifically, for a $3d$ SCFT with no free fields, superconformal representation theory forbids the appearance of negative terms of the form $x^s$ for $s<2$. Negative terms, however, are allowed for $s\geq 2$, where for $s=2$ they can only come from global symmetry currents. As such the $x^2$ term in the index counts the number of marginal operators minus conserved currents. Indeed, at that order we see in (\ref{Indexexp}) the $3g-3$ marginal operators coming from the complex structure moduli of the torus, the $(1+\chi_{adj} (\tilde{G}))g$ marginal operators coming from holonomies and the $-(1+\chi_{adj} (\tilde{G}))$ coming from the conserved currents of the global symmetry. This also implies that there is a stark difference between the first and last terms in (\ref{Indexexp}). As we mentioned the first terms should have R-charge smaller than $2$, and so negative terms cannot appear. This also forbids cancellations, implying that these index contributions faithfully count the number of such operators. This is not true for the last term, where the index, and as such also our higher dimensional expectations, only gives information on the difference between families of operators.

\section{Studying basic models}
\label{2t}

In the previous sections we have motivated several $3d$ models as the results of the compactification of the $5d$ $E_{N_f+1}$ SCFTs on a sphere with two punctures in the presence of flux in its global symmetry. We have also summarized the various expectations for the lower dimensional theory due to its higher dimensional origin. In the next few sections, we shall put these expectations to the test by explicitly studying the resulting $3d$ models.

This will serve two purposes. First, it will provide evidence for our claim that these models have a higher dimensional origin. Second, there are several parts in the derivation that were determined through such tests. These include the flux of the basic tube, the Chern-Simons level of the gauge groups and the presence of monopole superpotentials. As such, we shall show that the proposed $3d$ models indeed give results consistent with the proposed higher dimensional origin, justifying our previous claims.

As previously noted, for our tests we shall need to consider theories associated with torus compactifications. To do this we shall take some combination of the tubes we previously discussed and glue them together to form a theory that we associate with a torus. As we know the flux associated with each tube, by summing it up we shall get the flux associated with the full surface. We can then check the global symmetry, conformal manifold and the existence of the operators expected from the $5d$ broken currents in these theories.

Finally, we want to comment on the issue of fractional flux. In general the flux needs to be quantized so as to be an integer. However, in many cases it is possible to have fractional fluxes if these are accompanied by a non-trivial background for some of the preserved non-abelian symmetries. This background has the effect of breaking part of this symmetry. This is manifested in the gluing process by the fact that the gluing will break part of the $5d$ originated global symmetry. For a detailed discussion of this, we refer the reader to appendix C of \cite{KRVZ}. We shall initially avoid this, and concentrate only on cases where the full $5d$ global symmetry that commutes with the flux is preserved.     

\subsection{Gluing basic tubes together}

The first test we can consider is related to the theories that we can build by gluing several copies of the basic tube, the one in figure \ref{BTube}. This should lead to the $3d$ theory associated with the compactification of the $5d$ $E_{N_f+1}$ SCFTs on a torus with flux $n(\frac{\sqrt{8-N_f}}{4};\frac{1}{4},\frac{1}{4},...,\frac{1}{4})$, where $n$ is the number of tubes used. This flux corresponds to $\frac{n}{2}$ units of flux in a $U(1)$ Cartan of an $SU(2)$ subgroup of the $E_{N_f+1}$ global symmetry. As such the preserved symmetry, in addition to the $U(1)$ Cartan of the $SU(2)$, is the commutant of $SU(2)$ in $E_{N_f+1}$. These are: $E_7$ for $N_f=7$, $SO(12)$ for $N_f=6$, $SU(6)$ for $N_f=5$, $SU(2)\times SU(4)$ for $N_f=4$, $U(1)\times SU(3)$ for $N_f=3$, $U(1)\times SU(2)$ for $N_f=2$ and $U(1)$ for $N_f=1$. We refer the reader to appendix \ref{App:flux} for an explanation on how the global symmetry can be read from the flux.

For $n$ odd, the flux is fractional and the global symmetry is further reduced. This can be seen in the tube as in that case the two $SU(2) \times SU(N_f)$ bifundamentals at the two edges that are glued together are in conjugate representations under $SU(N_f)$ so performing the gluing will break this symmetry to a real subgroup. As such, for now we shall concentrate on the case of $n$ even, the simplest case being $n=2$, which is case we consider in this section.

\begin{figure}
\center
\includegraphics[width=0.75\textwidth]{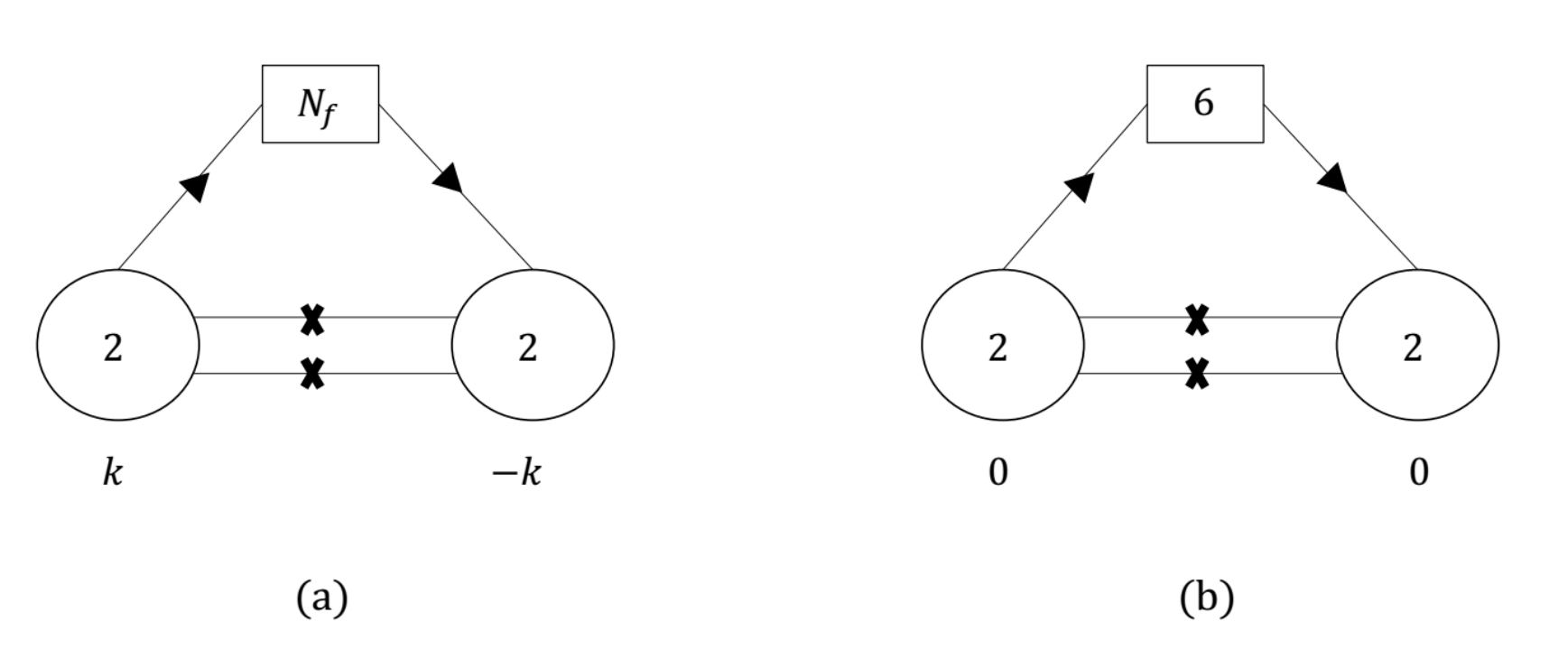} 
\caption{The $3d$ quiver theory associated with the compactification of the $5d$ $E_{N_f+1}$ SCFTs on a torus with flux $\frac{1}{2}(\sqrt{8-N_f};1,1,...,1)$. The theory in (a) shows the case of generic $N_f$, with $k=\frac{6-N_f}{2}$, while (b) is specialized to the $N_f=6$ case. The theory also contains various superpotential interactions. Notably, there are cubic superpotentials running along the triangle as well as the flipping superpotentials.}
\label{3dquivers}
\end{figure}

We then proceed by taking two of the tubes in figure \ref{BTube} and gluing them together following the previously outlined procedure. Specifically, we glue the tubes together by gauging the puncture $SU(2)$ of each of them with an $SU(2)$ vector multiplet and $N_f$ chiral fields, which are connected by a linear superpotential to the $SU(2) \times SU(N_f)$ bifundamentals connected to the gauged puncture $SU(2)$ groups. The first gluing leads to the theory shown in figure \ref{DTubes} (a). The second one is done by gluing the two $SU(2)$ groups at the edges of the quiver. The resulting quiver gauge theory for arbitrary $N_f$ is depicted in figure in figure \ref{3dquivers} (a). There are also cubic superpotential terms that the theory inherits from the tubes, notably, one connecting the fundamental and bifundamental chiral fields, and one connecting the two singlet chiral fields to quadratic invariants made solely from the bifundamentals. Besides these there may also be superpotentials involving monopole operators, on which we shall for the moment remain agnostic about.    

Finally, we need to consider the possibility of Chern-Simons terms. As we shall soon see, for the case of $N_f=6$, it appears that the Chern-Simons term should be zero for both groups. For $N_f$ smaller than that we expect to have the Chern-Simons term $\frac{6-N_f}{2}$ for one group and $\frac{N_f-6}{2}$ for the other\footnote{In order for the Chern-Simons term to be gauge invariant we need to have an integrally quantized Chern-Simons renormalized level. This is obtained by summing to the bare one the 1-loop contribution of the fermions. For $3d$ $\mathcal{N}=2$ theories, when the number of chirals in the fundamental of $SU(2)$ is even the bare Chern-Simons level has to be integer in order for this to happen, while when the number of fundamental chirals is odd the Chern-Simons level has to be half-integer \cite{Aharony:1997bx,Intriligator:2013lca}. In our cases this condition is always satisfied since we take the bare Chern-Simons levels to be $\pm\frac{N_f-6}{2}$.}. The case of $N_f=7$ presents some issue, on which we will comment later on.

An interesting aspect of $5d$ SCFTs compared to $6d$ SCFTs is that the former can be connected via mass deformations. Specifically, we can flow from SCFTs with one value of $N_f$ to those with a smaller value of $N_f$ using supersymmetry preserving mass deformations. These mass deformations are specified by giving a vev to a real scalar in a background vector multiplet coupled to the flavor symmetry. It is natural to relate this to $3d$ mass deformations. In that context, the relation between theories with different values of $N_f$ seems natural: when we integrate out some of the flavors with a mass term with the same sign we produce some Chern-Simons level. In principle, it should be possible to also integrate out the flavors using mass deformations with different signs, and it is an interesting question what happens in that case, which we shall nevertheless reserve for future study.

Next we shall explore the behavior of these models, for various values of $N_f$.        

\subsubsection{$N_f=6$}
\label{2DWNf6}

The first case we consider is the $N_f=6$ one. The model is given in figure \eqref{3dquivers} (b). There are two $U(1)$ global symmetries, which we denote as $U(1)_a$ and $U(1)_b$, as well as an $SU(6)$ global symmetry. The charges for the fields are chosen such that the $6$ chiral fields transforming under the left $SU(2)$ carry charge $1$ under $U(1)_b$, while those transforming under the right $SU(2)$ carry charge $-1$, with the rest uncharged. Under $U(1)_a$, the fundamental chiral fields carry charge $1$, the bifundamentals charge $-2$ and the singlet fields charge $4$. 
Additionally, there is a $U(1)_R$ R-symmetry group, that we will take to be that under which all fields carry R-charge $\frac{2}{3}$. The actual superconformal R symmetry, denoted by $U\left(1\right)_{\hat{R}}$, is obtained using F-maximization (see appendix B) and is given by a mixing of $U(1)_R$ and $U(1)_a$. Under this symmetry, the singlet fields turn out to violate the unitarity bound (having R charge smaller than 0.5), and the value of $U\left(1\right)_{\hat{R}}$ in the theory obtained after removing them is as follows, 
\begin{equation}
\hat{R}=R-0.085q_{a}\,.
\end{equation}

Let us compute the index of the model with the singlet fields using the reference R-symmetry $U(1)_R$. We find:

\bea \label{IndexE7SO12n2}
\mathcal{I}&=&1+2\,a^4x^{\frac{2}{3}}+\left(\frac{2}{a^4}+a^2(b^6+\frac{1}{b^6}+b^2{\bf 15}+\frac{1}{b^2}{\bf \overline{15}})+3a^8\right)x^{\frac{4}{3}}+\nn\\
&+&\left(4a^{12}+2a^6(b^6+\frac{1}{b^6}+b^2{\bf 15}+\frac{1}{b^2}{\bf \overline{15}})\right)x^2+\left(-\frac{1}{a^2}(b^6+\frac{1}{b^6}+b^2{\bf 15}+\frac{1}{b^2}{\bf \overline{15}})+\right.\nn\\
&+&\left.\frac{4}{a^8}+5a^{16}+3a^{10}(b^6+\frac{1}{b^6}+b^2{\bf 15}+\frac{1}{b^2}{\bf \overline{15}})+a^4(b^{12}+\frac{1}{b^{12}}+b^4{\bf \overline{105}}+\frac{1}{b^4}{\bf 105}+\right.\nn\\
&+&\left.b^8{\bf 15}+\frac{1}{b^8}{\bf \overline{15}}+{\bf 189}-{\bf 35})\right)x^{\frac{8}{3}}+\cdots\,.
\eea

Here the terms at order $x^{\frac{2}{3}}$ come from the singlet fields. The terms at order $x^{\frac{4}{3}}$ with charges $\frac{1}{a^4}$ come from the unflipped quadratic invariant made from the bifundamentals and the $(1,1)$ monopole\footnote{Remember that the lattice of magnetic fluxes for $SU(2)$ is $\mathbb{Z}_+$. Here and for the rest of the paper, by ``$(1,1)$ monopole" we mean the monopole operators corresponding to a unit of magnetic flux for two adjacent $SU(2)$ gauge nodes in the quiver and zero flux for the others.}. The terms with charge $a^2$ come from the quadratic invariant made from the fundamentals and the $(1,0)$ and $(0,1)$ monopoles. 

Recall our claim, that this theory is the result of the compactification of the $5d$ rank $1$ $E_7$ SCFT on a torus with flux. Specifically, the flux is in a $U(1)$ whose commutant in $E_7$ is $SO(12)$, and its value is $1$ in a normalization such that the minimal charge is $1$. There are several pieces of evidence supporting this. First we note that the index forms characters of $SO(12)$, where the embedding is such that:

\be
\label{emb12}
\bold{12} \rightarrow \frac{1}{b^2}\bold{6} \oplus b^2\overline{\bold{6}} \,.
\ee 
In terms of characters of $SO(12)$, the index reads:
\bea \label{IndE7}
\mathcal{I}&=& 1 + 2 a^4 x^{\frac{2}{3}} + \left( \frac{2}{a^4} + a^2 \bold{32} + 3a^8\right)x^{\frac{4}{3}}  +  \left(4a^{12} + 2a^6 \bold{32} \right)x^2 + \nn\\
&+& \left(\frac{4}{a^8} - \frac{1}{a^2}\bold{32} + a^4 (\bold{462}-\bold{66}) + 3 a^{10} \bold{32} + 5a^{16}\right)x^{\frac{8}{3}} + \cdots\, .
\eea

Moreover, the terms in the index follow the pattern expected from compactification of higher dimensional theories. Specifically, as previously noted, we expect to get marginal operators, under the R-symmetry given by the Cartan of the $5d$ $SU(2)$ R-symmetry, coming from the $5d$ stress-energy tensor and conserved current. The $5d$ stress-energy tensor should give $3g-3$ such operators, which in our case is zero. The conserved current should contribute $g-1+q F$, where $F$ is the value of the flux, and $q$ is the charge of the specific state under the symmetry with the flux. Here we adopt the normalization convention where the minimal charge is $1$.

The states coming from the conserved current in the adjoint representation of $E_7$ split into representations of the $SO(12)\times U(1)$ subgroup according to the branching rule
\be\label{E7adjBR}
{\bf 133}\to{\bf 1}^0\oplus {\bf 66}^0\oplus{\bf 1}^{\pm2}\oplus{\bf 32}^{\pm1}\,.
\ee
In our case, the $5d$ R-symmetry should be such that the bifundamental have R-charge $0$ and the fundamentals have R-charge $1$. This is expected from analogy with the case of $6d$ compactifications, where this can be checked explicitly using 't Hooft anomalies, see \cite{KRVZ}. This can also be argued as follows. Recall that the fundamentals come from bulk $4d$ hypermultiplets receiving Neumann boundary conditions at the boundaries. As such, these should have R-charge of $\pm 1$, if we use the normalization where the minimal charge is $1$. Here, we will choose to normalize this R-symmetry so that the charges of these fields are all $1$. The charges of the bifundamental and the flip fields is then dictated by the superpotentials. 

This $5d$ originated R-symmetry can be related to the R-symmetry we used in our computation with the shift $a\to a\,x^{\frac{1}{3}}$ in the index. Moreover, the $U(1)$ for which we turned on a unit of flux is related to the $U(1)_a$ of our model by a normalization of $\frac{1}{2}$.  With this dictionary, we can immediately identify in the index \eqref{IndE7} the states ${\bf 1}^{2}$ and ${\bf 32}^{\pm1}$ in the decomposition \eqref{E7adjBR}: they are the contributions $2\,a^4x^{\frac{2}{3}}$, $a^2{\bf 32}x^{\frac{4}{3}}$ and $-a^{-2}{\bf 32}x^{\frac{8}{3}}$. The state ${\bf 1}^{-2}$ contributes $-2\,a^{-4}x^{\frac{10}{3}}$ to the $3d$ index so it appears above the highest order we reported in \eqref{IndE7}, but we checked for its presence. 

Another evidence for the proposed enhancement of the $SU\left(6\right)\times U\left(1\right)_{b}$ part of the symmetry to $SO(12)$ comes from examining the central charges of these symmetries. Following the discussion in appendix B (see in particular Eq. \eqref{d2F}), we compute numerically the real part of the free energy of the model as a function of the mixing coefficients of certain $U(1)$ symmetries with the R symmetry. Then, calculating the second derivatives at the superconformal point (that is, where the mixing coefficients take the values corresponding to the IR R symmetry) yields the central charges of these $U(1)$ symmetries. Applying this procedure for $U(1)_b$ and for the Cartan $\textrm{diag}\left(1,0,0,0,0,-1\right)$ of $SU(6)$, which we denote by $C$, we find the following values for the central charges, 
\begin{equation}
C_{b}=28.4407\,\,\,,\,\,\,C_{C}=2.36998\,.
\end{equation}
The value of the ratio of these charges is
\begin{equation}
\label{ratioC}
\frac{C_{b}}{C_{C}}=12.0004,
\end{equation}
which exactly matches our expectations from the proposed symmetry enhancement and the embedding \eqref{emb12}. To understand why this is the case, let us consider more generally the embedding of a group $H$ into a group $G$ under which the representation $\boldsymbol{R}_{G}$ of $G$ decomposes into representations of $H$ in the following way: 
\begin{equation}
\boldsymbol{R}_{G}\rightarrow\sum_{i}\boldsymbol{R}_{H}^{(i)}\,.
\end{equation}
Then, the central charges of $G$ and $H$ are related as follows, 
\begin{equation}
C_{H}=I_{H\rightarrow G}C_{G}\,,
\end{equation}
where $I_{H\rightarrow G}$ is known as the embedding index and is given by 
\begin{equation}
\label{embI}
I_{H\rightarrow G}=\frac{\sum_{i}T_{\boldsymbol{R}_{H}^{(i)}}}{T_{\boldsymbol{R}_{G}}}
\end{equation}
where $T_{\boldsymbol{R}}$ is the Dynkin index of the representation $\boldsymbol{R}$. In the case where the group $H$ is a $U(1)$, $T_{\boldsymbol{R}_{H}^{(i)}}$ in \eqref{embI} is replaced by $q^2_i$ where $q_i$ are the charges of the states in the representation $\boldsymbol{R}_{G}$ under that $U(1)$.

Applying this to the embedding $SU\left(6\right)\times U\left(1\right)_{b}\rightarrow SO\left(12\right)$ under which \eqref{emb12} is satisfied, we find the embedding indices 
\begin{equation}
I_{SU\left(6\right)\rightarrow SO\left(12\right)}=1\,,\,\,\,I_{U\left(1\right)_{b}\rightarrow SO\left(12\right)}=48\,,\,\,\,I_{U\left(1\right)_{C}\rightarrow SU\left(6\right)}=4\,,
\end{equation}
implying the following relations between the various central charges, 
\begin{equation}
C_{SU\left(6\right)}=C_{SO\left(12\right)}\,,\,\,\,C_{b}=48C_{SO\left(12\right)}\,,\,\,\,C_{C}=4C_{SU\left(6\right)}\,.
\end{equation}
This, in turn, results in the ratio
\begin{equation}
\label{CbCc12}
\frac{C_{b}}{C_{C}}=12
\end{equation}
which agrees with our numerical result \eqref{ratioC}.

Finally, we want to consider the conformal manifold. While we have seen that the index is consistent with our $5d$ expectations, it is illuminating to actually consider the marginal operators and study the structure of the conformal manifold expected from them. As we noted the number of marginal operators minus conserved currents can be read from the terms at order $x^2$ in the index. Looking at \eqref{IndexE7SO12n2}, we see that that the $x^2$ term under the superconformal R-symmetry vanishes. This follows as all terms appearing there are charged under $U(1)_a$, which mixes with the R-symmetry. 

As we should have the $U(1)_a \times U(1)_b \times SU(6)$ conserved currents, there should be $37$ marginal operators transforming in the adjoint of the $U(1)_a \times U(1)_b \times SU(6)$ global symmetry that cancel them in the index. This should lead to a $7$ dimensional conformal manifold along which the global symmetry is broken to its Cartan subalgebra. This would fit our expectations from $5d$, and we would further speculate that the symmetry enhances to $U(1)\times SO(12)$ on a $1d$ subspace on the conformal manifold, as expected from $5d$, which is consistent with the index and central charges.

However, there is a problem with this picture, that we shall now address. To see it we need to consider the origin of the various marginal operators, which are two singlets and an $SU(6)$ adjoint. The latter comes from the cubic superpotential along the triangle, when the indices are contracted so as to be in the adjoint of the global symmetry\footnote{There are two of these differing by the $SU(2)\times SU(2)$ bifundamental used, but one of them can recombine with the broken $SU(6)$, acting on the two pairs of six fundamentals in the same representation, to form a long multiplet.}. The former comes from the operator made from the product of the $(1,1)$ monopole operator and one of the two flip fields. However, we have noted that the flip fields go below unitarity, and so are expected to decouple in the IR. This leads to an enhanced symmetry acting on them, and as the two marginal operators we mentioned contain these, they are also charged under this enhanced symmetry. As such, these operators are actually marginally irrelevant in the IR theory.

 Looking at the conformal manifold of the interacting theory then, we see that we only have one marginal operator in the adjoint of the $SU(6)$. This leads to a $5$ dimensional conformal manifold, on a generic point of which only $U(1)_a$, $U(1)_b$ and the Cartan subalgebra of the $SU(6)$ is preserved. The change in the dimension of the conformal manifold also implies that the theory cannot have an enhanced $SO(12)$ at some point on it. The argument is as follows. Say such a point exist, then at that point there must be $SO(12)$ conserved currents, leading to a contribution in the index. However, as the index is invariant, and we do not see the contribution of the currents in the index, then these must be canceled by a marginal operator in the adjoint of the $SO(12)$. However, the latter would imply that there is at least a six dimensional conformal manifold, which contradicts the structure we observed from the Lagrangian description\footnote{This is under the assumption that there are no additional accidental symmetries in the IR limit of the theory in figure \ref{3dquivers} (b). The enhancement to $SO(12)$ might be consistent, however, if one relaxes this assumption. Specifically, if we assume the existence of an accidental $U(1)$ symmetry, then similar arguments would force an additional marginal operator, raising the expected dimension of the conformal manifold to six, which would be consistent with a point with enhanced $SO(12)$ global symmetry.}. Nevertheless, subgroups of $SO(12)$ with rank $5$ or less, preserving all the Cartan symmetries are consistent. Therefore, there can be points with $SO(10)$, $SU(2)\times SO(8)$, $SU(4)\times SU(3)$ and $SU(4)\times SU(2)^2$ global symmetries.

This provides an example of the loophole we previously mentioned, in which the existence of accidental symmetries in the IR leads to part of the conformal manifold expected from the higher dimensional construction becoming unacceptable. Unfortunately, we do not have any good understanding as to why this occurs for this particular case, though we shall soon see that this is a special feature of the $N_f=6$ case with minimal flux preserving $U(1)\times SO(12)$, with more generic cases behaving as expected. It would be interesting if this exceptional behavior can be better understood, but we would not pursue this here.

Finally, we note that despite this, the index and central charges are all compatible with the existence of the enhanced symmetry. This is quite remarkable and consistent with the higher dimensional construction. Specifically, we would expect the operator spectrum to be in characters of this symmetry as it originates from operators of a higher dimensional theory possessing this symmetry. It is somewhat stranger for the central charges. However, in analogous situations in $4d$, like the example in \cite{KRVZ}, the compatibility of the central charges is understood as these are related to the anomalies of the $4d$ theory, that are in turn determined by the anomalies of the $6d$ SCFTs. It is tempting to think that similar though more complicated relations might exist between the central charges of the $3d$ theory and the parent $5d$ theory.

There is another interesting interpretation of this. While we mentioned that the enhancement to $SO(12)$ seems incompatible, enhancements to rank $5$ subgroups of it, which can also be expected from the $5d$ construction, seem consistent. If points with these enhanced symmetries indeed exist, then the compatibility with $SO(12)$ can be understood as stemming from the need of the index and central charges to be simultaneously compatible with all these different symmetries.  

\subsubsection{$N_f=5$}
\label{2DWNf5}

We next turn to consider the case of $N_f=5$. The symmetries here are the same as in the previous case (except the $SU(6)$ part which is now $SU(5)$), and the superconformal R-charge is given by 
\begin{equation}
\hat{R}=R-0.04068q_{a}
\end{equation}
where $U(1)_R$ is the reference R-symmetry under which all the fields carry R-charge $\frac{2}{3}$, as before. Notice that in contrast to the previous case of $N_{f}=6$, here the singlet fields do not violate the unitarity bound. Moreover, we also take the Chern-Simons levels to be $(\frac{1}{2},-\frac{1}{2})$.

Let us now compute the index of the model using the reference R-symmetry $U(1)_R$. We find:
\bea
\mathcal{I} &=& 1 + 2 a^4 x^{\frac{2}{3}} +  \left( \frac{3}{a^4} + a^2 (b^2 \bold{10} + \frac{1}{b^2}\overline{\bold{10}} ) + 3a^8 \right)x^{\frac{4}{3}} +  \left(4a^{12} + 2a^6 (b^2 \bold{10} + \frac{1}{b^2}\overline{\bold{10}}) \right)x^2\nn\\
&+&\left(-a^4(1+b^4{\bf \overline{5}}+\frac{1}{b^4}{\bf 5}+{\bf 24})-\frac{1}{a^2}(b^2{\bf 10}+\frac{1}{b^2}{\bf \overline{10}})+\frac{6}{a^8}+5a^{16}+3a^{10}(b^2{\bf 10}+\frac{1}{b^2}{\bf \overline{10}})+\right.\nn\\
&+&\left.a^4(b^4{\bf \overline{50}}+\frac{1}{b^4}{\bf 50}+{\bf 75})\right)x^{\frac{8}{3}} +\cdots \,.
\eea
We claim that this theory is the result of the compactification of the $5d$ rank $1$ $E_6$ SCFT on a torus with flux. Specifically, the flux is in a $U(1)$ whose commutant in $E_6$ is $SU(6)$, and its value is $1$ in a normalization such that the minimal charge is $1$. There are several pieces of evidence supporting this. First we note that the index forms characters of $SU(6)$, where the embedding is such that:
\be
\label{emb6}
\bold{6} \rightarrow \frac{1}{b^{\frac{2}{3}}}\bold{5} \oplus b^{\frac{10}{3}}\,.
\ee 
In terms of characters of $SU(6)$, the index reads:
\bea \label{IndE6}
\mathcal{I} &=& 1 + 2 a^4 x^{\frac{2}{3}} + \left ( \frac{3}{a^4} + a^2 \bold{20} + 3a^8\right)x^{\frac{4}{3}} + \left (4a^{12} + 2a^6 \bold{20} \right)x^2 +\nn\\
&+& \left(\frac{6}{a^8} - \frac{1}{a^2}\bold{20} + a^4 (\bold{175}-\bold{35}) + 3 a^{10} \bold{20} + 5a^{16}\right)x^{\frac{8}{3}} + \cdots \,.
\eea
Here we note that despite the fact that all of the terms in characters in the index come from the perturbative states, the monopole operators are essential to get the character structure. Specifically, the $(1,0)$ and $(0,1)$ monopoles contribute terms at order $x^{\frac{8}{3}}$ in the $\bold{5}$ and $\overline{\bold{5}}$, which cancel similar terms coming from the perturbative states. Without this cancellation the index would not form $SU(6)$ characters. Here also the Chern-Simons terms are important.  

Like in the previous case, the structure of the index follows the expectations from $5d$ compactifications. The states coming from the conserved current in the adjoint representation of $E_6$ split into representations of the $SU(6)\times U(1)$ subgroup according to the branching rule
\be\label{E6adjBR}
{\bf 78}\to{\bf 1}^0\oplus{\bf 35}^0\oplus{\bf 1}^{\pm2}\oplus{\bf 20}^{\pm1}\,.
\ee
Using again the dictionary we worked out in the previous section between $5d$ and $3d$ symmetries, we can immediately identify in the index \eqref{IndE6} the states ${\bf 1}^{2}$ and ${\bf 20}^{\pm1}$ in the decomposition \eqref{E6adjBR}: they are the contributions $2\,a^4x^{\frac{2}{3}}$, $a^2{\bf 20}x^{\frac{4}{3}}$ and $-a^{-2}{\bf 20}x^{\frac{8}{3}}$. The state ${\bf 1}^{-2}$ contributes $-2\,a^{-4}x^{\frac{10}{3}}$ to the $3d$ index so it appears above the highest order we reported in \eqref{IndE6}, but we checked its presence.

Another evidence for the proposed enhancement of the $SU(5)\times U(1)_{b}$ part of the symmetry to $SU(6)$ comes from examining the central charges of these symmetries, as before. Following the discussion in appendix B, we compute numerically the real part of the free energy of this model, and use it to find the following values for the central charges of $U(1)_b$ and of the Cartan $\textrm{diag}\left(1,0,0,0,-1\right)$ of $SU(5)$, which we denote again by $C$,
\begin{equation}
C_{b}=11.3734\,\,\,,\,\,\,C_{C}=1.70603\,.
\end{equation}
The value of the ratio of these charges is
\begin{equation}
\label{ratioCNf5}
\frac{C_{b}}{C_{C}}=6.66663,
\end{equation}
which again matches our expectations from the proposed symmetry enhancement and the embedding \eqref{emb6}. Indeed, the corresponding embedding indices are 
\begin{equation}
I_{SU\left(5\right)\rightarrow SU\left(6\right)}=1\,,\,\,\,I_{U\left(1\right)_{b}\rightarrow SU\left(6\right)}=\frac{80}{3}\,,\,\,\,I_{U\left(1\right)_{C}\rightarrow SU\left(5\right)}=4\,,
\end{equation}
implying the following relations between the various central charges, 
\begin{equation}
C_{SU\left(5\right)}=C_{SU\left(6\right)}\,,\,\,\,C_{b}=\frac{80}{3}C_{SU\left(6\right)}\,,\,\,\,C_{C}=4C_{SU\left(5\right)}\,.
\end{equation}
This, in turn, results in the ratio
\begin{equation}
\label{cbccsu6}
\frac{C_{b}}{C_{C}}=\frac{20}{3}
\end{equation}
which agrees with our numerical result \eqref{ratioCNf5}.

Finally, we can again consider the conformal manifold. The analysis closely follows the one done for $N_f=6$. From the superconformal index we again see that we have one marginal operator in the adjoint of $SU(5)$, and two which are singlets. Their origin in the quiver theory is similar to the ones in the previous section, with one difference, with the ones associated with the two singlets. These come from the product of the $(1,1)$ monopole and each of the two flip fields, but as in this case we have a Chern-Simons term the $(1,1)$ monopole is gauge charged so the operator that actually appears in this product is the gauge invariant made of the $(1,1)$ monopole and a state in the $SU(2)\times SU(2)$ bifundamental.   

These marginal operators should lead to a $6$ dimensional conformal manifold along which the global symmetry is broken to its Cartan subalgebra. This fits our expectations from $5d$, and we would further speculate that the symmetry enhances to $U(1)\times SU(6)$ on a $1d$ subspace on the conformal manifold, as expected from $5d$, which is consistent with the index and central charges. Note that in this case the flip fields are above the unitarity bound so we do not expect accidental symmetries in the IR. 

\subsubsection{$N_f=4$}  

We next consider the case of $N_f<5$. Here the situation is more subtle, since we start having monopole operators with non-positive R-charges. Specifically, for $N_f = 4$ the $(1,1)$ monopoles carry charge $0$ under all symmetries, including the R-symmetry for whatever R-charge we assign to the chirals. For smaller $N_f$ we can't find an R-symmetry such that both the $(1,1)$ monopole and the chirals have positive R-charge. One may think that this would cause some problems in the computation of the index, but it turns out that, because of the Chern-Simons interaction, all of the gauge invariant monopoles have positive R-charge if we assign R-charge $\frac{2}{3}$ to all the chirals as we did so far. For example, turning on Chern-Simons terms with levels  $(\frac{6-N_f}{2},\frac{N_f-6}{2})$ for the two $SU(2)$ gauge nodes, the $(1,1)$ monopole can be made gauge invariant by dressing it with $6-N_f$ bifundamentals.

For $N_f=4$ we take the Chern-Simons levels to be $(1,-1)$. Using the same parametrization of the global symmetries of the previous cases, we find the following index:
\bea  \label{IndE5}
\mathcal{I}&=&1+2a^4x^{\frac{2}{3}}+\left(\frac{4}{a^4}+3a^8+a^2(b^2+\frac{1}{b^2}){\bf 6}\right)x^{\frac{4}{3}}+\nn\\
&+&\left(4a^{12}+2a^6(b^2+\frac{1}{b^2}){\bf 6}\right)x^2+\left(-\frac{1}{a^2}(b^2+\frac{1}{b^2}){\bf 6}+\frac{8}{a^8}+5a^{16}+\right.\nn\\
&+&\left.a^4((1+b^4+\frac{1}{b^4})({\bf 20}-1)-{\bf 15})+3a^{10}(b^2+\frac{1}{b^2}){\bf 6}\right)x^{\frac{8}{3}}+\cdots\,.
\eea
We claim that this theory is the result of the compactification of the $5d$ rank $1$ $E_5=SO(10)$ SCFT on a torus with flux. Specifically, the flux is in a $U(1)$ whose commutant in $SO(10)$ is $SU(4)\times SU(2)$, and its value is $1$ in a normalization such that the minimal charge is $1$. There are several pieces of evidence supporting this. First we note that the index forms characters of $SU(4)\times SU(2)\times U(1)$, where the embedding for the $SU(2)$ is such that:
\be
\bold{2} \rightarrow b^2\oplus b^{-2}\,.
\ee 
Moreover, like in the previous cases, the structure of the index follows the expectations from $5d$ compactifications. The states coming from the conserved current in the adjoint representation of $SO(10)$ split into representations of the $SU(4)\times SU(2)\times U(1)$ subgroup according to the branching rule
\be\label{E5adjBR}
{\bf 45}\to({\bf 1},{\bf 1})^0\oplus({\bf 15},{\bf 1})^0\oplus({\bf 1},{\bf 3})^0\oplus({\bf 1},{\bf 1})^{\pm2}\oplus({\bf 6},{\bf 2})^{\pm1}\,.
\ee
We can then identify in the index \eqref{IndE5} the states $({\bf 1},{\bf 1})^{2}$ and $({\bf 6},{\bf 2})^{\pm1}$ in the decomposition \eqref{E5adjBR}: they are the contributions $2\,a^4x^{\frac{2}{3}}$, $a^2(b^2+b^{-2}){\bf 6}x^{\frac{4}{3}}$ and $-a^{-2}(b^2+b^{-2}){\bf 6}x^{\frac{8}{3}}$. The state $({\bf 1},{\bf 1})^{-2}$ contributes $-2\,a^{-4}x^{\frac{10}{3}}$ to the $3d$ index so it appears above the highest order we reported in \eqref{IndE5}, but we checked its presence.

Finally, we can again consider the conformal manifold. The analysis is similar to the one done in the previous case. From the superconformal index we again see that we have one marginal operator in the adjoint of $SU(4)$, and two which are singlets. These should lead to a $5$ dimensional conformal manifold along which the global symmetry is broken to its Cartan subalgebra. This fits our expectations from $5d$, and we would further speculate that the symmetry enhances to $U(1)\times SU(2)\times SU(4)$ on a $1d$ subspace on the conformal manifold, as expected from $5d$, which is consistent with the index. Again this is up to the potential loophole of accidental symmetries in the IR, like ones that would arise if the flip fields decouple in the IR.

\subsubsection{$N_f=3$}

For $N_f=3$ we take the Chern-Simons terms to be $(\frac{3}{2},-\frac{3}{2})$. Using the same parametrization of the global symmetries of the previous cases, we find the following index:
\bea  \label{IndE4}
\mathcal{I}&=&1+2a^4x^{\frac{2}{3}}+\left(\frac{5}{a^4}+3a^8+a^2(b^2{\bf \overline{3}}+\frac{1}{b^2}{\bf 3})\right)x^{\frac{4}{3}}+\nn\\
&+&\left(4a^{12}+2a^6(b^2{\bf \overline{3}}+\frac{1}{b^2}{\bf 3})\right)x^2+\left(-\frac{1}{a^2}(b^2{\bf \overline{3}}+\frac{1}{b^2}{\bf 3})+\frac{10}{a^8}+5a^{16}+\right.\nn\\
&+&\left.a^4(b^4{\bf 6}+\frac{1}{b^4}{\bf \overline{6}}-(1+{\bf 8}))+3a^{10}(b^2{\bf \overline{3}}+\frac{1}{b^2}{\bf 3})\right)x^{\frac{8}{3}}+\cdots\,.
\eea
We claim that this theory is the result of the compactification of the $5d$ rank $1$ $E_4=SU(5)$ SCFT on a torus with flux. Specifically, the flux is in a $U(1)$ whose commutant in $SU(5)$ is $SU(3)\times  U(1)$, and its value is $1$ in a normalization such that the minimal charge is $1$. There are several pieces of evidence supporting this. First we note that the global symmetry preserved by the flux is manifest in our Lagrangian description of the model. 
Moreover, like in the previous cases, the structure of the index follows the expectations from $5d$ compactifications. The states coming from the conserved current in the adjoint representation of $SU(5)$ split into representations of the $SU(3)\times  U(1)^2$ subgroup according to the branching rule
\be\label{E4adjBR}
{\bf 24}\to2\times{\bf 1}^{(0,0)}\oplus{\bf 8}^{(0,0)}\oplus{\bf 1}^{(\pm2,0)}\oplus {\bf 3}^{(\pm1,-1)}\oplus{\bf \overline{3}}^{(\pm1,1)}\,.
\ee
where, up to some normalization, the first entry of the exponent is related to the $U(1)_a$ symmetry, while the second entry is related to the $U(1)_b$ symmetry.
We can then identify in the index \eqref{IndE4} the states ${\bf 1}^{(2,0)}$, ${\bf 3}^{(\pm1,-1)}$ and ${\bf \overline{3}}^{(\pm1,1)}$ in the decomposition \eqref{E4adjBR}: they are the contributions $2\,a^4x^{\frac{2}{3}}$, $a^2(b^2{\bf \overline{3}}+b^{-2}{\bf 3})x^{\frac{4}{3}}$ and $-a^{-2}(b^2{\bf \overline{3}}+b^{-2}{\bf 3})x^{\frac{8}{3}}$. The state ${\bf 1}^{(-2,0)}$ contributes $-2\,a^{-4}x^{\frac{10}{3}}$ to the $3d$ index so it appears above the highest order we reported in \eqref{IndE4}, but we checked its presence.

 The structure of the conformal manifold also fits our expectations. Similarly to the previous cases, the superconformal index suggests that we have one marginal operator in the adjoint of $SU(3)$, and two which are singlets. These should lead to a $4$ dimensional conformal manifold along which the global symmetry is broken to its Cartan subalgebra. This fits our expectations from $5d$. Again this is up to the potential loophole of accidental symmetries in the IR, like ones that would arise if the flip fields decouple in the IR.

\subsubsection{$N_f=2$}

For $N_f=2$ we take the Chern-Simons terms to be $(2,-2)$. Using the same parametrization of the global symmetries of the previous cases, we find the following index:
\bea  \label{IndE3}
\mathcal{I}&=&1+2a^4x^{\frac{2}{3}}+\left(\frac{6}{a^4}+3a^8+a^2(b^2+\frac{1}{b^2})\right)x^{\frac{4}{3}}+\left(4a^{12}+2a^6(b^2+\frac{1}{b^2})\right)x^2+\nn\\
&+&\left(-\frac{1}{a^2}\left(b^2+\frac{1}{b^2}\right)+\frac{12}{a^8}+5a^{16}+a^4(b^4+\frac{1}{b^4}-(1+2\times{\bf 3}))+3a^{10}(b^2+\frac{1}{b^2})\right)x^{\frac{8}{3}}+\cdots\,.\nn\\
\eea
We claim that this theory is the result of the compactification of the $5d$ rank $1$ $E_3=SU(3)\times SU(2)$ SCFT on a torus with flux. Specifically, considering the $SU(2)\times U(1)$ subgroup of the $SU(3)$ factor in the $5d$ global symmetry, the flux is for the $U(1)$ inside the $SU(2)$, and its value is $1$ in a normalization such that the minimal charge is $1$. There are several pieces of evidence supporting this. First we note that the $SU(2)\times U(1)^2$ global symmetry preserved by the flux is manifest in our Lagrangian description of the model. 
Moreover, like in the previous cases, the structure of the index follows the expectations from $5d$ compactifications. The states coming from the conserved current in the adjoint representation of $SU(3)\times SU(2)$ split into representations of the $SU(2)\times  U(1)^2$ subgroup according to the branching rule
\bea \label{E3adjBR}
&&({\bf 8},{\bf 1})\to 2\times{\bf 1}^{(0,0)}\oplus{\bf 1}^{(\pm2,0)}\oplus{\bf 1}^{(\pm1,1)}\oplus{\bf 1}^{(\pm1,-1)}\nn\\
&&({\bf 1},{\bf 3})\to{\bf 3}^{(0,0)}\,.
\eea
where, up to some normalization, the first entry of the exponent is related to the $U(1)_a$ symmetry, while the second entry is related to the $U(1)_b$ symmetry.
We can then identify in the index \eqref{IndE3} the states ${\bf 1}^{(2,0)}$, ${\bf 1}^{(\pm1,1)}$ and ${\bf 1}^{(\pm1,-1)}$ in the decomposition \eqref{E3adjBR}: they are the contributions $2\,a^4x^{\frac{2}{3}}$, $a^2(b^2+b^{-2})x^{\frac{4}{3}}$ and $-a^{-2}(b^2+b^{-2})x^{\frac{8}{3}}$. The state ${\bf 1}^{(-2,0)}$ contributes $-2\,a^{-4}x^{\frac{10}{3}}$ to the $3d$ index so it appears above the highest order we reported in \eqref{IndE3}, but we checked its presence.

The structure of the conformal manifold also fits our expectations. Similarly to the previous cases, the superconformal index suggests that we have one marginal operator in the adjoint of $SU(2)$, and two which are singlets. These should lead to a $3$ dimensional conformal manifold along which the global symmetry is broken to its Cartan subalgebra. This fits our expectations from $5d$. Again this is up to the potential loophole of accidental symmetries in the IR, like ones that would arise if the flip fields decouple in the IR.

\subsubsection{$N_f=1$}

For $N_f=1$ we take the Chern-Simons terms to be $(\frac{5}{2},-\frac{5}{2})$. Using the same parametrization of the global symmetries of the previous cases, we find the following index:
\bea  \label{IndE2}
\mathcal{I}&=&1+2a^4x^{\frac{2}{3}}+\left(\frac{7}{a^4}+3a^8\right)x^{\frac{4}{3}}+4a^{12}x^2+\left(-2a^4+\frac{14}{a^8}+5a^{16}\right)x^{\frac{8}{3}}+\nn\\
&+&\left(-\frac{2}{a^4}-a^2(b^2+\frac{1}{b^2})-4a^8+6a^{20}\right)x^{\frac{10}{3}}+\cdots\,.
\eea
We claim that this theory is the result of the compactification of the $5d$ rank $1$ $E_2=SU(2)\times U(1)$ SCFT on a torus with flux. Specifically, the flux is for the $U(1)$ inside the $SU(2)$ factor of the $5d$ global symmetry, and its value is $1$ in a normalization such that the minimal charge is $1$. There are several pieces of evidence supporting this. First we note that the $U(1)^2$ global symmetry preserved by the flux is manifest in our Lagrangian description of the model. 
Moreover, like in the previous cases, the structure of the index follows the expectations from $5d$ compactifications. The states coming from the conserved current in the adjoint representation of $SU(2)\times U(1)$ split into representations of the $ U(1)^2$ subgroup according to the branching rule
\bea \label{E2adjBR}
&&{\bf 3}^0\to(0,0)\oplus(\pm2,0)\nn\\
&&{\bf 1}^0\to (0,0)\,.
\eea
where, up to some normalization, the first entry of the terms on the r.h.s.~is related to the $U(1)_a$ symmetry, while the second entry is related to the $U(1)_b$ symmetry.
We can then identify in the index \eqref{IndE2} the states $(\pm2,0)$ in the decomposition \eqref{E2adjBR}: they are the contributions $2\,a^4x^{\frac{2}{3}}$ and $-2\,a^{-4}x^{\frac{10}{3}}$. 

The structure of the conformal manifold also fits our expectations. Similarly to the previous cases, the superconformal index suggests the presence of two marginal operators, which are singlets of the global symmetry. These should lead to a $2$ dimensional conformal manifold along which the global symmetry is preserved. This fits our expectations from $5d$. Again this is up to the potential loophole of accidental symmetries in the IR, like ones that would arise if the flip fields decouple in the IR.

\subsubsection{$N_f=7$}

Finally, we return to the case of $N_f=7$. We take the Chern-Simons terms to be $(-\frac{1}{2},\frac{1}{2})$. As in the case of $N_f=6$, the singlet fields turn out to violate the unitarity bound, and the value of the superconformal R-charge in the theory obtained after removing them is as follows, 
\begin{equation}
\hat{R}=R-0.11q_{a}\,,
\end{equation}
where $U(1)_R$ is the usual reference R symmetry under which all the fields carry R charge $\frac{2}{3}$. Using this symmetry, the index of the model with the singlet fields contain only perturbative contributions up to order $x^2$. These are:

\be
I = 1 + 2 a^4 x^{\frac{2}{3}} + x^{\frac{4}{3}} ( \frac{1}{a^4} + a^2 ( b^2\bold{21} + \frac{1}{b^2}\overline{\bold{21}} ) + 3a^8) + x^2 (4a^{12} + 2a^6 ( b^2\bold{21} + \frac{1}{b^2}\overline{\bold{21}} ) -2 ) + ... .
\ee 

However, this does not follow the previous pattern. Specifically, we would expect this theory to correspond to the compactification of the rank $1$ $E_8$ $5d$ SCFT on a torus with flux preserving its $U(1)\times E_7$ subgroup. However, the index does not form characters of $E_7$, contradicting this.

It appears that in this case, our assumptions regarding the domain wall are not accurate, and the matter fields living on them are different. Nevertheless, it is still possible for the theory we presented in this case to have an higher dimensional interpretation, albeit not as a direct compactification, but rather as a compactification followed by a deformation. This has been observed to occur in some cases in the compactification of $6d$ SCFTs to $4d$, see for instance \cite{KRVZ,Zafrir:2018hkr}. In those cases, these claims could be tested by comparing anomalies. Unfortunately, it is not clear how to test such a possibility with our currently available tools.  

\section{Studying more general models}
\label{4t}

In this section we consider some more general cases, which involve four domain walls. The resulting models consequently possess more gauge nodes and a slightly more complicated structure, but they all pass our tests for being correct compactifications of the $5d$ rank 1 $E_{N_f+1}$ SCFTs. In these more general cases we will also encounter the new feature of monopole superpotentials. For definiteness, we will focus on cases with $N_f=5,6$.

\subsection{$N_f=6$}

\subsubsection{$U(1)\times SO(12)$ case}
\label{E7flux2}

\begin{figure}
\center
\includegraphics[width=0.4\textwidth]{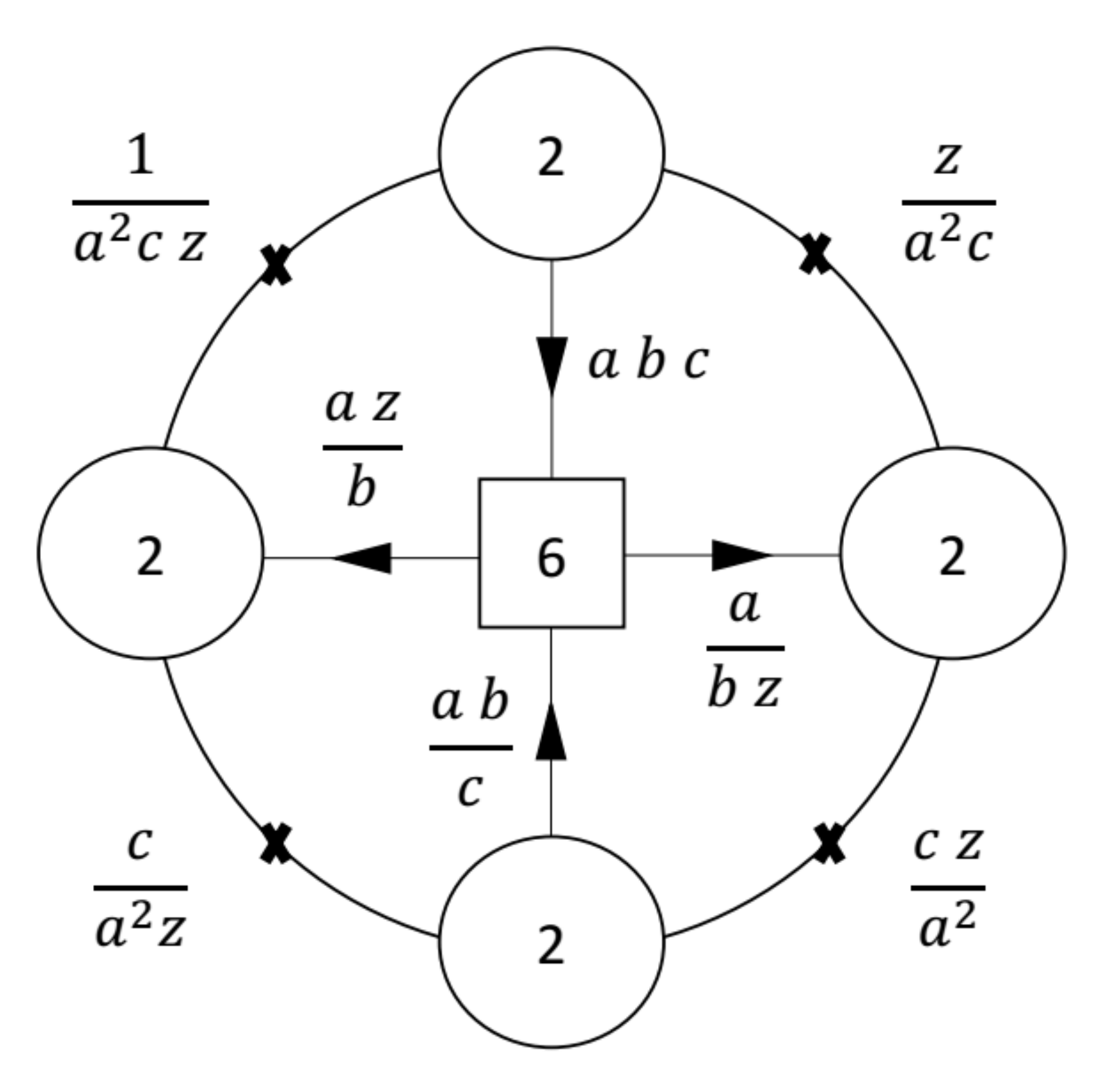} 
\caption{The $3d$ quiver theory associated with the compactification of the $5d$ $E_7$ SCFT on a torus with flux $(\sqrt{8-N_f};1,1,...,1)$. The theory also contains various superpotential interactions. Notably, there are cubic superpotentials running along the triangles as well as the flipping superpotentials.}
\label{QuiverE7F2}
\end{figure}

The first model we consider is obtained by gluing four copies of the tube in figure \ref{BTube}, and so it is expected to be associated with flux two.
The quiver of the resulting theory is shown in figure \ref{QuiverE7F2}. The superpotential consists of cubic interactions for each triangle of the quiver and the standard flipping terms.
This preserves four $U(1)$ global symmetries. Additionally, there are the $SU(6)$ global symmetry associated with the four collections of six flavors, and the $U(1)_R$ R-symmetry, which we choose to be the same as the one we previously used. The singlet fields turn out to violate the unitarity bound, and the value of the superconformal R-charge in the theory obtained after removing them is given by
\begin{equation}
\hat{R}=R-0.083q_{a}\,.
\end{equation} 

We can evaluate the index of this theory finding (using the usual R-symmetry and including the singlet fields):

\bea
\mathcal{I} & = & 1 + a^4(z^2 + \frac{1}{z^2})(c^2 + \frac{1}{c^2}) x^{\frac{2}{3}} +  \left( a^2 ( b^6 (z^2 + \frac{1}{z^2}) + \frac{1}{b^6} (c^2 + \frac{1}{c^2}) + b^2 (c^2 + \frac{1}{c^2})\bold{15} \right.\nonumber \\ & + & \left. \frac{1}{b^2} (z^2+\frac{1}{z^2})\overline{\bold{15}} ) + a^8\left(1+(z^4 + 1 + \frac{1}{z^4})(c^4 + 1 + \frac{1}{c^4})\right)\right)x^{\frac{4}{3}} \nonumber\\  & + &  \left((z^2 + \frac{1}{z^2})(c^2 + \frac{1}{c^2})(1+(z^4 + \frac{1}{z^4})(c^4 + \frac{1}{c^4}))a^{12}\right.\nonumber \\ & + & \left. a^6 (z^2 + \frac{1}{z^2})(c^2 + \frac{1}{c^2}) ( b^6 (z^2 + \frac{1}{z^2}) + \frac{1}{b^6} (c^2 + \frac{1}{c^2}) + b^2 (c^2 + \frac{1}{c^2})\bold{15} \right.\nonumber \\ & + & \left.  \frac{1}{b^2} (z^2+\frac{1}{z^2})\overline{\bold{15}} ) -4 + (z^4 + \frac{1}{z^4})(c^4 + \frac{1}{c^4})\right)x^2 + \cdots\,.
\eea
The last four terms come from the $(1,1,0,0)$, $(0,1,1,0)$, $(0,0,1,1)$ and $(1,0,0,1)$ monopoles, where each entry in these vectors denotes the magnetic flux under each of the four $SU(2)$ gauge nodes.  The additional abelian symmetries here seem to prevent the enhancement. However, we can break them by introducing superpotentials associated with the four monopole operators we mentioned\footnote{These operators have the R-charge of marginal operators under the superconformal R-symmetry. Furthermore, they have a non-trivial Kahler quotient under the global symmetry, and so contain two exactly marginal directions, which are the ones we are turning on.}. These force $c=z=1$. With this, the index forms characters of $SO(12)$, which is enhanced from $SU(6)\times U(1)_b$ where the embedding is the same \eqref{emb12} we had in the case of subsection \ref{2DWNf6}.
Specifically, the index can be rewritten as
\bea
\mathcal{I} & = & 1 + 4a^4 x^{\frac{2}{3}} +  \left( 2a^2 \bold{32} + 10a^8\right)x^{\frac{4}{3}} +  \left(20a^{12} + 8a^6 \bold{32}\right)x^2 + \cdots\, .
\eea
This result conforms to our expectations. Indeed, we can see not only the symmetry preserved by the flux of the $5d$ compactification, but also the spectrum of operators expected from $5d$. This is the same as in the discussion around equation \eqref{E7adjBR}, but this time the multiplicities of all the operators are doubled since the flux is two.
We then see that in the quiver theories, in addition to the triangle superpotentials, monopole superpotentials associated with adjacent gauge groups should also be introduced. 

Another evidence for the proposed enhancement of the $SU(6)\times U(1)_{b}$ part of the symmetry to $SO(12)$ comes from examining the central charges of these symmetries, as before. Following the discussion in appendix B, we compute numerically the real part of the free energy of this model, and use it to find the following values for the central charges of $U(1)_b$ and of the Cartan $\textrm{diag}\left(1,0,0,0,0,-1\right)$ of $SU(6)$, which we denote again by $C$,
\begin{equation}
C_{b}=56.7\,\,\,,\,\,\,C_{C}=4.7\,.
\end{equation}
The value of the ratio of these charges is
\begin{equation}
\label{ratioA}
\frac{C_{b}}{C_{C}}=12.04,
\end{equation}
which matches our expectations from the proposed symmetry enhancement and the embedding \eqref{emb12}. Indeed, exactly as in Eq. \eqref{CbCc12}, this ratio is expected to be equal to 12, which agrees with our numerical result \eqref{ratioA} within the accuracy of the calculation. 

Finally, we can consider the structure of the conformal manifold. Similarly to the case with minimal flux, we can read them from the superconformal index, and see that we need to have one again in the adjoint of $SU(6)$ and two singlets. Unlike the previous case, though, here the singlets come directly from $(1,1)$ monopole operators of adjacent $SU(2)$ groups\footnote{More correctly, these are the two exactly marginal deformations formed from the four $(1,1)$ monopole operators of adjacent $SU(2)$ groups. The remaining two combine with the two $U(1)$ global symmetries that the monopole superpotentials break to form long multiplets.} without involving the flip fields. As such these remain exactly marginal even should the flip fields decouple, which has we showed, is expected to happen in this case. Like the previous case, the structure of the marginal operators leads to a $7$ dimensional conformal manifold on a generic point of which the $SU(6)$ is broken down to its Cartan subalgebra, in agreement with our $5d$ expectations. These then also lead us to expect that there is a $1d$ subspace along which the global symmetry is enhanced to $U(1)\times SO(12)$. The index and central charges are both consistent with this. Unlike the case of minimal flux, the decoupling of the flip fields does not lead to an inconsistency. Therefore, it appears that indeed the deviation in the behavior of the minimal flux case is an exceptional feature, which is not shared in more generic cases.

Next we consider similar quivers, but with the flavors allocated differently.

\subsubsection{$U(1)\times SU(2)\times SU(6)$ case}

\begin{figure}
\center
\includegraphics[width=0.6\textwidth]{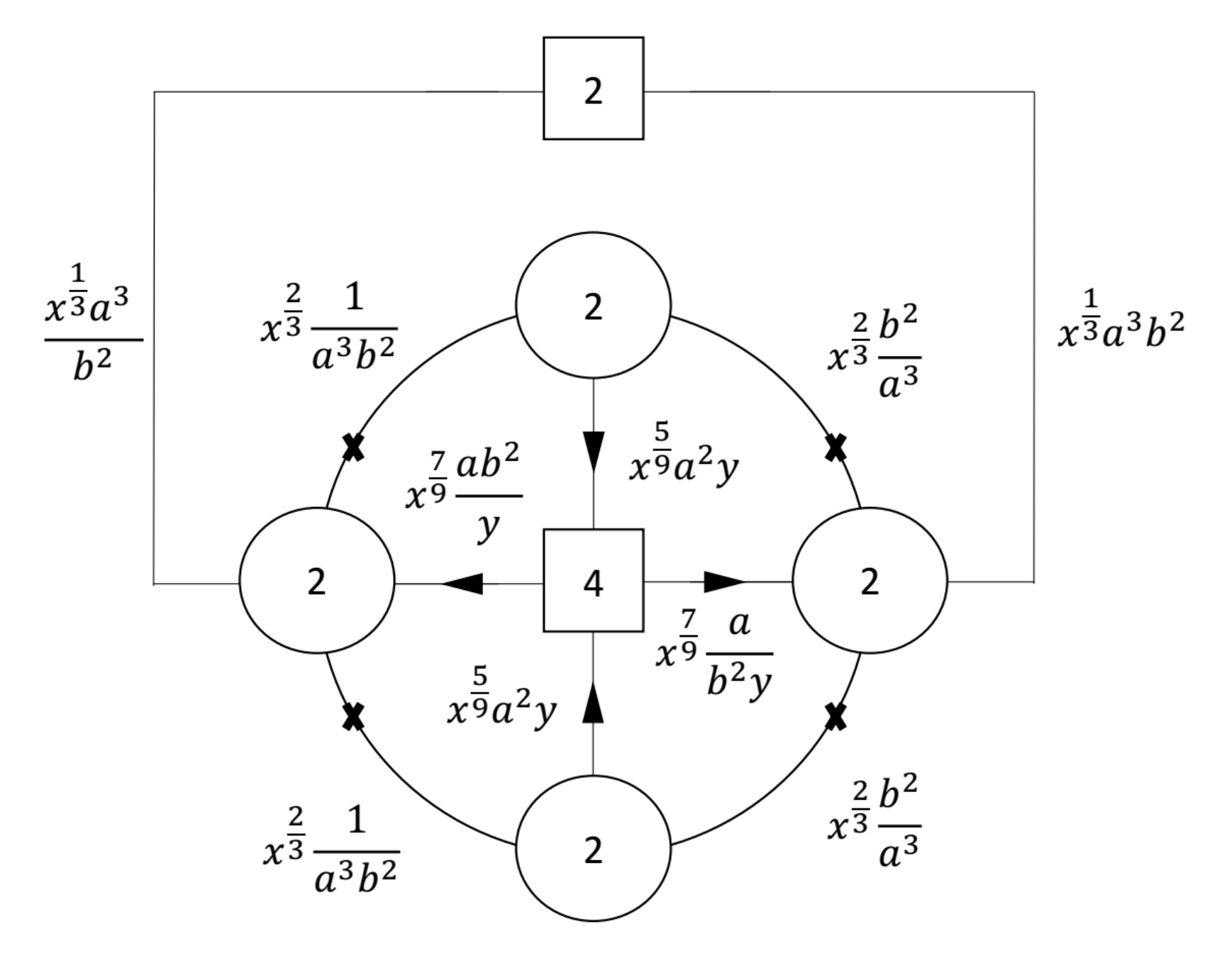} 
\caption{The $3d$ quiver theory associated with the compactification of the $5d$ $E_{7}$ SCFT on a torus with flux $(\sqrt{2};1,1,1,1,0,0)$. In the figure we have used fugacities to represent the charges of the fields under the global symmetries, specifically, the fugacities $a$, $b$ and $y$ for the three $U(1)$ global symmetries and $x$ for the R-symmetry. The theory also contains various superpotential interactions, which corresponds to all operators build from the fields that carry R-charge $2$ and are global symmetry singlets. For completeness, we note that these include the four cubic superpotentials running along the triangles, the flipping superpotentials and two quadratic superpotentials running along the two bifundamentals connected to the global $SU(2)$ in the upper square and the upper and lower semi-circles. Finally, there are four monopole superpotential of the type $(1,1)$ for all adjacent $SU(2)$ groups.}
\label{QuiverE742}
\end{figure}

We next consider the model shown in figure \ref{QuiverE742}. This is generated by gluing two of the tubes of figure \ref{DTubes} (b) for $N_f=6$ and $x=2$. As such we expect the resulting theory to be associated with the flux $(\sqrt{2};1,1,1,1,0,0)$. This corresponds to the minimal value of flux preserving the $U(1)\times SU(2)\times SU(6)$ subgroup of $E_7$, that is flux of value $1$ in the $U(1)$ group, where we use a normalization where the minimal charge is $1$.

The charges under the three $U(1)$ groups, the $SU(2)\times SU(4)$ non-abelian symmetry groups, and the R-symmetry that we shall use are shown in the figure. For latter convenience, we have slightly changed the charges under various symmetries compared to the previous sections so $U(1)_a$, $U(1)_b$ and the R-symmetry are not the same as the ones we used previously. We note that here the model includes superpotentials involving all four triangles, the two loops involving the bifundamentals charged under the global $SU(2)$ with the upper and lower semi-circles, and, following the discussion in the previous section, also all four $(1,1)$ monopoles for adjacent groups. The latter is responsible for breaking a possible $U(1)$ symmetry. The superconformal R-symmetry is given by the following expression in terms of the symmetries appearing in the figure: 
\begin{equation}
\hat{R}=R-0.009q_{a}\,.
\end{equation}

We can compute the index of this model (using the R-symmetry in the figure) finding:  

\bea
\mathcal{I} & = & 1 + 3 a^{6} \left(b^4 + \frac{1}{b^4}\right) x^{\frac{2}{3}} + 2 a^{4} \left(y^2\bold{6}_{SU(4)} + \frac{1}{y} \bold{4}_{SU(4)} \bold{2}_{SU(2)} + \frac{1}{y^4}\right) x^{\frac{10}{9}} \nn\\ \nonumber & + & 3 a^{12} \left(2b^8 + 3 + \frac{2}{b^8}\right) x^{\frac{4}{3}} + a^{2} \left(b^4 + \frac{1}{b^4}\right) \left(\frac{1}{y^2}\bold{6}_{SU(4)} + y \overline{\bold{4}}_{SU(4)} \bold{2}_{SU(2)} + y^4 \right) x^{\frac{14}{9}} \\ \nonumber & + & 6 a^{10} \left(b^4 + \frac{1}{b^4}\right) \left(y^2\bold{6}_{SU(4)} + \frac{1}{y} \bold{4}_{SU(4)} \bold{2}_{SU(2)} + \frac{1}{y^4}\right) x^{\frac{16}{9}} + \nn\\
&+&2 a^{18} \left(5b^{12} + 9b^4 + \frac{9}{b^4} + \frac{5}{b^{12}}\right) x^{2} + \cdots\,.
\eea
We note that the index forms characters of $SU(2)\times SU(6)$, where the embedding is such that
\be\label{emb6y42}
\bold{2}_{SU(2)} = b^4 \oplus \frac{1}{b^4},\qquad \bold{6}_{SU(6)} = y \bold{4}_{SU(4)} \oplus \frac{1}{y^2}\bold{2}_{SU(2)}\,.
\ee
In terms of characters of these groups the index reads:   
\bea\label{indNf6x2}
\mathcal{I} & = & 1 + 3 a^{6} \bold{2}_{SU(2)} x^{\frac{2}{3}} + 2 a^{4} \bold{15}_{SU(6)} x^{\frac{10}{9}} + 3 a^{12} (1+2\times \bold{3}_{SU(2)}) x^{\frac{4}{3}} + a^{2} \bold{2}_{SU(2)} \overline{\bold{15}}_{SU(6)} x^{\frac{14}{9}} +\nonumber \\ & + & 6 a^{10} \bold{2}_{SU(2)} \bold{15}_{SU(6)} x^{\frac{16}{9}} + 2 a^{18} (5\times \bold{4}_{SU(2)} + 4\times \bold{2}_{SU(2)}) x^{2} + \cdots\,.
\eea
We again see that the spectrum of states is consistent with the $5d$ expectations, namely it organizes into representations of the $U(1)\times SU(2)\times SU(6)$ symmetry preserved by the flux and we can also see in the index the contributions of relevant and irrelevant operators coming from the broken currents of the $5d$ global symmetry. In this case, the current in the adjoint representation of $E_7$ splits under the $U(1)\times SU(2)\times SU(6)$ subgroup according to the branching rule
\be
{\bf 133}\to({\bf 1},{\bf 1})^0\oplus({\bf 3},{\bf 1})^0\oplus({\bf 1},{\bf 35})^0\oplus ({\bf 2},{\bf 1})^{\pm3}\oplus({\bf 1},{\bf 15})^2\oplus({\bf 1},{\bf \overline{15}})^{-2}\oplus ({\bf 2},{\bf \overline{15}})^1\oplus ({\bf 2},{\bf 15})^{-1}\,.
\ee
In this case the $5d$ R-symmetry is related to the $3d$ R-symmetry that we used for our computation by a mixing with $U(1)_a$ that can be implemented in the index by the shift $a\to ax^{\frac{2}{9}}$. Moreover, the $U(1)_a$ symmetry is related to the $U(1)$ symmetry inside the $5d$ global symmetry for which we turned on a flux by a normalization of $\frac{1}{2}$. With this dictionary, we can immediately identify the states $({\bf 2},{\bf 1})^3$, $({\bf 1},{\bf 15})^2$, $({\bf 2},{\bf \overline{15}})^1$ in the index \eqref{indNf6x2}: they are represented by the contributions $3 a^{6} \bold{2}_{SU(2)} x^{\frac{2}{3}}$, $2 a^{4} \bold{15}_{SU(6)} x^{\frac{10}{9}}$, $a^{2} \bold{2}_{SU(2)} \overline{\bold{15}}_{SU(6)} x^{\frac{14}{9}}$. Notice that, as expected, their multiplicities are given by their $U(1)$ charges times the $U(1)$ flux, which in this case is one. The other states appear at higher orders, but we checked their presence up to order $x^3$.

Another evidence for the proposed enhancement of the $SU(4)\times SU(2)\times U(1)_{y}$ part of the symmetry to $SU(6)$ comes from examining the central charges of these symmetries, as before. We compute numerically the real part of the free energy of this model, and use it to find the following values for the central charges of $U(1)_y$ and of the Cartans $\textrm{diag}\left(1,0,0,-1\right)$ and $\textrm{diag}\left(1,-1\right)$ of $SU(4)$ and $SU(2)$, which we denote by $C_4$ and $C_2$,
\begin{equation}
C_{y}=23.2\,\,\,,\,\,\,C_{C_{4}}=C_{C_{2}}=3.9.
\end{equation}
The values of the ratios of these charges are
\begin{equation}
\label{ratioB}
\frac{C_{C_{4}}}{C_{C_{2}}}=1\,\,\,,\,\,\,\frac{C_{y}}{C_{C_{2}}}=5.95,
\end{equation}
which again match our expectations from the proposed symmetry enhancement and the embedding \eqref{emb6y42}. Indeed, the corresponding embedding indices are 
\begin{equation}
\begin{split}
I_{SU(4)\rightarrow SU(6)}=I_{SU(2)\rightarrow SU(6)}=1,&\,\,\,I_{U(1)_{C_{4}}\rightarrow SU(4)}=I_{U(1)_{C_{2}}\rightarrow SU(2)}=4,\\
I_{U(1)_{y}\rightarrow SU(6)}&=24,
\end{split}
\end{equation}
implying the following ratios of central charges, 
\begin{equation}
\frac{C_{C_{4}}}{C_{C_{2}}}=1\,\,\,,\,\,\,\frac{C_{y}}{C_{C_{2}}}=6
\end{equation}
which agree with our numerical results \eqref{ratioB} within the accuracy of the calculation. 

Finally, we can consider the structure of the conformal manifold. Similarly to the previous cases, the superconformal index suggests the presence of several marginal operators, now in the adjoint of the $SU(4)$, $SU(2)$ and three singlets. These come from the cubic superpotentials along the triangles, the quartic superpotential involving the global $SU(2)$ bifundamentals and the upper and lower semi-circles and the $(1,1)$ monopole superpotentials for adjacent groups. More specifically, they are the combination of these that don't recombine with global symmetries broken by the presence of these and similar superpotentials.

These should lead to a $7$ dimensional conformal manifold along which the global symmetry is broken to its Cartan subalgebra. This fits our expectations from $5d$. Like in the previous cases, the $5d$ expectations lead us to expect that the symmetry be enhanced to $U(1)\times SU(2)\times SU(6)$ at some $1d$ subspace on the conformal manifold. Our results for the superconformal index and central charges are consistent with this. Again this is up to the potential loophole of accidental symmetries in the IR, like ones that would arise if the flip fields decouple in the IR.

\subsubsection{$U(1)\times SU(2)\times SO(10)$ case}

\begin{figure}
\center
\includegraphics[width=0.6\textwidth]{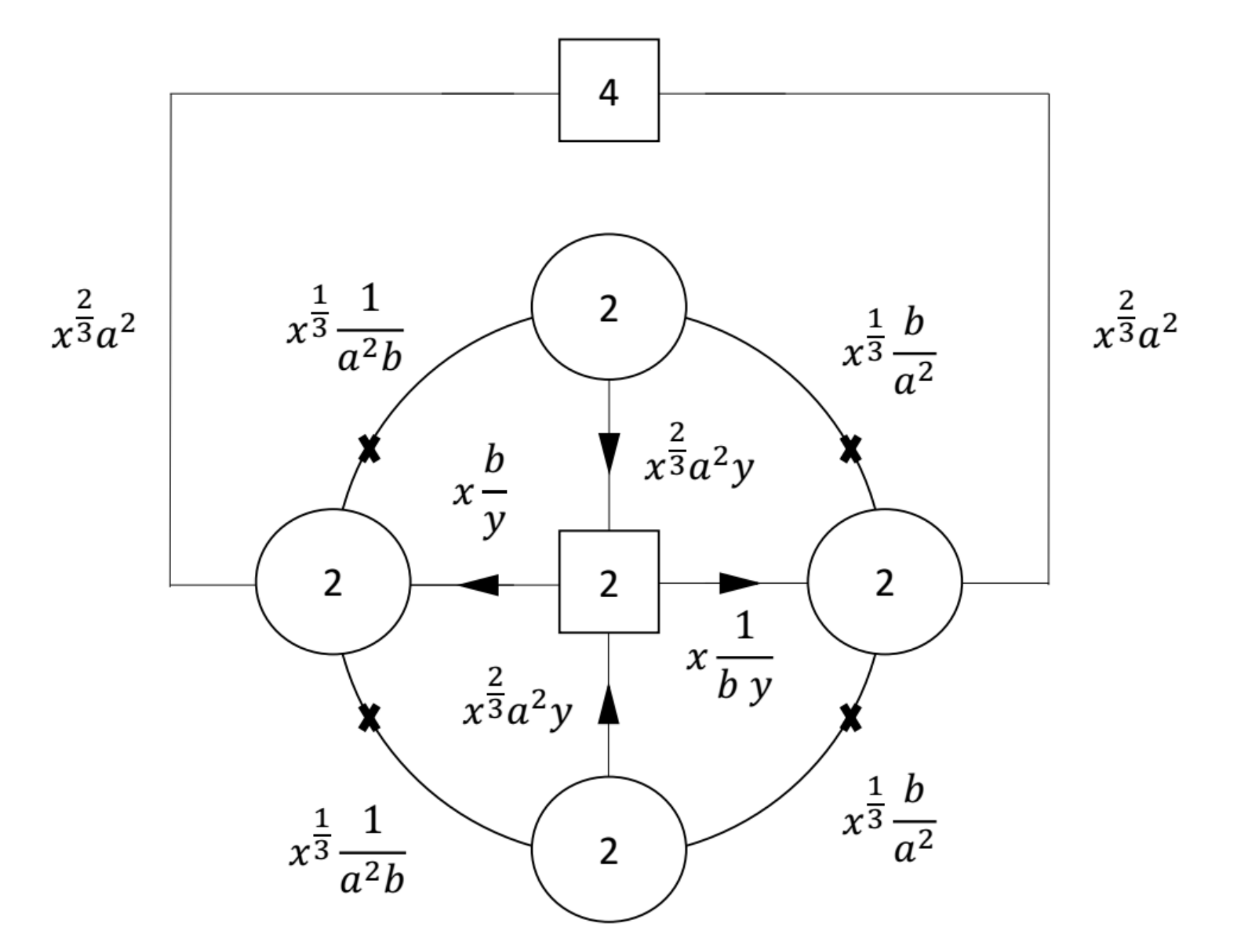} 
\caption{The $3d$ quiver theory associated with the compactification of the $5d$ $E_{7}$ SCFT on a torus with flux $(\sqrt{2};1,1,0,0,0,0)$. In the figure we have used fugacities to represent the charges of the fields under the global symmetries, specifically, the fugacities $a$, $b$ and $y$ for the three $U(1)$ global symmetries and $x$ for the R-symmetry. The theory also contains various superpotential interactions, which correspond to all operators build from the fields that carry R-charge $2$ and are global symmetry singlets. For completeness, we note that these include the four cubic superpotentials running along the triangles, the flipping superpotentials and two quadratic superpotentials running along the two bifundamentals connected to the global $SU(4)$ in the upper square and the upper and lower semi-circles. Finally, there are four monopole superpotential of the type $(1,1)$ for all adjacent $SU(2)$ groups.}
\label{QuiverE724}
\end{figure}

We next consider the model shown in figure \ref{QuiverE724}. This is generated by gluing two of the tubes of figure \ref{DTubes} (b) for $N_f=6$ and $x=4$. As such we expect the resulting theory to be associated with the flux $(\sqrt{2};1,1,0,0,0,0)$. This corresponds to the minimal value of flux preserving the $U(1)\times SU(2)\times SO(10)$ subgroup of $E_7$.

The charges under the three $U(1)$ groups, the $SU(2)\times SU(4)$ non-abelian symmetry groups, and the R-symmetry that we shall use are shown in the figure. For latter convenience, we have slightly changed the charges under various symmetries compared to the previous sections. We note that here the model includes superpotentials involving all four triangles, the two loops involving the bifundamentals charged under the global $SU(4)$ with the upper and lower semi-circles, and, following the discussion in the previous sections, also all four $(1,1)$ monopoles for adjacent groups. The latter is responsible for breaking a possible $U(1)$ symmetry. The superconformal R-symmetry is given by the following expression in terms of the symmetries appearing in the figure: 
\begin{equation}
\hat{R}=R-0.126q_{a}\,.
\end{equation}

We can compute the index of this model (using the R-symmetry in the figure) finding:  

\bea
\mathcal{I} & = & 1 + 2 \left(\frac{1}{a^8} + a^4 ( b^2 + \frac{1}{b^2} + y^2 + \frac{1}{y^2} + \bold{6}_{SU(4)} )\right) x^{\frac{4}{3}}+ a^{2} \bold{2}_{SU(2)} \left((\frac{b}{y}+\frac{y}{b})\bold{4}_{SU(4)} + \right.\nn\\
&+&\left.(b y+\frac{1}{b y})\overline{\bold{4}}_{SU(4)}\right) x^{\frac{5}{3}} -\frac{1}{a^2}{\bf 2}_{SU(2)} \left((\frac{b}{y}+\frac{y}{b})\overline{\bold{4}}_{SU(4)} +(b y+\frac{1}{b y})\bold{4}_{SU(4)}\right)x^{\frac{7}{3}}+\nn\\
&+&\left(-\frac{2}{a^4}( b^2 + \frac{1}{b^2} + y^2 + \frac{1}{y^2} + \bold{6}_{SU(4)} )+\frac{4}{a^{16}}+a^8(3(b^4+\frac{1}{b^4}+y^4+\frac{1}{y^4})+\right.\nn\\
&+&\left.4(b^2+\frac{1}{b^2}+y^2+\frac{1}{y^2}){\bf 6}_{SU(4)}+4(b^2+\frac{1}{b^2})(y^2+\frac{1}{y^2})+3\times{\bf 20}_{SU(4)}+{\bf 15}+7)\right)x^{\frac{8}{3}}\cdots\,.\nn\\
\eea
We note that the index forms characters of $SU(2)\times SO(10)$, where the embedding is such that
\be
\label{embbyc}
\bold{10}_{SO(10)} = b^2 \oplus \frac{1}{b^2} \oplus y^2 \oplus \frac{1}{y^2} \oplus \bold{6}_{SU(4)}\,.
\ee
In terms of characters of these groups the index reads:   
\bea
\mathcal{I} &=& 1 + 2 \left(\frac{1}{a^8} + a^4 \bold{10}_{SO(10)}\right) x^{\frac{4}{3}} + a^{2} \bold{2}_{SU(2)} \bold{16}_{SO(10)} x^{\frac{5}{3}}-\frac{1}{a^2}{\bf 2}_{SU(2)}{\bf \overline{16}}_{SO(10)}x^{\frac{7}{3}} +\nn\\
&+&\left(-\frac{2}{a^4}{\bf 10}_{SO(10)}+\frac{4}{a^{16}}+a^8(3\times{\bf 54}+{\bf 45}-1)\right)x^{\frac{8}{3}} +\cdots\,.
\eea
We again see that the spectrum of states is consistent with the $5d$ expectations, namely it organizes into representations of the $U(1)\times SU(2)\times SO(10)$ symmetry preserved by the flux and we can also see in the index the contributions of relevant and irrelevant operators coming from the broken currents of the $5d$ global symmetry. In this case, the current in the adjoint representation of $E_7$ splits under the $U(1)\times SU(2)\times SO(10)$ subgroup according to the branching rule
\be\label{indNf6x4}
{\bf 133}\to({\bf 1},{\bf 1})^0\oplus({\bf 3},{\bf 1})^0\oplus({\bf 1},{\bf 45})^0\oplus ({\bf 1},{\bf 10})^{\pm 2}\oplus({\bf 2},{\bf 16})^{1}\oplus ({\bf 2},{\bf \overline{16}})^{-1}\,.
\ee
In this case the $5d$ R-symmetry is related to the $3d$ R-symmetry that we used for our computation by a mixing with $U(1)_a$ that can be implemented in the index by the shift $a\to ax^{\frac{1}{6}}$. Moreover, the $U(1)_a$ symmetry is related to the $U(1)$ symmetry inside the $5d$ global symmetry for which we turned on a flux by a normalization of $\frac{1}{2}$. With this dictionary, we can immediately identify the states $({\bf 1},{\bf 10})^{\pm2}$, $({\bf 2},{\bf 16})^1$ and $({\bf 2},{\bf \overline{16}})^{-1}$ in the index \eqref{indNf6x4}: they are represented by the contributions $2a^4 \bold{10}_{SO(10)}x^{\frac{4}{3}}$, $a^{2} \bold{2}_{SU(2)} \bold{16}_{SO(10)} x^{\frac{5}{3}}$, $-a^{-2} \bold{2}_{SU(2)} \bold{\overline{16}}_{SO(10)} x^{\frac{7}{3}}$ and $-2a^{-4} \bold{10}_{SO(10)}x^{\frac{8}{3}}$. Notice that, as expected, their multiplicities are given by their $U(1)$ charges times the $U(1)$ flux, which in this case is one. 

Another evidence for the proposed enhancement of the $SU(4)\times U(1)_{b}\times U(1)_{y}$ part of the symmetry to $SO(10)$ comes from examining the central charges of these symmetries, as before. We compute numerically the real part of the free energy of this model, and use it to find the following values for the central charges of $U(1)_b$, $U(1)_y$ and of the Cartan $\textrm{diag}\left(1,0,0,-1\right)$ of $SU(4)$, which we denote by $C$,
\begin{equation}
C_{b}=8.5\,\,\,,\,\,\,C_{y}=8.8\,\,\,,\,\,\,C_{C}=4.2.
\end{equation}
The values of the ratios of these charges are
\begin{equation}
\label{ratio1210}
\frac{C_{y}}{C_{b}}=1.04\,\,\,,\,\,\,\frac{C_{b}}{C_{C}}=2.02,
\end{equation}
which again match our expectations from the proposed symmetry enhancement and the embedding \eqref{embbyc}. Indeed, the corresponding embedding indices are 
\begin{equation}
\begin{split}
I_{U(1)_{b}\rightarrow SO(10)}=&I_{U(1)_{y}\rightarrow SO(10)}=8,\\
I_{SU(4)\rightarrow SO(10)}=1\,\,,&\,\,I_{U(1)_{C}\rightarrow SU(4)}=4,
\end{split}
\end{equation}
implying the following ratios of central charges, 
\begin{equation}
\frac{C_{y}}{C_{b}}=1\,\,\,,\,\,\,\frac{C_{b}}{C_{C}}=2
\end{equation}
which agree with our numerical results \eqref{ratio1210} within the accuracy of the calculation.

Finally, we can consider the structure of the conformal manifold. This can be analyzed as done in the previous cases, and we find a $7$ dimensional conformal manifold along which the global symmetry is broken to its Cartan subalgebra. This fits our expectations from $5d$. Like in the previous cases, the $5d$ expectations leads us to expect that the symmetry is enhanced to $U(1)\times SU(2)\times SO(10)$ at some $1d$ subspace on the conformal manifold, and our results for the superconformal index and central charges are consistent with this.

\subsubsection{$U(1)\times SO(12)$ case and duality}

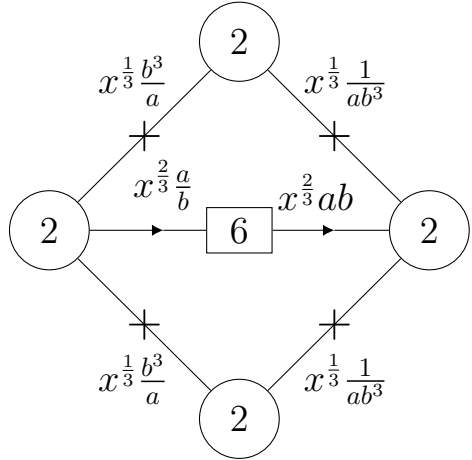
\begin{figure}
\center
\begin{tikzpicture}[baseline=0, font=\scriptsize]
\node[draw, circle] (a1) at (0,0) {\large $\,\, 2\,\,$};
\node[draw, circle] (a2) at (2.5,2.5) {\large $\,\, 2\,\,$};
\node[draw, circle] (a3) at (5,0) {\large $\,\, 2\,\,$};
\node[draw, circle] (a4) at (2.5,-2.5) {\large $\,\, 2\,\,$};
\node[draw, rectangle] (a5) at (2.5,0) {\large $\,\, 6\,\,$};
\draw[draw, solid] (a1)--(a2);
\draw[draw, solid] (a2)--(a3);
\draw[draw, solid] (a3)--(a4);
\draw[draw, solid] (a4)--(a1);
\draw[draw, solid,->] (a1)--(1.5,0);
\draw[draw, solid] (1.5,0)--(a5);
\draw[draw, solid,->] (a5)--(3.75,0);
\draw[draw, solid] (3.75,0)--(a3);
\node[thick,rotate=45] at (1.25,1.25) {\Large$\times$};
\node[thick,rotate=45] at (3.75,1.25) {\Large$\times$};
\node[thick,rotate=45] at (3.75,-1.25) {\Large$\times$};
\node[thick,rotate=45] at (1.25,-1.25) {\Large$\times$};

\node[above] at (1.5,0.125) {\large $x^{\frac{2}{3}}\frac{a}{b}$};
\node[above] at (3.5,0.125) {\large $x^{\frac{2}{3}}ab$};
\node[above] at (1.1,1.5) {\large $x^{\frac{1}{3}}\frac{b^3}{a}$};
\node[above] at (3.9,1.5) {\large $x^{\frac{1}{3}}\frac{1}{ab^3}$};
\node[below] at (1.1,-1.5) {\large $x^{\frac{1}{3}}\frac{b^3}{a}$};
\node[below] at (3.9,-1.5) {\large $x^{\frac{1}{3}}\frac{1}{ab^3}$};

\end{tikzpicture}
\caption{The $3d$ quiver theory associated with the compactification of the $5d$ $E_{7}$ SCFTs on a torus with flux $(\sqrt{2};0,0,0,0,0,0)$. In the figure we have used fugacities to represent the charges of the fields under the global symmetries, specifically, the fugacities $a$ and $b$ for the two $U(1)$ global symmetries and $x$ for the R-symmetry. The theory also contains various superpotential interactions, which corresponds to all operators build from the fields that carry R-charge $2$ and are global symmetry singlets. For completeness, we note that these include the two quartic superpotentials running along the upper and lower triangles, the flipping superpotentials and the four monopole superpotential of the type $(1,1)$ for all adjacent $SU(2)$ groups.}
\label{QuiverE760}
\end{figure}

We consider the last possible case corresponding to four domain walls, which is generated by gluing two of the tubes of figure \ref{DTubes} (b) for $N_f=x=6$. As such we expect the resulting theory to be associated with the flux $(\sqrt{2};0,0,0,0,0,0)$. The resulting model is represented in figure \ref{QuiverE760}. The superpotential consists of a quartic interaction between the bifundamentals and the $SU(6)$ fundamentals for each triangle in the quiver, as well as the standard flipping terms. Moreover, in analogy with the previous cases, all the $(1,1)$ monopoles associated with adjacent $SU(2)$ groups are turned on in the superpotential. This preserves two $U(1)$ symmetries which, together with the R-symmetry, we shall parameterize as written in figure \ref{QuiverE760}.

The flux vector for this case preserves $SO(12)\times U(1)$, which is expected to get enhanced from the manifest $SU(6)\times U(1)_b$ of the quiver, and turns out to be related by an element of the Weyl group of $E_7$ to the flux $\frac{1}{2}(\sqrt{2};1,1,1,1,1,1)$ that we considered in subsection 3.1.2 (see table \ref{FluxE72} in appendix A). We thus expect the compactifications associated to these two fluxes to lead to the same $3d$ $\mathcal{N}=2$ SCFT. From the three-dimensional perspective, this means that the theories in figures \ref{QuiverE7F2} and \ref{QuiverE760} are dual in the IR. We indeed checked that the indices of the two theories agree up to order $x^{3}$. In the following, we are going to argue that this duality that we can predict from the $5d$ point of view is nothing but an instance of Aharony duality in $3d$ \cite{Aharony:1997gp}.

Let us denote by $P_i$ and $F_i$ with $i=1,\cdots,4$ the bifundamentals and the flipping fields respectively, and by $Q$, $\tilde{Q}$ the fundamentals. Moreover, we will label the monopole operators with a set of fluxes ordered such that the first entry corresponds to the left gauge node and the subsequent entries to the nodes obtained from it and moving clockwise in the quiver, with the notation  that a $\bullet$ corresponds to a unit of flux under the associated node and a zero corresponds to no flux. The superpotential of the theory in figure \ref{QuiverE760} then explicitly reads
\bea\label{superpotE760}
\mathcal{W}=\mathfrak{M}^{\bullet,\bullet,0,0}+\mathfrak{M}^{0,\bullet,\bullet,0}+\mathfrak{M}^{0,0,\bullet,\bullet}+\mathfrak{M}^{\bullet,0,0,\bullet}+\sum_{i=1}^4F_iP_i^2+Q\,\widetilde{Q}(P_1P_2+P_3P_4)\,.
\eea
We can dualize the upper $SU(2)$ node using Aharony duality, which confines since it only sees 4 chirals. In appendix C we give more details on the effect of the confining Aharony duality in a quiver with $SU(2)$ gauge nodes and on how the gauge invariant operators are mapped across such duality, see in particular \eqref{monopolemap} for the map of the monopole operators. Here we shall just state the main results that we need for our current purposes. Specifically, the effect of the duality is to remove the upper $SU(2)$ node, leaving behind an $SU(2)\times SU(2)$ bifundamental $B_1$ for the left and right nodes and three gauge singlets $S_1$, $\alpha$ and $\beta$ with the cubic interaction $S_1(B_1^2+\alpha\beta)$. Moreover, we need to know how the operators appearing in the superpotential \eqref{superpotE760} are mapped across the duality
\bea
&&\mathfrak{M}^{\bullet,\bullet,0,0}\leftrightarrow\mathfrak{M}^{\bullet,0,0},\quad \mathfrak{M}^{0,\bullet,\bullet,0}\leftrightarrow\mathfrak{M}^{0,\bullet,0},\quad \mathfrak{M}^{0,0,\bullet,\bullet}\leftrightarrow\alpha\,\mathfrak{M}^{0,\bullet,\bullet},\quad \nn\\
&&\mathfrak{M}^{\bullet,0,0,\bullet}\leftrightarrow\beta\,\mathfrak{M}^{\bullet,0,\bullet},\quad P_1^2\leftrightarrow\alpha,\quad P_2^2\leftrightarrow\beta,\quad P_1P_2\leftrightarrow B_1
\eea
Using these results, we find that the dual frame is described by the quiver
\bea\label{torusint}
\begin{tikzpicture}[baseline=0, font=\scriptsize]
\node[draw, circle] (a1) at (0,0) {2};
\node[draw, circle] (a2) at (2,-2) {2};
\node[draw, circle] (a3) at (4,0) {2};
\node[draw, rectangle] (a4) at (2,2) {6};
\draw[draw, solid] (a1)--(a2);
\draw[draw, solid] (a2)--(a3);
\draw[draw, solid] (a1)--(a3);
\draw[draw, solid,->] (a1)--(1,1);
\draw[draw, solid] (1,1)--(a4);
\draw[draw, solid,->] (a4)--(3,1);
\draw[draw, solid] (3,1)--(a3);
\node[thick] at (2,0) {\Large$\times$};
\node[thick,rotate=45] at (1,-1) {\Large$\times$};
\node[thick,rotate=45] at (3,-1) {\Large$\times$};
\node[below left] at (0.5,-0.5) {$P_4$};
\node[below right] at (3.5,-0.5) {$P_3$};
\node[above] at (1,1.2) {$Q$};
\node[above] at (3,1.2) {$\widetilde{Q}$};
\node[below] at (1,-1.2) {$F_3$};
\node[below] at (3,-1.2) {$F_4$};
\node[above] at (2,0.1) {$S_1$};
\node[above] at (1.2,0.1) {$B_1$};
\end{tikzpicture}
\eea
with the superpotential
\bea
\mathcal{W}&=&\mathfrak{M}^{\bullet,0,0}+\mathfrak{M}^{0,\bullet,0}+\alpha\,\mathfrak{M}^{0,\bullet,\bullet}+\beta\,\mathfrak{M}^{\bullet,0,\bullet}+\nn\\
&+&\alpha\,F_1+\beta\,F_2+S_1(B_1^2+\alpha\,\beta)+\sum_{i=3}^4F_iP_i^2+Q\,\widetilde{Q}(B_1+P_3P_4)\,.
\eea
In the drawing we are not representing the two original flipping fields $F_1$ and $F_2$ and the two singlets $\alpha$ and $\beta$ produced by the duality, while we are representing the singlet $S_1$ with a cross over the bifundamental $B_1$. From the superpotential we can see that all of the 4 singlets $F_1$, $F_2$, $\alpha$ and $\beta$ are massive and can be integrated out. Once we do that, all the remaining matter content is the one depicted in \eqref{torusint} and the superpotential is
\bea
\mathcal{W}=\mathfrak{M}^{\bullet,0,0}+\mathfrak{M}^{0,\bullet,0}+S_1B_1^2+\sum_{i=3}^4F_iP_i^2+Q\,\widetilde{Q}(B_1+P_3P_4)\,.
\eea

Now we can also dualize the lower $SU(2)$ node, which again confines. We apply the same strategy of using the operator map \eqref{monopolemap} worked out in the appendix  to understand what are the new quiver and superpotential. The quiver is
\bea\label{torus1}
\begin{tikzpicture}[baseline=0, font=\scriptsize]
\node[draw, circle] (a1) at (0,0) {2};
\node[draw, circle] (a3) at (4,0) {2};
\node[draw, rectangle] (a4) at (2,2) {6};
\draw[draw, solid] (a1) edge [out=30,in=150,loop,looseness=1] (a3);
\draw[draw, solid] (a1) edge [out=-30,in=-150,loop,looseness=1] (a3);
\draw[draw, solid,->] (a1)--(1,1);
\draw[draw, solid] (1,1)--(a4);
\draw[draw, solid,->] (a4)--(3,1);
\draw[draw, solid] (3,1)--(a3);
\node[thick] at (2,0.65) {\Large$\times$};
\node[thick] at (2,-0.65) {\Large$\times$};
\node[above] at (1,1.2) {$Q$};
\node[above] at (3,1.2) {$\widetilde{Q}$};
\node[above] at (2,0) {$S_1$};
\node[above] at (1.2,0) {$B_1$};
\node[below] at (2,-0.8) {$S_2$};
\node[below] at (2.8,-0.6) {$B_2$};
\end{tikzpicture}
\eea
and the superpotential is
\bea
\mathcal{W}=\alpha\,\mathfrak{M}^{0,\bullet}+\beta\,\mathfrak{M}^{\bullet,0}+\alpha\,F_3+\beta\,F_4+S_2\alpha\,\beta+\sum_{i=1}^2S_iB_i^2+Q\,\widetilde{Q}(B_1+B_2)\,.
\eea
Again we are not representing the flipping fields $F_3$, $F_4$, $\alpha$, $\beta$ which turn out to be massive. Integrating them out we obtain the theory whose matter content is completely represented in \eqref{torus1} and whose superpotential is
\bea
\mathcal{W}=\sum_{i=1}^2S_iB_i^2+Q\widetilde{Q}(B_1+B_2)\,.
\eea
This is precisely the torus model with flux $\frac{1}{2}(\sqrt{2};1,1,1,1,1,1)$ that we analyzed in subsection \ref{2DWNf6}.

\subsection{$N_f=5$}

\subsubsection{$U(1)\times SU(6)$ case}

\begin{figure}
\center
\begin{tikzpicture}[baseline=0, font=\scriptsize]
\node[draw, circle] (a1) at (0,0) {\large $\,\, 2\,\,$};
\node[draw, circle] (a2) at (2.5,2.5) {\large $\,\, 2\,\,$};
\node[draw, circle] (a3) at (5,0) {\large $\,\, 2\,\,$};
\node[draw, circle] (a4) at (2.5,-2.5) {\large $\,\, 2\,\,$};
\node[draw, rectangle] (a5) at (2.5,0) {\large $\,\, 5\,\,$};
\draw[draw, solid] (a1)--(a2);
\draw[draw, solid] (a2)--(a3);
\draw[draw, solid] (a3)--(a4);
\draw[draw, solid] (a4)--(a1);
\draw[draw, solid] (a1)--(1.25,0);
\draw[draw, solid,<-] (1.25,0)--(a5);
\draw[draw, solid,->] (a2)--(2.5,1);
\draw[draw, solid] (2.5,1)--(a5);
\draw[draw, solid,->] (a5)--(3.75,0);
\draw[draw, solid] (3.75,0)--(a3);
\draw[draw, solid] (a5)--(2.5,-1);
\draw[draw, solid,<-] (2.5,-1)--(a4);
\node[thick,rotate=45] at (1.25,1.25) {\Large$\times$};
\node[thick,rotate=45] at (3.75,1.25) {\Large$\times$};
\node[thick,rotate=45] at (3.75,-1.25) {\Large$\times$};
\node[thick,rotate=45] at (1.25,-1.25) {\Large$\times$};

\node[above] at (1.5,0.125) {\large $ab$};
\node[above] at (3,0.5) {\large $\frac{a}{b}$};
\node[above] at (3.5,-0.7) {\large $ab$};
\node[above] at (2,-1.25) {\large $\frac{a}{b}$};
\node[above] at (1.1,1.5) {\large $\frac{1}{a^2}$};
\node[above] at (3.9,1.5) {\large $\frac{1}{a^2}$};
\node[below] at (1.1,-1.5) {\large $\frac{1}{a^2}$};
\node[below] at (3.9,-1.5) {\large $\frac{1}{a^2}$};

\end{tikzpicture}
\caption{The $3d$ quiver theory associated with the compactification of the $5d$ $E_6$ SCFT on a torus with flux $(\sqrt{3};1,1,1,1,1)$. There are Chern-Simons terms of levels $(\frac{1}{2},-\frac{1}{2},\frac{1}{2},-\frac{1}{2})$. The theory also contains eight cubic superpotential interactions, four of which running along the triangles, and the other four being the flipping superpotentials. Additionally, there are four monopole superpotential of the type $(1,1)$ for all adjacent $SU(2)$ groups, properly dressed by $SU(2)\times SU(2)$ bifundamentals to form a gauge invariant.}
\label{QuiverE605}
\end{figure}
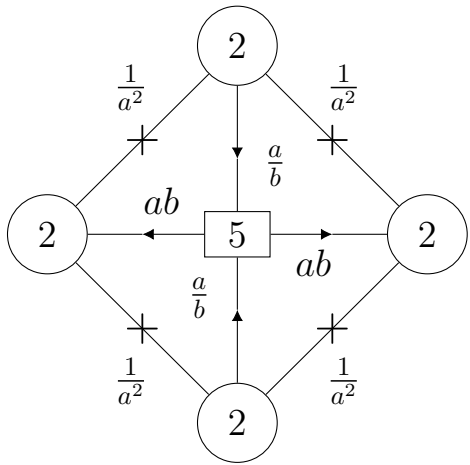

Here we concentrate on the $N_f = 5$ case. The first model we consider is obtained by gluing four copies of the tube in figure \ref{BTube}, and so it is expected to be associated with the flux $(\sqrt{3};1,1,1,1,1)$, which is equivalent to flux two in a $U(1)$ whose commutant in $E_6$ is $SU(6)$. The quiver associated with this theory is shown in figure \ref{QuiverE605}. There are cubic superpotentials associated with each triangle, and we have also turned on superpotentials associated with the $(1,1)$ monopoles for adjacent groups. Since we also have Chern-Simons terms at level $(\frac{1}{2},-\frac{1}{2},\frac{1}{2},-\frac{1}{2})$, these monopole operators should be dressed with one power of the corresponding bifundamentals to make them gauge invariant. There are two $U(1)$ global symmetry groups consistent with the superpotentials. Additionally, there are the $SU(5)$ global symmetry associated with the four collections of five flavors, and the $U(1)_R$ R-symmetry, which we choose such that all the chirals have R-charge $\frac{2}{3}$. In terms of this symmetry, the superconformal value of the R-symmetry is given by 
\begin{equation}
\hat{R}=R-0.03846q_{a}\,.
\end{equation}

We can evaluate the index of this theory (using $U(1)_R$) finding:
\bea
\mathcal{I} &=& 1 + 4 a^4 x^{\frac{2}{3}} + 2 \left(a^2 (b^2 \bold{10} + \frac{1}{b^2}\overline{\bold{10}} ) + 5a^8 \right)x^{\frac{4}{3}} + 4 \left(5a^{12} + 2a^6 (b^2 \bold{10} + \frac{1}{b^2}\overline{\bold{10}}) \right)x^2 + \nn\\
&+&\left(-\frac{2}{a^2}(b^2 \bold{10} + \frac{1}{b^2}\overline{\bold{10}} )+a^4(3(b^4{\bf \overline{50}}+b^{-4}{\bf 50})+b^4{\bf \overline{45}}+b^{-4}{\bf 45}+4\times{\bf 75}-(b^4{\bf \overline{5}}+b^{-4}{\bf 5}))+\right.\nn\\
&+&\left.\frac{3}{a^8}+35a^{16}+20a^{10}(b^2 \bold{10} + \frac{1}{b^2}\overline{\bold{10}} )\right)x^{\frac{8}{3}}+ \cdots\,.
\eea
The index forms characters of $SU(6)$, which is enhanced from $SU(5)\times U(1)_b$ where the embedding is the same one, \eqref{emb6}, that we used in subsection \ref{2DWNf5}. Specifically, the index can be rewritten in characters of $SU(6)$ as
\bea
\mathcal{I} &=& 1 + 4 a^4 x^{\frac{2}{3}} + 2 (a^2 \bold{20} + 5a^8)x^{\frac{4}{3}} + 4 (5a^{12} + 2a^6 \bold{20} )x^2+\nn\\
&+& \left(-\frac{2}{a^2}{\bf 20}+a^4(3\times{\bf 175}+{\bf 189}+1-{\bf 35})+\frac{3}{a^8}+35a^{16}+20a^{10}{\bf 20}\right)x^{\frac{8}{3}}+\cdots\,.
\eea
This result is consistent with our expectations. Indeed, we can see not only the symmetry preserved by the flux of the $5d$ compactification, but also the spectrum of operators expected from $5d$. This is the same as discussed around equation \eqref{E6adjBR}, but this time the multiplicities of all the operators are doubled since the flux is two. 

Similarly, the structure of the conformal manifold is the same as the one for the theory with minimal flux. There is one difference though, and that is that the marginal operators coming from the monopole superpotentials now are associated directly with the dressed $(1,1)$ monopole operators for adjacent groups without involving the flip fields. This, however, does not seem to have any effect on this discussion. Like in the previous case, we expect a $1d$ subspace where the symmetry enhances to $U(1)\times SU(6)$, and the superconformal index appears consistent with this. 

Finally, another evidence for the proposed enhancement of the $SU(5)\times U(1)_{b}$ part of the symmetry to $SU(6)$ comes from examining the central charges of these symmetries, as before. We compute numerically the real part of the free energy of this model, and use it to find the following values for the central charges of $U(1)_b$ and of the Cartan $\textrm{diag}\left(1,0,0,0,-1\right)$ of $SU(5)$, which we denote by $C$,
\begin{equation}
C_{b}=20\,\,\,,\,\,\,C_{C}=3\,.
\end{equation}
The value of the ratio of these charges is
\begin{equation}
\label{ratiogfd}
\frac{C_{b}}{C_{C}}=\frac{20}{3},
\end{equation}
which matches our expectations from the proposed symmetry enhancement and the embedding \eqref{emb6}. Indeed, exactly as in Eq. \eqref{cbccsu6}, this ratio is expected to be equal to $20/3$, which agrees with our numerical result \eqref{ratiogfd}. 

Next we consider similar quivers, but with the flavors allocated differently.

\subsubsection{$U(1)\times SU(2)\times SU(3)^2$ case}

\begin{figure}[t]
\center
\includegraphics[width=0.6\textwidth]{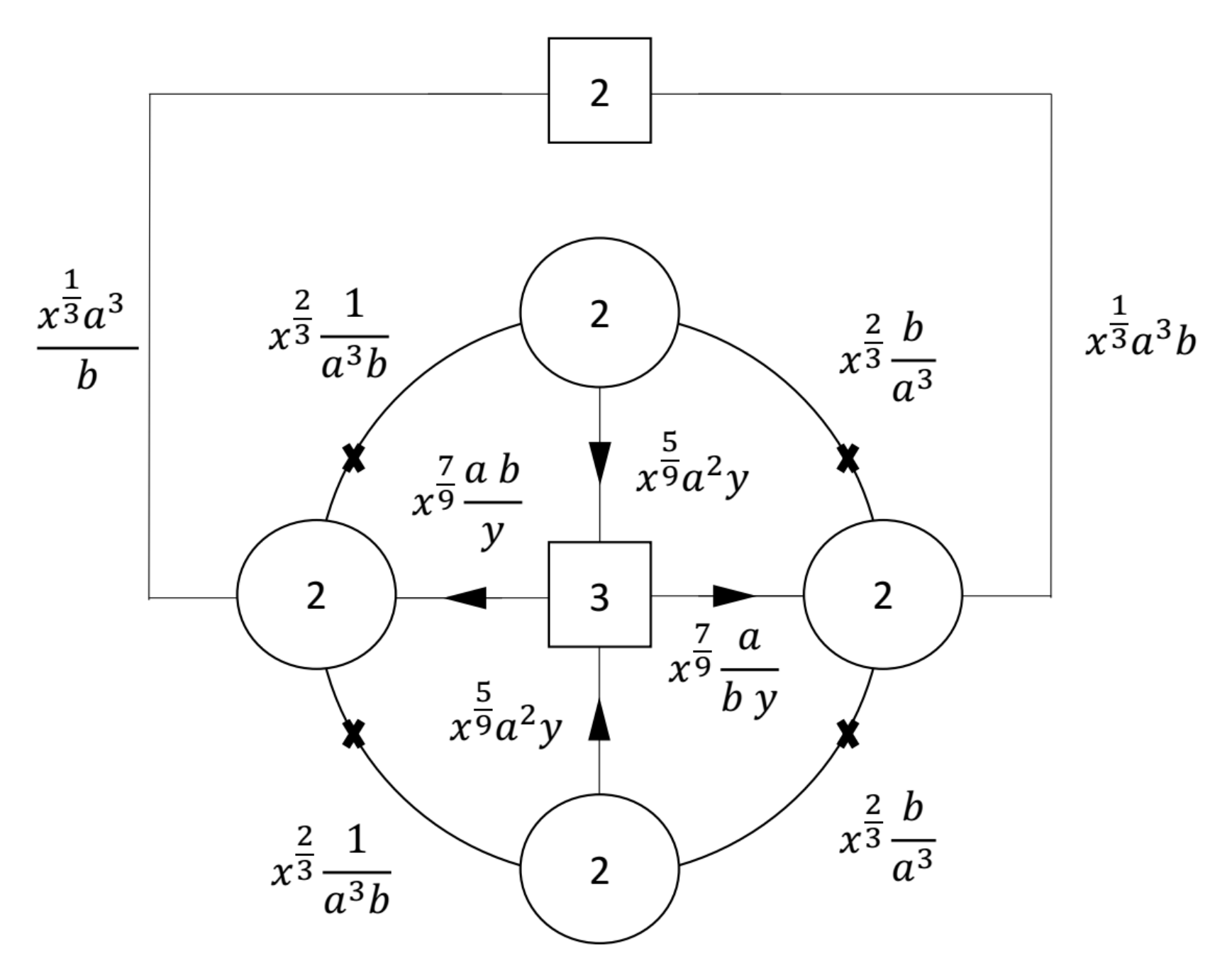} 
\caption{The $3d$ quiver theory associated with the compactification of the $5d$ $E_6$ SCFT on a torus with flux $(\sqrt{3};1,1,1,0,0)$. There are Chern-Simons terms of levels $(\frac{1}{2},-\frac{1}{2},\frac{1}{2},-\frac{1}{2})$. The theory also contains cubic superpotential interactions associated with the four triangles, and the flipping superpotentials. Additionally, there are two quartic superpotential involving the bifundamentals charged under the global $SU(2)$ group with the upper and lower semi-circles, and four monopole superpotential of the type $(1,1)$ for all adjacent $SU(2)$ groups, properly dressed by $SU(2)\times SU(2)$ bifundamentals to form a gauge invariant.}
\label{QuiverE632}
\end{figure}

We next consider the model shown in figure \ref{QuiverE632}. This is generated by gluing two of the tubes of figure \ref{DTubes} (b) for $N_f=5$ and $x=2$. As such we expect the resulting theory to be associated with the flux $(\sqrt{3};1,1,1,0,0)$. This corresponds to the minimal value of flux preserving the $U(1)\times SU(2)\times SU(3)^2$ subgroup of $E_6$.

The charges under the three $U(1)$ groups, the $SU(2)\times SU(3)$ non-abelian symmetry groups, and the R-symmetry that we shall use are shown in the figure. For latter convenience, we have slightly changed the charges under various symmetries compared to the previous sections. We note that here the model includes superpotentials involving all four triangles, and the two loops involving the bifundamentals charged under the global $SU(2)$ with the upper and lower semi-circles. Furthermore, following the discussion in the previous sections, also all four $(1,1)$ monopoles for adjacent groups, which should be dressed with the bifundamentals because of the Chern-Simons levels $(\frac{1}{2},-\frac{1}{2},\frac{1}{2},-\frac{1}{2})$. The latter is responsible for breaking a possible $U(1)$ symmetry. The superconformal R-symmetry is given by the following expression in terms of the symmetries appearing in the figure: 
\begin{equation}
\hat{R}=R-0.0009q_{a}\,.
\end{equation} 

We can compute the index for this model (using $U(1)_R$) finding:
\bea
\mathcal{I} & = & 1 + 3 a^{6} \left(b^2 + \frac{1}{b^2}\right) x^{\frac{2}{3}} + 2 a^{4} \left(y^2 + \frac{1}{y}\bold{2}_{SU(2)}\right) \bold{3}_{SU(3)} x^{\frac{10}{9}} +\\ \nonumber & + & 3 a^{12} \left(2b^4 + 3 + \frac{2}{b^4}\right) x^{\frac{4}{3}} + a^{2} \left((b^2 + \frac{1}{b^2}\right) \left(\frac{1}{y^2} + y\bold{2}_{SU(2)}\right) \overline{\bold{3}}_{SU(3)} x^{\frac{14}{9}} +\\ \nonumber & + & 6 a^{10} \left(b^2 + \frac{1}{b^2}\right) \left(y^2 + \frac{1}{y}\bold{2}_{SU(2)}\right) \bold{3}_{SU(3)} x^{\frac{16}{9}} + 2 a^{18} \left(5b^{6} + 9b^2 + \frac{9}{b^2} + \frac{5}{b^{6}}\right) x^{2} + \cdots\,.
\eea
We note that the index forms characters of $SU(2)\times SU(3)^2$, where the embedding is such that
\be
\label{emb1233}
\bold{2}_{SU(2)} = b^2 \oplus \frac{1}{b^2},\quad \bold{3}_{SU(3)_2} = y^2 \oplus \frac{1}{y}\bold{2}_{SU(2)}\,.
\ee
In terms of characters of these groups the index reads:   
\bea\label{indNf5x2}
I & = & 1 + 3 a^{6} \bold{2}_{SU(2)} x^{\frac{2}{3}} + 2 a^{4} \bold{3}_{SU(3)_1} \bold{3}_{SU(3)_2} x^{\frac{10}{9}} + 3 a^{12} (1+2\times\bold{3}_{SU(2)}) x^{\frac{4}{3}} +\nn\\
&+& a^{2} \bold{2}_{SU(2)} \overline{\bold{3}}_{SU(3)_1} \overline{\bold{3}}_{SU(3)_2} x^{\frac{14}{9}} + 6 a^{10} \bold{2}_{SU(2)} \bold{3}_{SU(3)_1} \bold{3}_{SU(3)_2} x^{\frac{16}{9}} +\nn\\
&+& 2 a^{18} (5\times\bold{4}_{SU(2)} + 4\times\bold{2}_{SU(2)}) x^{2} + ...
\eea
We again see that the spectrum of states is consistent with the $5d$ expectations, namely it organizes into representations of the $U(1)\times SU(2)\times SU(3)^2$ symmetry preserved by the flux and we can also see in the index the contributions of relevant and irrelevant operators coming from the broken currents of the $5d$ global symmetry. In this case, the current in the adjoint representation of $E_6$ splits under the $U(1)\times SU(2)\times SU(3)^2$ subgroup according to the branching rule
\bea
{\bf 78}&\to&({\bf 1},{\bf 1},{\bf 1})^0\oplus({\bf 3},{\bf 1},{\bf 1})^0\oplus({\bf 1},{\bf 8},{\bf 1})^0\oplus({\bf 1},{\bf 1},{\bf 8})^0\oplus({\bf 2},{\bf 1},{\bf 1})^{\pm3}\oplus\nn\\
&&\oplus({\bf 1},{\bf 3},{\bf 3})^2\oplus({\bf 1},{\bf \overline{3}},{\bf \overline{3}})^{-2}\oplus({\bf 2},{\bf \overline{3}},{\bf \overline{3}})^1\oplus({\bf 2},{\bf 3},{\bf 3})^{-1}\,.
\eea
The $5d$ R-symmetry is related to the $3d$ R-symmetry that we used for our computation by a mixing with $U(1)_a$ that can be implemented in the index by the shift $a\to ax^{\frac{2}{9}}$. Moreover, the $U(1)_a$ symmetry is related to the $U(1)$ symmetry inside the $5d$ global symmetry for which we turned on a flux by a normalization of $\frac{1}{2}$. With this dictionary, we can immediately identify the states $({\bf 2},{\bf 1},{\bf 1})^3$, $({\bf 1},{\bf 3},{\bf 3})^2$, $({\bf 2},{\bf \overline{3}},{\bf \overline{3}})^1$ in the index \eqref{indNf5x2}: they are represented by the contributions $3 a^{6} \bold{2}_{SU(2)} x^{\frac{2}{3}}$, $2 a^{4} \bold{3}_{SU(3)_1} \bold{3}_{SU(3)_2} x^{\frac{10}{9}}$, $a^{2} \bold{2}_{SU(2)} \bold{\overline{3}}_{SU(3)_1} \bold{\overline{3}}_{SU(3)_2} x^{\frac{14}{9}}$. Notice that, as expected, their multiplicities are given by their $U(1)$ charges times the $U(1)$ flux, which in this case is one. The other states appear at higher orders, but we checked their presence up to order $x^3$.

Another evidence for the proposed enhancement of the $SU(2)\times U(1)_{y}$ part of the symmetry to $SU(3)_2$ comes from examining the central charges of these symmetries, as before. We compute numerically the real part of the free energy of this model, and use it to find the following ratio between the central charges of $U(1)_y$ and the Cartan $\textrm{diag}\left(1,-1\right)$ of $SU(2)$, which we denote by $C$,
\begin{equation}
\label{ratio1233}
\frac{C_{y}}{C_{C}}=3.3.
\end{equation}
This matches our expectations from the proposed symmetry enhancement and the embedding \eqref{emb1233}. Indeed, the corresponding embedding indices are 
\begin{equation}
I_{SU(2)\rightarrow SU(3)_{2}}=1\,,\,\,\,I_{U(1)_{C}\rightarrow SU(2)}=4\,,\,\,\,I_{U(1)_{y}\rightarrow SU(3)_{2}}=12\,,
\end{equation}
implying the following ratio of central charges, 
\begin{equation}
\frac{C_{y}}{C_{C}}=3
\end{equation}
which agrees with our numerical result \eqref{ratio1233} within the accuracy of the calculation. 

\subsubsection{$U(1)^2\times SO(8)$ case}

\begin{figure}[t]
\center
\begin{tikzpicture}[baseline=0, font=\scriptsize]
\node[draw, circle] (a1) at (0,0) {\large $\,\, 2\,\,$};
\node[draw, circle] (a2) at (2.5,2.5) {\large $\,\, 2\,\,$};
\node[draw, circle] (a3) at (5,0) {\large $\,\, 2\,\,$};
\node[draw, circle] (a4) at (2.5,-2.5) {\large $\,\, 2\,\,$};
\node[draw, rectangle] (a5) at (2.5,0) {\large $\,\, 1\,\,$};
ù\node[draw, rectangle] (a6) at (2.5,4) {\large $\,\, 4\,\,$};
\draw[draw, solid] (a1)--(a2);
\draw[draw, solid] (a2)--(a3);
\draw[draw, solid] (a3)--(a4);
\draw[draw, solid] (a4)--(a1);
\draw[draw, solid] (a1)--(1.25,0);
\draw[draw, solid,<-] (1.25,0)--(a5);
\draw[draw, solid,->] (a2)--(2.5,1);
\draw[draw, solid] (2.5,1)--(a5);
\draw[draw, solid,->] (a5)--(3.75,0);
\draw[draw, solid] (3.75,0)--(a3);
\draw[draw, solid] (a5)--(2.5,-1);
\draw[draw, solid,<-] (2.5,-1)--(a4);

\draw[draw,solid,->] (a1) -- (0,2.5);
\draw[draw,solid] (0,2.5) -- (0,4);
\draw[draw,solid] (0,4) -- (a6);
\draw[draw,solid] (a6) -- (5,4);
\draw[draw,solid,->] (5,4) -- (5,2.5);
\draw[draw,solid] (5,2.5) -- (a3);

\node[thick,rotate=45] at (1.25,1.25) {\Large$\times$};
\node[thick,rotate=45] at (3.75,1.25) {\Large$\times$};
\node[thick,rotate=45] at (3.75,-1.25) {\Large$\times$};
\node[thick,rotate=45] at (1.25,-1.25) {\Large$\times$};

\node[above] at (1.5,0.125) {\normalsize $xaby^{-1}$};
\node[above] at (3,0.5) {\normalsize $x^{\frac{2}{3}}ay$};
\node[above] at (3.7,-0.7) {\normalsize $xab^{-1}y^{-1}$};
\node[above] at (2,-1.25) {\normalsize $x^{\frac{2}{3}}ay$};
\node[above] at (1.1,1.5) {\large $\frac{x^{\frac{1}{3}}}{a^2b}$};
\node[above] at (3.9,1.5) {\large $\frac{x^{\frac{1}{3}}b}{a^2}$};
\node[below] at (1.1,-1.5) {\large $\frac{x^{\frac{1}{3}}}{a^2b}$};
\node[below] at (3.9,-1.5) {\large $\frac{x^{\frac{1}{3}}b}{a^2}$};
\node[left] at (0,2.5) {\normalsize $x^{\frac{2}{3}}a^2$};
\node[right] at (5,2.5) {\normalsize $x^{\frac{2}{3}}a^2$};

\end{tikzpicture}
\caption{The $3d$ quiver theory associated with the compactification of the $5d$ $E_6$ SCFT on a torus with flux $(\sqrt{3};1,0,0,0,0)$. There are Chern-Simons terms of levels $(\frac{1}{2},-\frac{1}{2},\frac{1}{2},-\frac{1}{2})$. The theory also contains cubic superpotential interactions associated with the four triangles, and the flipping superpotentials. Additionally, there are two quartic superpotential involving the bifundamentals charged under the global $SU(4)$ group with the upper and lower semi-circles, and four monopole superpotential of the type $(1,1)$ for all adjacent $SU(2)$ groups, properly dressed by $SU(2)\times SU(2)$ bifundamentals to form a gauge invariant.}
\label{QuiverE641}
\end{figure}
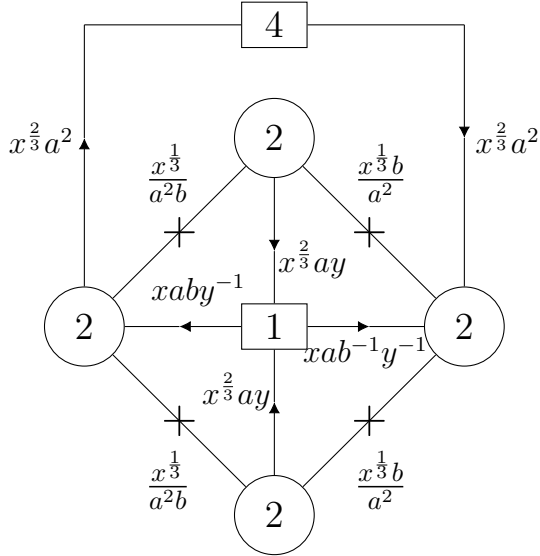

We next consider the model in figure \ref{QuiverE641}. This is generated by gluing two of the tubes of figure \ref{DTubes} (b) for $N_f=5$ and $x=4$. As such we expect the resulting theory to be associated with the flux $(\sqrt{3};1,0,0,0,0)$. This corresponds to the minimal value of flux preserving $U(1)^2\times SO(8)$. Specifically, we have one unit of flux for a $U(1)$ inside the $SO(10)$ contained in the subgroup $U(1)\times SO(10)$ of $E_6$, which breaks it to $U(1)\times SO(8)$.

The charges under the three $U(1)$ groups, the $SU(4)$ non-abelian symmetry, and the R-symmetry that we shall use are shown in the figure. For latter convenience, we have slightly changed the charges under the various symmetries compared to the previous sections. We note that here the model includes superpotentials involving all four triangles, the two loops involving the bifundamentals charged under the global $SU(4)$ and the upper and lower semi-circles, and, following the discussion in the previous sections, also all four $(1,1)$ monopoles for adjacent groups, which should be dressed with the bifundamentals because of the Chern-Simons levels $(\frac{1}{2},-\frac{1}{2},\frac{1}{2},-\frac{1}{2})$. The latter is responsible for breaking a possible $U(1)$ symmetry. The superconformal R-symmetry is given by the following expression in terms of the symmetries appearing in the figure: 
\begin{equation}
\hat{R}=R-0.1026q_{a}\,.
\end{equation}

We can compute the index of this theory (using $U(1)_R$) finding:
\bea
\mathcal{I}&=&1+\left(\frac{3}{a^8}+2a^4(b^2+\frac{1}{b^2}+{\bf 6})\right)x^{\frac{4}{3}}+\left(\frac{a^3}{y}(b{\bf 4}+\frac{1}{b}{\bf \overline{4}})+ay(b{\bf \overline{4}}+\frac{1}{b}{\bf 4})\right)x^{\frac{5}{3}}+\nn\\
&-&\left(\frac{y}{a^3}(b{\bf 4}+\frac{1}{b}{\bf \overline{4}})+\frac{1}{ay}(b{\bf \overline{4}}+\frac{1}{b}{\bf 4})\right)x^{\frac{7}{3}}+\left(-\frac{2}{a^4}(b^2+\frac{1}{b^2}+{\bf 6})+\frac{6}{a^{16}}+\right.\nn\\
&+&\left.a^8(3(b^4+\frac{1}{b^4})+4(b^2+\frac{1}{b^2}){\bf 6}+3\times{\bf 20}+{\bf 15}+3)\right)x^{\frac{8}{3}}+\cdots\,.
\eea
We note that the index forms characters of $SO(8)$, where the embedding can be chosen up to an element of the triality automorphism group of $SO(8)$ to be such that
\be
{\bf 8_v}\to b^2\oplus\frac{1}{b^2}\oplus{\bf 6}\,.
\ee
In terms of characters of this group, the index reads:
\bea\label{indNf5x4}
\mathcal{I}&=&1+\left(\frac{3}{a^8}+2a^4{\bf 8_v}\right)x^{\frac{4}{3}}+\left(\frac{a^3}{y}{\bf 8_c}+ay{\bf 8_s}\right)x^{\frac{5}{3}}-\left(\frac{y}{a^3}{\bf 8_c}+\frac{1}{ay}{\bf 8_s}\right)x^{\frac{7}{3}}+\nn\\
&+&\left(-\frac{2}{a^4}{\bf 8_v}+\frac{6}{a^{16}}+a^8(3\times{\bf 35_v}+{\bf 28}-1)\right)x^{\frac{8}{3}}+\cdots\,.
\eea
We again see that the spectrum of states is consistent with the $5d$ expectations, namely it organizes into representations of the $U(1)^2\times SO(8)$ symmetry preserved by the flux and we can also see in the index the contributions of relevant and irrelevant operators coming from the broken currents of the $5d$ global symmetry. 
In this case, the current in the adjoint representation of $E_6$ splits under the $U(1)_1\times U(1)_2\times SO(8)$ subgroup according to the branching rule
\be\label{E6adjBRSO8}
{\bf 78}\to2\times{\bf 1}^{(0,0)}\oplus{\bf 28}^{(0,0)}\oplus{\bf 8_v}^{(0,\pm2)}\oplus{\bf 8_c}^{(-3,1)}\oplus{\bf 8_c}^{(3,-1)}\oplus{\bf 8_s}^{(3,1)}\oplus{\bf 8_s}^{(-3,-1)}\,.
\ee
In our notation, the $U(1)$ for which we are turning on a unit of flux is $U(1)_2$, whose charges correspond to the second entry in the upper index of each state.
Since in this case we have two abelian symmetries, the dictionary between $5d$ and $3d$ symmetries needed to identify such states in our index computation is slightly more involved. More precisely, the $5d$ R-symmetry is related to the $3d$ R-symmetry that we used for our computation by a mixing with $U(1)_a$ and $U(1)_y$ that can be implemented in the index by the simultaneous shifts $a\to ax^{\frac{1}{6}}$ and $y\to yx^{\frac{1}{6}}$. Moreover, the $U(1)_a$ and $U(1)_y$ symmetries are related to the $U(1)_1$ and $U(1)_2$ symmetries by
\be 
U(1)_a=-\frac{1}{3}U(1)_1+2U(1)_2,\quad U(1)_y=\frac{1}{3}U(1)_1\,.
\ee
With this dictionary, we can immediately identify in the index \eqref{indNf5x4} all the states appearing in the branching rule \eqref{E6adjBRSO8} with the correct multiplicity given by their $U(1)_2$ charge times the flux, which in this case is one. Specifically, we can see the contributions $2a^4{\bf 8_v}x^{\frac{4}{3}}$, $a^3y^{-1}{\bf 8_c}x^{\frac{5}{3}}$, $ay{\bf 8_s}x^{\frac{5}{3}}$, $a^{-3}y{\bf 8_c}x^{\frac{7}{3}}$, $a^{-1}y^{-1}{\bf 8_s}x^{\frac{7}{3}}$ and $-2a^{-4}{\bf 8_v}x^{\frac{8}{3}}$.

\section{Additional cases}
\label{addc}

Finally, we want to consider several additional cases of interest. This includes cases involving more than four domain walls as well as cases where the gluing breaks part of the global symmetry originating from $5d$. 

\subsection{A dual of the $U(1)\times SU(2)\times SU(6)$ case}

We have previously noted that the tube in figure \ref{DTubes} (b) for $x=N_f=6$ is naively dual to the one in figure \ref{BTube}. This has been interpreted has stemming from the fact that the flux associated with both of these tubes is related via an $E_7$ Weyl transformation. However, the flux vector associated with each of these, in our chosen basis, is different. This means that while the theories we get while gluing the tubes to themselves should be equivalent, when glued to other tubes the theories will be different. Specifically, if we glue the two types of tubes together then the resulting theory does not preserve the $SO(12)$ subgroup of $E_7$, even-though each tube individually preserves it but they are embedded differently inside $E_7$. However, the resulting theory is naively equivalent to just gluing one tube to itself, which should have this symmetry. In this section we shall explain how this works, and in the process also encounter an interesting duality.

The basic loophole that allows us to get different theories from gluing the two tubes is the way the monopole operators are mapped. We noted previously that when gluing tubes we also need to introduce monopole superpotentials associated with two adjacent $SU(2)$ groups. However, as illustrated in appendix \ref{App:AharonyDuality}, monopole operators are mapped non-trivially between the two theories. As such, while the resulting theories have the same matter, they may differ by the monopole superpotential. Next we shall illustrate this with an example.

\begin{figure}
\center
\includegraphics[width=0.8\textwidth]{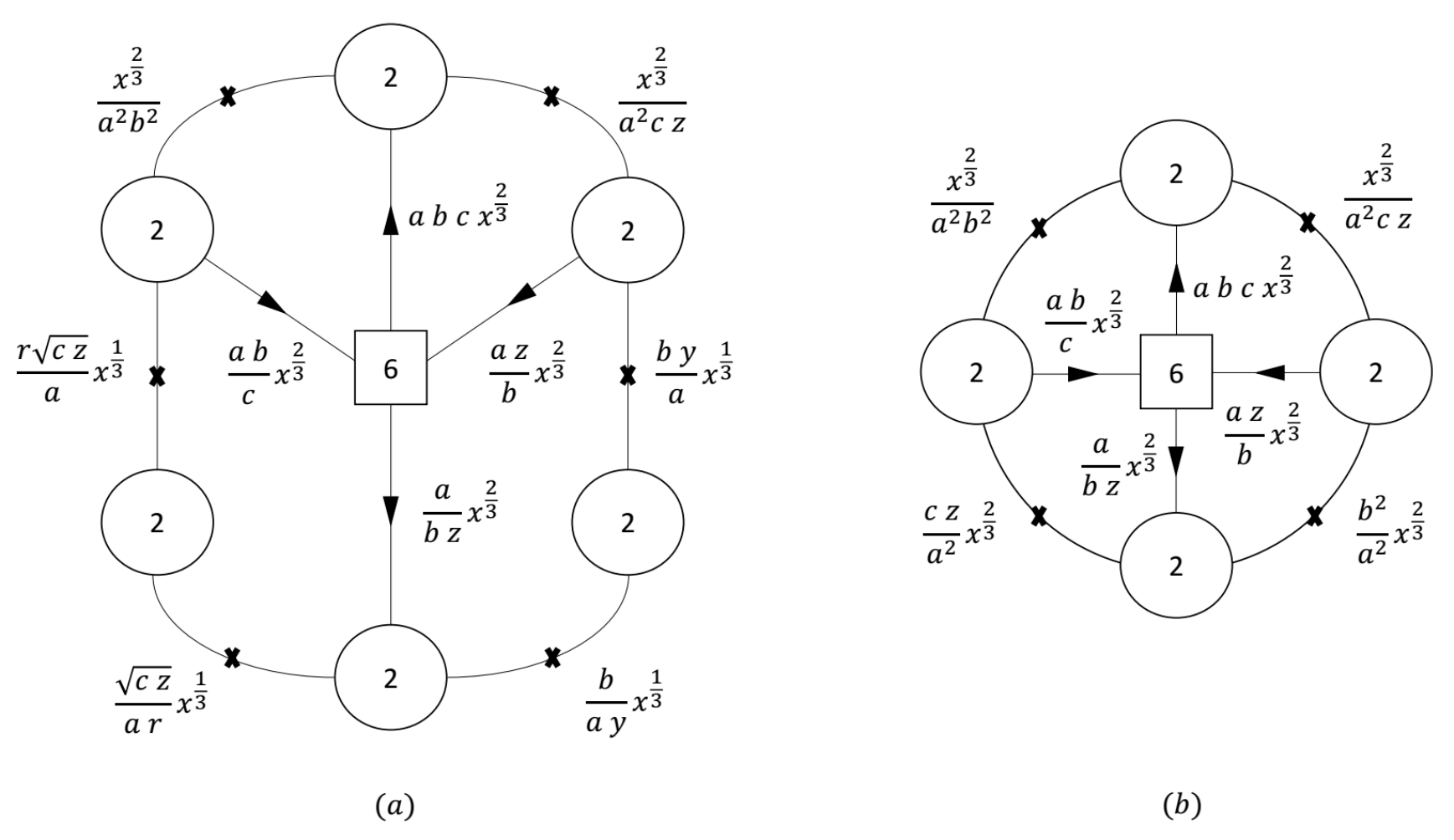} 
\caption{(a) The $3d$ quiver theory we get by gluing three of the tubes in figure \ref{DTubes} (b), two with $x=6$ and one with $x=0$. There are cubic and quartic superpotentials running along the perimeters of the internal faces. (b) The theory we get from (a) after performing Aharony duality on the bottom left and right $SU(2)$ groups. Again, there are cubic superpotentials running along the perimeters of the internal faces.}
\label{QuiverE7060660}
\end{figure}

If we want to avoid breaking symmetries, then the simplest case we can take involves gluing three of the tubes in figure \ref{DTubes} (b), two with $x=6$ and one with $x=0$, which is given by connecting two basic tubes. The resulting theory is shown in figure \ref{QuiverE7060660} (a). For the time being, we shall ignore the monopole superpotentials, which we shall introduce later on. In the figure we have also denoted the charges of all the fields under the global symmetries of the theory via fugacities. Additionally, we have denoted the charges under a specific R-symmetry using the fugacity $x$.   

The flux associated with this theory should be given by summing up the fluxes of the constituent tubes: $2(\frac{\sqrt{2}}{2};0,0,0,0,0,0)+(\frac{\sqrt{2}}{2};\frac{1}{2},\frac{1}{2},\frac{1}{2},\frac{1}{2},\frac{1}{2},\frac{1}{2}) = (\frac{3\sqrt{2}}{2};\frac{1}{2},\frac{1}{2},\frac{1}{2},\frac{1}{2},\frac{1}{2},\frac{1}{2})$. This flux in turn is the minimal one preserving $U(1)\times SU(2)\times SU(6)$, and so the resulting theory should be equivalent to the one in figure \ref{QuiverE742}. 

We can perform Aharony duality on the two nodes seeing four doublets to get the theory in figure \ref{QuiverE7060660} (b). As expected this theory is equivalent to the one in figure \ref{QuiverE7F2}. However, the resulting theory is not the same as the one considered in section \ref{E7flux2}. Recall that in that theory we also included monopole superpotentials under each two adjacent $SU(2)$ groups. We would expect similar superpotentials to be needed also in the theory in figure \ref{QuiverE7060660} (a), although we have refrained from turning them on for now. However, these are not mapped to one another under the duality. Specifically, looking at the charges of the four monopole operators we had in the theory in section \ref{E7flux2}, we see that they originate from the monopole operators with charges: $(0,1,1,1,1,0)$, $(0,0,1,1,1,1)$, $(1,0,0,0,1,1)$ and $(1,1,1,0,0,0)$, where we start from the top $SU(2)$ group in figure \ref{QuiverE7060660} (a), and continue clockwise. These operators differ from the monopole superpotentials that we expect are needed in figure \ref{QuiverE7060660} (a) from our experience so far.

Now let us consider the six monopole operators under each adjacent $SU(2)$ groups in figure \ref{QuiverE7060660} (a). Their charges are (in the theory without the monopole superpotential):

\be \label{Opcharges}
\frac{c z}{a^2 b^8 r^2} x^{\frac{8}{3}}, \frac{b^2}{a^2 y^2 c^4 z^4} x^{\frac{8}{3}}, \frac{r^2 b^2 z^5}{c} x^{2}, \frac{y^2 b^4 z^4}{c^2} x^{2}, \frac{a^2 b^2 c^2}{z^4} x^{\frac{4}{3}}, \frac{a^2 c^4}{b^2 z^2} x^{\frac{4}{3}},
\ee  
where the terms are ordered as: $(1,0,0,0,0,1)$, $(1,1,0,0,0,0)$, $(0,0,1,1,0,0)$, $(0,0,0,1,1,0)$, $(0,1,1,0,0,0)$, $(0,0,0,0,1,1)$. We note that the first four terms are charged under $U(1)_y$ and $U(1)_r$ and have $U(1)_R$ charges bigger or equal to $2$. This is not a coincidence. First we note that $U(1)_y$ and $U(1)_r$ are not present in the dual model in figure \ref{QuiverE7060660} (b). As such, BPS operators charged under them are expected to come in combinations that can form a long multiplet. Indeed the index appears to be independent of these two $U(1)$ groups, implying there are additional operators carrying the same charges, but contributing negatively to the index. There is a result proved in \cite{Razamat:2016gzx} from superconformal representation theory that BPS operators contributing at orders below $x^2$, that is relevant operators, must contribute with a positive sign. As such the last two operators in \eqref{Opcharges}, that are relevant with respect to the superconformal R-symmetry determined in section \ref{E7flux2}, must be uncharged under these symmetries.

\begin{figure}
\center
\includegraphics[width=0.45\textwidth]{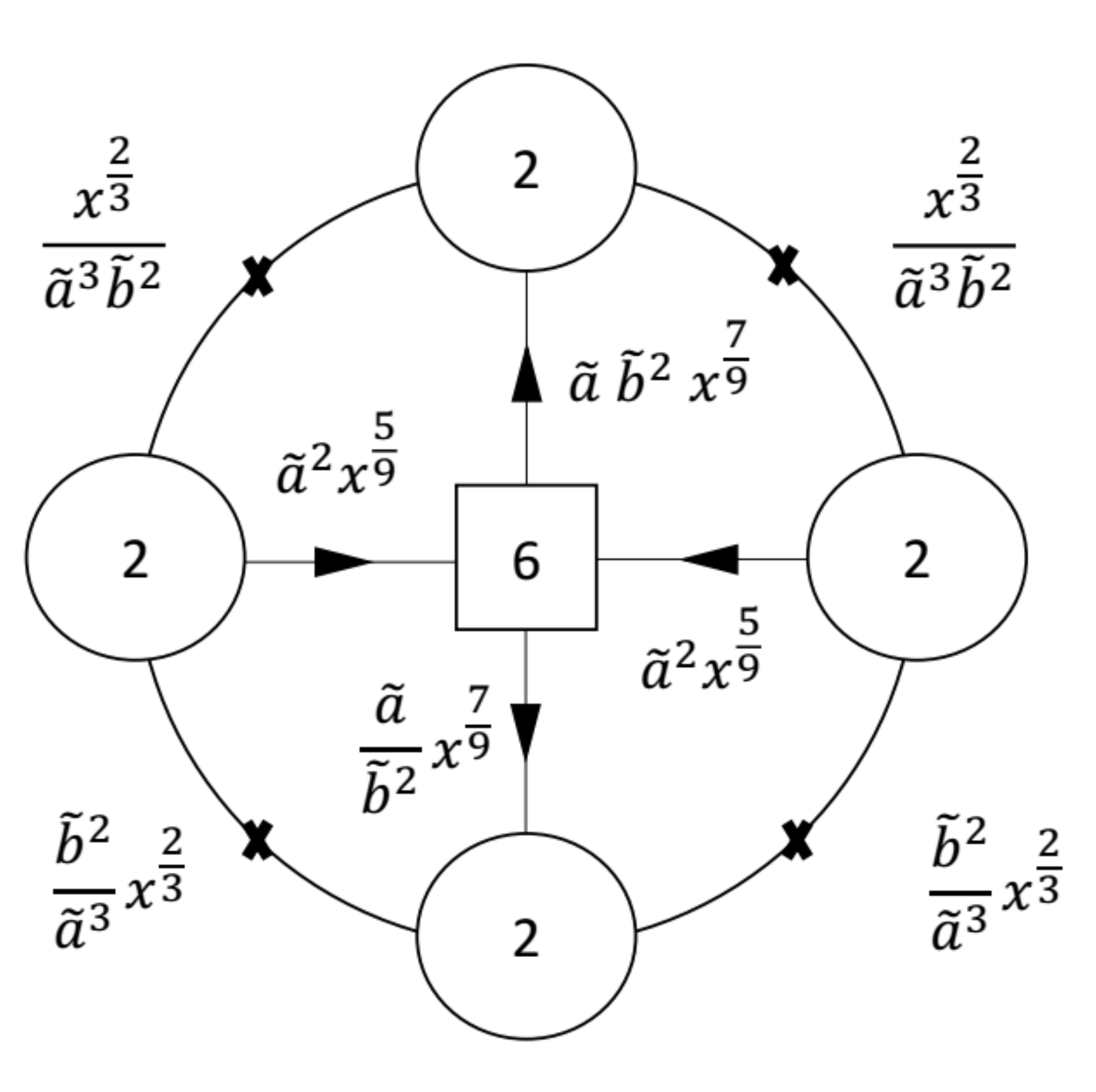} 
\caption{The model in figure (b) after the introduction of monopole superpotentials associated with the basic monopole operators of the left and right $SU(2)$ groups. Additionally, there are cubic superpotentials running along the perimeters of the internal faces.}
\label{QuiverAD}
\end{figure}

We can now consider introducing the six monopole superpotentials to the theory in figure \ref{QuiverE7060660} (b). From the previous discussion we see that the first four of these are canceled in the index, implying they can merge with other short representations to form long multiplets. This suggests that these are irrelevant as their dimension can increase above two. This leaves us with the last two. In the model in figure \ref{QuiverE7060660} (b) these correspond to the basic monopole operators of the left and right $SU(2)$ groups. From our analysis in section \ref{E7flux2}, we see that these are expected to be relevant with respect to the superconformal R-symmetry, and as such turning them on will initiate a flow to a new fixed point. Introducing these operators is expected to set $b = \sqrt{c z}$, $a=\frac{z^{\frac{3}{2}}}{c^{\frac{3}{2}}} x^{\frac{1}{3}}$. This leads to the theory in figure \ref{QuiverAD}, where we have redefined the two $U(1)$ symmetries as: $z=\tilde{b} \sqrt{\tilde{a}}$, $c=\frac{\tilde{b}}{\sqrt{\tilde{a}}}$. Additionally, we have defined a new R-symmetry as: $U(1)_R - \frac{2}{9} U(1)_{\tilde{a}}$, to avoid having fields with zero R-charges.

The claim we expect from the $5d$ picture is that the models in figure \ref{QuiverAD} and figure \ref{QuiverE742} are dual. Indeed, we can compute the index of the model in figure \ref{QuiverAD} and find it equal to the one in \ref{QuiverE742}, if we identify $\tilde{a}$ and $\tilde{b}$ with $a$ and $b$, to the order the index was evaluated. In addition, the $\mathbb{S}^{3}$ partition functions of the two models are equal. Therefore, we see that in this case the $5d$ picture leads to a non-trivial duality between two seemingly different $3d$ quiver gauge theories.

\subsection{$U(1)\times SU(2)\times SU(3)\times SU(4)$ case}

\begin{figure}
\center
\begin{tikzpicture}[baseline=0, font=\scriptsize]
\node[draw, rectangle] (a1) at (0,0) {\large $\,2\,$};
\node[draw, circle] (a2) at (1.5,0) {\large $2$};
\node[draw, circle] (a3) at (3,0) {\large $2$};
\node[draw, circle] (a4) at (4.5,0) {\large $2$};
\node[draw, rectangle] (a5) at (6,0) {\large $\,2\,$};
\node[draw, rectangle,red] (b1) at (3,2.5) {\large $\,2\,$};
\node[draw, rectangle,blue] (b2) at (3,-1.5) {\large $\,2\,$};
\node[draw, rectangle,RawSienna] (b3) at (3,-3) {\large $\,2\,$};

\draw[draw, solid] (a1)--(a2);
\draw[draw, solid] (a2)--(a3);
\draw[draw, solid] (a3)--(a4);
\draw[draw, solid] (a4)--(a5);
\draw[draw, solid] (a1)--(b1);
\draw[draw, solid] (a2)--(b1);
\draw[draw, solid] (a3)--(b1);
\draw[draw, solid] (a4)--(b1);
\draw[draw, solid] (a5)--(b1);
\draw[draw, solid] (a1)--(b2);
\draw[draw, solid] (a3)--(b2);
\draw[draw, solid] (a5)--(b2);
\draw[draw, solid] (a1)--(b3);
\draw[draw, solid] (a5)--(b3);
\draw[draw, solid] (a3) edge [out=225,in=135,loop,looseness=1] (b3);

\node[] at (7.5,0) {\Large $+$};

\node[draw, rectangle] (aa1) at (9,0) {\large $\,2\,$};
\node[draw, circle] (aa2) at (10.5,0) {\large $2$};
\node[draw, circle] (aa3) at (12,0) {\large $2$};
\node[draw, circle] (aa4) at (13.5,0) {\large $2$};
\node[draw, rectangle] (aa5) at (15,0) {\large $2$};
\node[draw, rectangle,red] (bb1) at (12,2.5) {\large $\,2\,$};
\node[draw, rectangle,blue] (bb2) at (12,-1.5) {\large $\,2\,$};
\node[draw, rectangle,RawSienna] (bb3) at (12,-3) {\large $\,2\,$};

\draw[draw, solid] (aa1)--(aa2);
\draw[draw, solid] (aa2)--(aa3);
\draw[draw, solid] (aa3)--(aa4);
\draw[draw, solid] (aa4)--(aa5);
\draw[draw, solid] (aa1)--(bb2);
\draw[draw, solid] (aa2)--(bb2);
\draw[draw, solid] (aa3)--(bb2);
\draw[draw, solid] (aa4)--(bb2);
\draw[draw, solid] (aa5)--(bb2);
\draw[draw, solid] (aa1)--(bb1);
\draw[draw, solid] (aa3)--(bb1);
\draw[draw, solid] (aa5)--(bb1);
\draw[draw, solid] (aa1)--(bb3);
\draw[draw, solid] (aa5)--(bb3);
\draw[draw, solid] (aa3) edge [out=-45,in=45,loop,looseness=1] (bb3);
\end{tikzpicture}
\caption{The $3d$ quiver representations of the tubes with flux $(\sqrt{2};1,1,0,0,0,0)$, on the left, and $(\sqrt{2};0,0,1,1,0,0)$, on the right. In each tube model, the four chirals forming the fundamental representation of the manifest $SU(4)$ symmetry has been split in two plus two, as needed to perform the gluing. We also depict with the same color the flavor $SU(2)$ that get identified in the two tubes after the gluing.}
\label{U1SU2SU3SU4tube}
\end{figure}
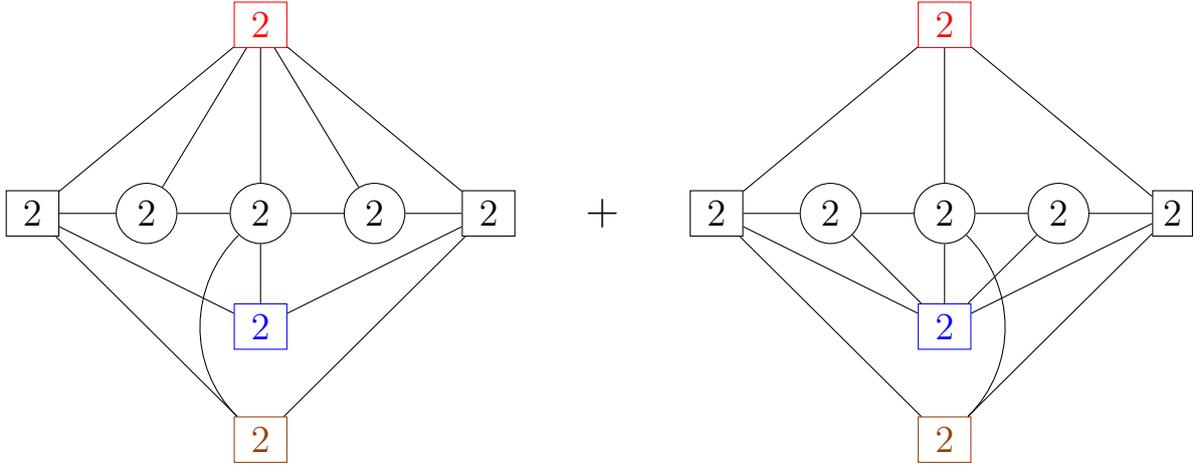

We consider now an example of torus compactification of the rank 1 $E_7$ SCFT with a flux preserving a $U(1)\times SU(2)\times SU(3)\times SU(4)$ subgroup of the $5d$ global symmetry. Such a symmetry can be achieved with a flux vector of the form $(2\sqrt{2};1,1,1,1,0,0)$, which in turn can be constructed by gluing several copies of the basic tube of figure \ref{DTubes} (b) for $x=4$, as we are now going to describe. The resulting model will be associated to eight domain walls and as such its structure will be more complicated than the previous examples, in particular the number of $SU(2)$ gauge nodes will be eight. Nevertheless, we will see that it still behaves as expected from its $5d$ origin.

In order to construct the model, we first observe that the desired flux vector can be obtained as follows:
\be
(2\sqrt{2};1,1,1,1,0,0)=(\sqrt{2};1,1,0,0,0,0)+(\sqrt{2};0,0,1,1,0,0)\,.
\ee
The first flux vector on the r.h.s.~can be simply obtained by gluing two copies of the tube model of figure \ref{DTubes} (b) for $x=4$. The second flux vector is simply obtained from the first one by acting with an element of the Weyl group of $SO(12)$. From our general discussion we have understood that this amounts to permuting the fields receiving different boundary conditions. Each of the two tubes preserves by itself a manifest $SU(4)\times SU(2)$ global symmetry, but because of this relative action of the $SO(12)$ Weyl when we perform the gluing the 4 chirals forming the fundamental of $SU(4)$ in each block should be broken into 2 plus 2, so that the manifest non-abelian symmetry of the resulting tube with flux $(2\sqrt{2};1,1,1,1,0,0)$ is only $SU(2)^3$. This procedure is schematically represented in figure \ref{U1SU2SU3SU4tube}. In the picture, we draw with the same color the $SU(2)$ flavor symmetries that get identified between the two tubes after performing the gluing.

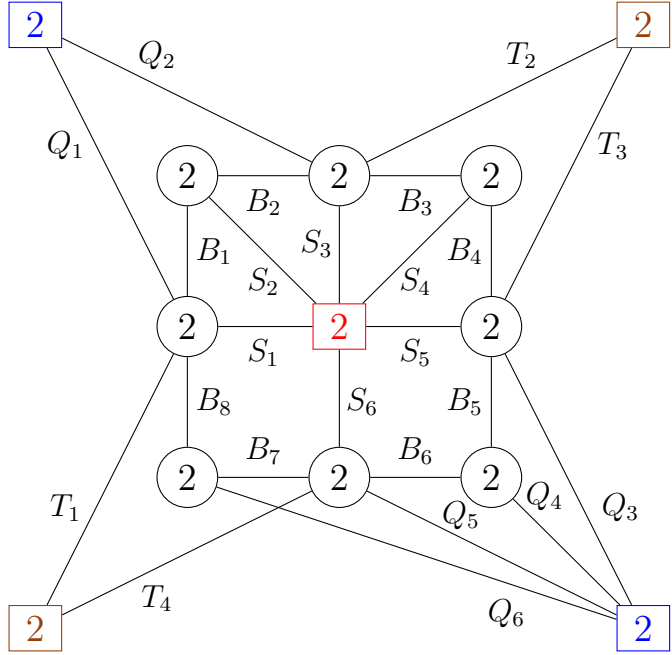
\begin{figure}
\center
\begin{tikzpicture}[baseline=0, font=\scriptsize]

\node[draw, rectangle, red] (b1) at (0,0) {\large $\,2\,$};
\node[draw, circle] (a1) at (-2,0) {\large $2$};
\node[draw, circle] (a2) at (-2,2) {\large $2$};
\node[draw, circle] (a3) at (0,2) {\large $2$};
\node[draw, circle] (a4) at (2,2) {\large $2$};
\node[draw, circle] (a5) at (2,0) {\large $2$};
\node[draw, circle] (a6) at (2,-2) {\large $2$};
\node[draw, circle] (a7) at (0,-2) {\large $2$};
\node[draw, circle] (a8) at (-2,-2) {\large $2$};
\node[draw, rectangle, blue] (b2a) at (-4,4) {\large $\,2\,$};
\node[draw, rectangle, blue] (b2b) at (4,-4) {\large $\,2\,$};
\node[draw, rectangle, RawSienna] (b3a) at (4,4) {\large $\,2\,$};
\node[draw, rectangle, RawSienna] (b3b) at (-4,-4) {\large $\,2\,$};

\draw[draw, solid] (a1)--(a2);
\draw[draw, solid] (a2)--(a3);
\draw[draw, solid] (a3)--(a4);
\draw[draw, solid] (a4)--(a5);
\draw[draw, solid] (a5)--(a6);
\draw[draw, solid] (a6)--(a7);
\draw[draw, solid] (a7)--(a8);
\draw[draw, solid] (a8)--(a1);
\draw[draw, solid] (a1)--(b1);
\draw[draw, solid] (a2)--(b1);
\draw[draw, solid] (a3)--(b1);
\draw[draw, solid] (a4)--(b1);
\draw[draw, solid] (a5)--(b1);
\draw[draw, solid] (a7)--(b1);
\draw[draw, solid] (a1)--(b2a);
\draw[draw, solid] (a3)--(b2a);
\draw[draw, solid] (a5)--(b2b);
\draw[draw, solid] (a6)--(b2b);
\draw[draw, solid] (a7)--(b2b);
\draw[draw, solid] (a8)--(b2b);
\draw[draw, solid] (a1)--(b3b);
\draw[draw, solid] (a3)--(b3a);
\draw[draw, solid] (a5)--(b3a);
\draw[draw, solid] (a7)--(b3b);

\node[] at (-1.65,1) {\normalsize $B_1$};
\node[] at (-1,1.65) {\normalsize $B_2$};
\node[] at (1.65,1) {\normalsize $B_4$};
\node[] at (1,1.65) {\normalsize $B_3$};
\node[] at (1.65,-1) {\normalsize $B_5$};
\node[] at (1,-1.65) {\normalsize $B_6$};
\node[] at (-1.65,-1) {\normalsize $B_8$};
\node[] at (-1,-1.65) {\normalsize $B_7$};
\node[] at (-1,-0.35) {\normalsize $S_1$};
\node[] at (-1,0.6) {\normalsize $S_2$};
\node[] at (-0.3,1.1) {\normalsize $S_3$};
\node[] at (1,0.6) {\normalsize $S_4$};
\node[] at (1,-0.35) {\normalsize $S_5$};
\node[] at (0.3,-1) {\normalsize $S_6$};
\node[] at (-3.6,2.4) {\normalsize $Q_1$};
\node[] at (-2.4,3.6) {\normalsize $Q_2$};
\node[] at (3.7,-2.4) {\normalsize $Q_3$};
\node[] at (2.7,-2.25) {\normalsize $Q_4$};
\node[] at (1.6,-2.5) {\normalsize $Q_5$};
\node[] at (2.2,-3.8) {\normalsize $Q_6$};
\node[] at (3.6,2.4) {\normalsize $T_3$};
\node[] at (2.4,3.6) {\normalsize $T_2$};
\node[] at (-3.6,-2.4) {\normalsize $T_1$};
\node[] at (-2.4,-3.6) {\normalsize $T_4$};
\end{tikzpicture}
\caption{The $3d$ quiver representation of the torus with flux $(2\sqrt{2};1,1,1,1,0,0)$. In order to avoid cluttering the drawing with too many intersecting lines, we depicted the quiver with the convention that $SU(2)$ flavor nodes of the same color should be identified. We also explicitly label each chiral field with a name: $B_i$ for the $SU(2)\times SU(2)$ gauge bifundamentals, $S_j$ for the fundamentals of $SU(2)_1$ (in red), $T_j$ for the fundamentals of $SU(2)_2$ (in blue) and $T_a$ for the fundamentals of $SU(2)_3$ (in brown).}
\label{U1SU2SU3SU4torus}
\end{figure}

Once we have the tube with flux $(2\sqrt{2};1,1,1,1,0,0)$, we can glue its two extremities together so to get a torus with the same value of flux, as desired. The final result is depicted in figure \ref{U1SU2SU3SU4torus}. As mentioned before, the model has eight $SU(2)$ gauge nodes. The superpotential of the theory is analogous to the ones of the previous examples and it consists of various terms. First, we have the usual flips of the quadratic gauge invariants constructed from the $SU(2)\times SU(2)$ bifundamentals of each pair of adjacent gauge nodes. Then we have cubic and quartic superpotentials constrcuted from the bifundamentals and the fundamentals charged under the $SU(2)$ flavor symmetries. These correspond to all the smallest possible closed loops that we can have involving two fundamentals charged under the same flavor $SU(2)$ and under subsequent gauge $SU(2)$. Finally, following the discussion of the previous models, all the $(1,1)$ monopoles for adjacent gauge groups are turned on in the superpotentials. Explicitly, using the notation summarized in figure \ref{U1SU2SU3SU4torus} and calling $F_i$ for $i=1,\cdots,8$ the flipping fields, the full superpotential is
\bea
\mathcal{W}&=&\sum_{i=0}^7\mathfrak{M}^{(i,i+1)}+\sum_{i=1}^8F_iB_i^2+B_1S_1S_2+B_2S_2S_3+B_3S_3S_4+B_4S_4S_5+
B_5S_5S_6B_6+\nn\\
&+&B_7S_6S_1B_8+ B_1Q_1Q_2B_2+B_3Q_2Q_3B_4+B_5Q_3Q_4+B_6Q_4Q_5+
B_7Q_5Q_6+B_8Q_6Q_1+\nn\\
&+&B_1T_1T_2B_2+B_3T_2T_3B_4+B_5T_3T_4B_6
+B_7T_4T_1B_8\,,
\eea
where $\mathfrak{M}^{(i,i+1)}$ is the monopole corresponding to a unit of flux for the $i$-th and the $(i+1)$-th gauge groups only, with the cyclic identification $\mathfrak{M}^{(0,1)}=\mathfrak{M}^{(8,1)}$.

\begin{figure}
\center
\begin{tikzpicture}[baseline=0, font=\scriptsize]

\node[draw, rectangle, red] (b1) at (0,0) {\large $\,2\,$};
\node[draw, circle] (a1) at (-2,0) {\large $2$};
\node[draw, circle] (a2) at (-2,2) {\large $2$};
\node[draw, circle] (a3) at (0,2) {\large $2$};
\node[draw, circle] (a4) at (2,2) {\large $2$};
\node[draw, circle] (a5) at (2,0) {\large $2$};
\node[draw, circle] (a6) at (2,-2) {\large $2$};
\node[draw, circle] (a7) at (0,-2) {\large $2$};
\node[draw, circle] (a8) at (-2,-2) {\large $2$};
\node[draw, rectangle, blue] (b2a) at (-4,4) {\large $\,2\,$};
\node[draw, rectangle, blue] (b2b) at (4,-4) {\large $\,2\,$};
\node[draw, rectangle, RawSienna] (b3a) at (4,4) {\large $\,2\,$};
\node[draw, rectangle, RawSienna] (b3b) at (-4,-4) {\large $\,2\,$};

\draw[draw, solid] (a1)--(a2);
\draw[draw, solid] (a2)--(a3);
\draw[draw, solid] (a3)--(a4);
\draw[draw, solid] (a4)--(a5);
\draw[draw, solid] (a5)--(a6);
\draw[draw, solid] (a6)--(a7);
\draw[draw, solid] (a7)--(a8);
\draw[draw, solid] (a8)--(a1);
\draw[draw, solid] (a1)--(b1);
\draw[draw, solid] (a2)--(b1);
\draw[draw, solid] (a3)--(b1);
\draw[draw, solid] (a4)--(b1);
\draw[draw, solid] (a5)--(b1);
\draw[draw, solid] (a7)--(b1);
\draw[draw, solid] (a1)--(b2a);
\draw[draw, solid] (a3)--(b2a);
\draw[draw, solid] (a5)--(b2b);
\draw[draw, solid] (a6)--(b2b);
\draw[draw, solid] (a7)--(b2b);
\draw[draw, solid] (a8)--(b2b);
\draw[draw, solid] (a1)--(b3b);
\draw[draw, solid] (a3)--(b3a);
\draw[draw, solid] (a5)--(b3a);
\draw[draw, solid] (a7)--(b3b);

\node[] at (-1.65,1) {\normalsize $\frac{x^{\frac{1}{3}}}{a^2}$};
\node[] at (-1,1.65) {\normalsize $\frac{x^{\frac{1}{3}}b}{a^2}$};
\node[] at (1.65,1) {\normalsize $\frac{x^{\frac{1}{3}}}{a^2}$};
\node[] at (0.9,1.65) {\normalsize $\frac{x^{\frac{1}{3}}b}{a^2}$};
\node[] at (1.65,-1) {\normalsize $\frac{x^{\frac{1}{3}}}{a^2}$};
\node[] at (1,-1.6) {\normalsize $\frac{x^{\frac{1}{3}}c}{a^2}$};
\node[] at (-1.65,-1) {\normalsize $\frac{x^{\frac{1}{3}}}{a^2}$};
\node[] at (-1,-1.6) {\normalsize $\frac{x^{\frac{1}{3}}c}{a^2}$};
\node[] at (-1,-0.4) {\normalsize $\frac{x^{\frac{2}{3}}a^2}{d}$};
\node[] at (-1,0.6) {\footnotesize $xd$};
\node[] at (-0.38,1.1) {\normalsize $\frac{x^{\frac{2}{3}}a^2}{bd}$};
\node[] at (1,0.6) {\footnotesize $xd$};
\node[] at (1,-0.4) {\normalsize $\frac{x^{\frac{2}{3}}a^2}{d}$};
\node[] at (0.45,-1) {\normalsize $\frac{x^{\frac{2}{3}}a^2d}{c}$};
\node[] at (-3.75,2.4) {\normalsize $\frac{x^{\frac{2}{3}}a^2c}{bd}$};
\node[] at (-2.4,3.6) {\normalsize $\frac{x^{\frac{2}{3}}a^2d}{c}$};
\node[] at (3.7,-2.4) {\normalsize $\frac{x^{\frac{2}{3}}a^2c}{bd}$};
\node[] at (2.7,-2.25) {\normalsize $\frac{xbd}{c}$};
\node[] at (1.24,-2.29) {\footnotesize $\frac{x^{\frac{2}{3}}a^2}{bd}$};
\node[] at (2.2,-3.8) {\normalsize $\frac{xbd}{c}$};
\node[] at (3.7,2.35) {\normalsize $\frac{x^{\frac{2}{3}}a^2d}{c}$};
\node[] at (2.3,3.6) {\normalsize $\frac{x^{\frac{2}{3}}a^2c}{bd}$};
\node[] at (-3.75,-2.4) {\normalsize $\frac{x^{\frac{2}{3}}a^2d}{c}$};
\node[] at (-2.3,-3.6) {\normalsize $\frac{x^{\frac{2}{3}}a^2}{d}$};
\end{tikzpicture}
\caption{The $3d$ quiver representation of the torus with flux $(2\sqrt{2};1,1,1,1,0,0)$ with a specific choice for the parametrization of the abelian symmetries and the R-symmetry.}
\label{U1SU2SU3SU4torus1}
\end{figure}

On top of the $SU(2)^3$ non-abelian symmetry, the superpotential also preserves four abelian symmetries. The parametrization for these symmetries, together with the one for the R-symmetry, that we choose to use is summarized in figure \ref{U1SU2SU3SU4torus1}. We conjecture that this theory corresponds to the compactification of the rank 1 $E_7$ SCFT on a torus with flux 1, as usual in a normalization where the minimal flux is 1, for a $U(1)$ whose commutant in $E_7$ is $SU(2)\times SU(3)\times SU(4)$. In order to validate this claim we can for example compute the superconformal index with the reference R-symmetry of the figure. The computation is rather involved because of the complexity of the quiver and because of the high value of the large number of gauge groups, which means that there are many integrations and summations to perform. The first few orders give
\bea
\mathcal{I}&=&1+4a^4\left(\frac{1}{b^4}+b^2d^2+\frac{b^2}{d^2}\right)x^{\frac{4}{3}}+3a^3\left(c{\bf 2}_{SU(2)_1}+c^{-1}{\bf 2}_{SU(2)_2}\right){\bf 2}_{SU(2)_3}x^{\frac{3}{2}}+\nn\\
&+&2a^2\left(b^4+\frac{d^2}{b^2}+\frac{1}{b^2d^2}\right)\left(c^2+\frac{1}{c^2}+{\bf 2}_{SU(2)_1}{\bf 2}_{SU(2)_2}\right)x^{\frac{5}{3}}+\cdots\,,
\eea
where in our notation $SU(2)_1$ stands for the flavor $SU(2)$ in red in figure \ref{U1SU2SU3SU4torus}, $SU(2)_2$ stands for the one in blue and $SU(2)_3$ stands for the one in brown.
We can observe that this result conforms to our expectations from $5d$. First of all we can see that the index forms characters of the $U(1)\times U(2)\times SU(3)\times SU(4)$ global symmetry preserved by the flux, where the embedding is
\be
{\bf 4}_{SU(4)}\to \frac{1}{c}{\bf 2}_{SU(2)_1}\oplus c\,{\bf 2}_{SU(2)_2},\qquad {\bf 3}_{SU(3)}\to \frac{1}{b^4}\oplus b^2d^2\oplus\frac{b^2}{d^2}\,.
\ee
In terms of characters of this global symmetry, the index indeed reads
\bea\label{indU1SU2SU3SU4}
\mathcal{I}&=&1+4a^4{\bf 3}_{SU(3)}x^{\frac{4}{3}}+3a^3{\bf 2}_{SU(2)}{\bf \overline{4}}_{SU(4)}x^{\frac{3}{2}}+2a^2{\bf \overline{3}}_{SU(3)}{\bf 6}_{SU(4)}x^{\frac{5}{3}}+\cdots\,.
\eea

Moreover, once again we see that the spectrum of states is consistent with the $5d$ expectations.
In this case, the current in the adjoint representation of $E_7$ splits under the $U(1)\times U(2)\times SU(3)\times SU(4)$ subgroup according to the branching rule
\bea\label{E7adjU1SU2SU3SU4}
{\bf 133}&\to&({\bf 1},{\bf 1},{\bf 1})^0\oplus({\bf 3},{\bf 1},{\bf 1})^0\oplus({\bf 1},{\bf 8},{\bf 1})^0\oplus({\bf 1},{\bf 1},{\bf 15})^0\oplus({\bf 2},{\bf 3},{\bf 4})^1\oplus({\bf 2},{\bf \overline{3}},{\bf \overline{4}})^{-1}\nn\\
&\oplus&({\bf 1},{\bf \overline{3}},{\bf 6})^2\oplus({\bf 1},{\bf 3},{\bf 6})^{-2}\oplus({\bf 2},{\bf 1},{\bf \overline{4}})^3\oplus({\bf 2},{\bf 1},{\bf 4})^{-3}\oplus({\bf 1},{\bf 3},{\bf 1})^4\oplus({\bf 1},{\bf \overline{3}},{\bf 1})^{-4}\,.\nn\\
\eea
In this case, as we can see from figure \ref{U1SU2SU3SU4torus1}, the $5d$ R-symmetry is related to the $3d$ R-symmetry that we used for our computation by a mixing with $U(1)_a$ that can be implemented in the index by the shift $a\to ax^{\frac{1}{6}}$. Moreover, the $U(1)_a$ symmetry is directly identified with the abelian symmetry for which we turned on a unit of flux.
With this dictionary, we can immediately identify all of the representations appearing in the index \eqref{indU1SU2SU3SU4} with some of the states in the branching rule \eqref{E7adjU1SU2SU3SU4} with the correct multiplicity given by their $U(1)_a$ charge times the flux, which in this case is one. Specifically, we can see the contributions $4a^4{\bf 3}_{SU(3)}x^{\frac{4}{3}}$, $3a^3{\bf 2}_{SU(2)}{\bf \overline{4}}_{SU(4)}x^{\frac{3}{2}}$ and $2a^2{\bf \overline{3}}_{SU(3)}{\bf 6}_{SU(4)}x^{\frac{5}{3}}$ which correspond to the states $({\bf 1},{\bf 3},{\bf 1})^4$, $({\bf 2},{\bf 1},{\bf \overline{4}})^3$ and $({\bf 1},{\bf \overline{3}},{\bf 6})^2$.

Unfortunately, due to the complexity of the model, we didn't manage to compute the index up to order $x^2$ and also a numerical calculation of the $\mathbb{S}^3$ partition function, which would allow us to find the superconformal R-symmetry and check if the flipping fields are below the bound or not, is out of reach. Because of these reasons, we are not able to analyze in detail the structure of the conformal manifold of the theory and compare it with the $5d$ expectations, but we expect this to follow the same discussion we did for the previous examples.

\subsection{Gluing the basic tube to itself}

In this section we consider the models obtained by self-gluing a single tube. These can be associated to compactifications on tori with flux $\frac{1}{2}$, in the normalization we used so far, for a $U(1)$ inside the $5d$ global symmetry. Remember that our normalization for the flux is such that the minimal flux allowed is $1$. Fractional fluxes are also allowed, provided that these are accompanied by a flux in the center of the non-abelian symmetry that is the  commutant of the $U(1)$ inside the full $5d$ global symmetry. The effect of this flux for the center group is to further break the non-abelian global symmetry to some subgroup. Such a flux for the center group can be generated by turning on two holonomies, one for each cycle of the torus, that are almost commuting, that is they commute up to an element of the center group\footnote{This is sometimes also refereed to as a non-trivial second Stiefel-Witteney class. See \cite{Ohmori:2018ona} and references therein, for some discussion of this aimed at physicists.}. The symmetry preserved by the flux corresponds then to the one preserved by the holonomies. For a more in depth analysis we refer the reader to appendix C of \cite{KRVZ}.

For a given group, there may be more than one such choices of holonomies, which preserve a different subgroup and which can be continuously connected one to the other. The field theory interpretation of this is that the theories obtained from the compactification, which in our case are three-dimensional, will possess a conformal manifold and their marginal deformations are related to the aforementioned holonomies. If we are on a point of the conformal manifold where the global symmetry is the one preserved by one choice of holonomies, we can move on a generic point where the symmetry is broken to its maximal torus $U(1)^r$, with $r$ the rank of the preserved global symmetry group, and then to a special point where these $U(1)$ symmetries reassemble into a different symmetry group associated to a different choice of holonomies.

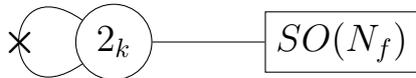
\begin{figure}
\center
\begin{tikzpicture}[baseline=0, font=\scriptsize]
\node[draw, circle] (a1) at (0,0) {\large $2_k$};
\node[draw, rectangle] (a2) at (3,0) {\large $SO(N_f)$};
\draw[draw, solid] (a1)--(a2);
\path[every node/.style={font=\sffamily\small,
  		fill=white,inner sep=1pt}]
(a1) edge [loop, out=145, in=215, looseness=6] (a1);
\node[thick] at (-1.24,0) {\large $\bm \times$};
\end{tikzpicture}
\caption{The theory we get when gluing the tube in figure \ref{BTube} to itself. Here we shall take $k=\frac{6-N_f}{2}$. Besides the flipping superpotential, there is also a cubic one involving the adjoint and two fundamentals. This breaks the symmetry in the square to $SO(N_f)$.}
\label{selfgluing}
\end{figure}

The theories obtained by compactifying the $5d$ rank $1$ $E_{N_f+1}$ SCFT on a torus with half-integer flux are obtained by gluing an odd number of the tube model of figure \ref{BTube}. We will focus on the theories associated with flux $\frac{1}{2}$, so we take only one copy of the tube model. Specifically, we gauge a diagonal combination of the two $SU(2)$ symmetries. We also add an $SU(2)\times SU(N_f)$ bifundamental that couples quadratically to both of the two sets of fundamentals in the tube model. The net effect is that one of these sets of chirals and the fields that we added become massive, so that we are left with one copy of $N_f$ fundamentals of the $SU(2)$ gauge group. What used to be the $SU(2)\times SU(2)$ bifundamental in the tube model now becomes an adjoint plus a singlet that couples with the quadratic operator constructed from the fundamentals. We also have the flipping field, which couples both with the adjoint and with the singlet squared. 
The resulting theory is summarized in figure \ref{selfgluing}.

Notice that in this construction the global symmetry of the original tube model has been broken from $SU(N_f)$ down to $SO(N_f)$, as expected because of the half-integer flux. Nevertheless, we will now see that this is not always the symmetry preserved by the almost commuting holonomies. We shall return to this point later, but first we should elaborate on the results. Here we shall concentrate on the case of $N_f=6$.

For $N_f=6$ the Chern-Simons level is zero. Our theory corresponds then to the $\mathcal{N}=4$ $SU(2)$ gauge theory with $6$ fundamental half-hypers deformed by the addition of two gauge singlets. The $SO(12)$ global symmetry that we had in the case of flux $1$ is broken to a subgroup in this case of flux $\frac{1}{2}$. There is a choice of almost commuting holonomies for which the preserved symmetry is $SU(4)$, which is also the manifest global symmetry in the theory of figure \ref{selfgluing} for $N_f=6$. On top of this, we have the usual $U(1)_a$, which corresponds to the $U(1)$ for which we turned on the flux. As in the previous examples, we parameterize it such that the fundamentals have charge $1$, the adjoint and the singlet have charge $-2$ and the flipping field has charge $4$. Also the R-symmetry can be taken again such that all the chirals have R-charge $\frac{2}{3}$. With these conventions, the index of the theory reads
\bea \label{selfgluingNf6}
\mathcal{I}&=&1+\left(\frac{2}{a^2}+a^4\right)x^{\frac{2}{3}}+\left(\frac{4}{a^4}+a^8+a^2({\bf 15}+1)\right)x^{\frac{4}{3}}+\left(\frac{6}{a^6}+a^{12}+a^6({\bf 15}+1)\right)x^{2}+\nn\\
&+&\left(-\frac{1}{a^2}({\bf 15}+1)+\frac{8}{a^8}+a^{16}+a^4({\bf 84}+1-{\bf 15})+a^{10}({\bf 15}+1)\right)x^{\frac{8}{3}}+\cdots\,,
\eea
where we can see indeed characters of the representations of $SU(4)$.

For $SO(12)$ there exists also a different choice of holonomies, which instead preserves a $USp(6)$ subgroup. The theory with $USp(6)$ symmetry can be reached as follows. Our theory with $SU(4)$ symmetry possesses marginal operators in the adjoint representation of $SU(4)$ and, consequently, a non-trivial conformal manifold\footnote{These marginal operators don't appear explicitly in the index \eqref{selfgluingNf6} since they are canceled by the conserved currents, which contribute with a negative sign to the order $x^2$ of the index.}. We can move on a generic point of this conformal manifold where the symmetry is broken to the Cartan $U(1)^3$ and then, from here, to the specific point where these $U(1)$ symmetries reassemble into $USp(6)$. For this to be possible, we should be able to rewrite the index \eqref{selfgluingNf6} in terms of characters of $USp(6)$ representations by just redefining properly the $SU(4)$ fugacities. This is indeed the case, as we can understand by noticing that the only $SU(4)$ representations appearing in \eqref{selfgluingNf6}, when decomposed under its $SU(3)\times U(1)$ subgroup, precisely reconstruct also $USp(6)$ representations decomposed under the same $SU(3)\times U(1)$ subgroup
\bea
&&{\bf 15}_{SU(4)}\to {\bf 14}_{USp(6)}+{\bf 1}_{USp(6)}\nn\\
&&{\bf 84}_{SU(4)} - {\bf 15}_{SU(4)}\to {\bf 90}_{USp(6)} - {\bf 21}_{USp(4)}\,.
\eea
Equivalently, if we parametrize the $SU(4)$ fugacities such that the character of the fundamental representation is $f_1+\frac{f_2}{f_1}+\frac{f_3}{f_2}+\frac{1}{f_3}$ and the $USp(6)$ fugacities such that the character of the fundamental representation is $y_1+\frac{1}{y_1}+y_2+\frac{1}{y_2}+y_3+\frac{1}{y_3}$, then upon the identification
\bea
f_1=\sqrt{\frac{y_2y_3}{y_1}},\quad f_2=\frac{1}{y_1},\quad f_3=\sqrt{\frac{y_2}{y_1y_3}}\,,
\eea
the index \eqref{selfgluingNf6} can be rewritten as
\bea
\mathcal{I}&=&1+\left(\frac{2}{a^2}+a^4\right)x^{\frac{2}{3}}+\left(\frac{4}{a^4}+a^8+a^2({\bf 14}+2)\right)x^{\frac{4}{3}}+\left(\frac{6}{a^6}+a^{12}+a^6({\bf 14}+2)\right)x^{2}+\nn\\
&+&\left(-\frac{1}{a^2}({\bf 14}+2)+\frac{8}{a^8}+a^{16}+a^4({\bf 90}+1-{\bf 21})+a^{10}({\bf 14}+2)\right)x^{\frac{8}{3}}+\cdots\,,
\eea
where now the characters are of $USp(6)$ representations.

So we indeed see that for the $N_f=6$ case, self-gluing of the tubes is consistent with our expectations. However, the story becomes more complicated when we consider the cases with $N_f<6$. The issue can be partly understood by looking at the tubes, specifically, at the way the symmetry changes along the two punctures. Consider the $SU(2)\times SU(N_f)$ bifundamentals on the two sides of the basic tubes. We have that their $U(1)_a$ charges are the same, but their $U(1)_b \times SU(N_f)$ charges are related by charge conjugation. This is what causes these symmetries to break to the invariant subgroup when we glue the tube to itself. For $N_f=6$ we have that $U(1)_b \times SU(N_f)$ are part of the $SO(12)$ group preserved by the flux. Note that $SO(12)$ is a real group, and so the charge conjugation action on $U(1)_b \times SU(N_f)$, that is an outer automorphism, becomes an inner automorphism in $SO(12)$. This is a common characteristic of center fluxes, as we mentioned that these involve two holonomies along the two cycles of the torus, which suggests that the symmetries on the two sides of the tube should be related by an inner automorphism.

However, when $N_f<6$, the $U(1)_b \times SU(N_f)$ is part of a complex $5d$ symmetry group so the action of charge conjugation on it is not mapped to an inner automorphism of the $5d$ symmetry group preserved by the flux, but rather an outer automorphism. This fits more with an holonomy in a discrete symmetry. However, it is not clear what is this discrete symmetry. Specifically, while $E_{N_f+1}$ is a complex group for $0<N_f<6$, the local operator spectrum for most cases appears to only be sensitive to the group $E_{N_f+1}/Z_{E_{N_f+1}}$, where $Z_{E_{N_f+1}}$ is the center of $E_{N_f+1}$. This can be seen for instance by looking at the superconformal index of these theories, evaluated in \cite{Kim:2012gu}, where only representations of $E_{N_f+1}/Z_{E_{N_f+1}}$ appear for $N_f>1$. The group $E_{N_f+1}/Z_{E_{N_f+1}}$ is invariant under charge conjugation implying that a symmetry with this action acts trivially on it\footnote{There might be extended operators that are in representations of $E_{N_f+1}$ that are not ones of $E_{N_f+1}/Z_{E_{N_f+1}}$. In this case such a symmetry can exist in the SCFT, though it would be a $1$-form symmetry rather than a $0$-form one.}.   

This may also be related to the issue that for $N_f\le 5$, we needed to turn on Chern-Simons terms with levels of alternating signs. This was sensible when the number of groups are even, but raises the question of the proper generalizations when the number of groups is odd. Due to the presence of these complications, we shall delay dealing with these cases to future work. We shall return to the issue of the Chern-Simons terms in the next section.

\section{Gluing rules}
\label{gr}

Having studied the various models, we wish to return to the question of gluing rules. These dictate how two punctures are to be glued. Here we adopted the viewpoint of the punctures as boundary conditions on the $4d$ theory we get when reducing the $5d$ theory on a circle potentially with an holonomy in its flavor symmetry. The data of the boundary conditions on the resulting $4d$ theory affects the properties of the punctures in various ways. Here we looked at the case of the so-called maximal punctures, where the $3d$ $\mathcal{N}=2$ vector component on the boundary of the $4d$ $\mathcal{N}=2$ bulk vector multiplet receives Dirichlet boundary conditions. This leads to the punctures having an associated global symmetry given by the bulk $4d$ gauge symmetry, which in this case is $SU(2)$.

Additionally, there are bulk $4d$ hypermultiplets, and when approaching the boundary these are also given boundary conditions. Specifically, we decompose the $4d$ hypermultiplet close to the boundary to two $3d$ $\mathcal{N}=2$ chiral fields in opposite representation and give Dirichlet boundary conditions to one group and Neumann boundary conditions to the other. The chiral fields receiving the Neumann boundary conditions are expected to survive and contribute to the $3d$ theory, and give a special class of operators associated with the punctures. These are usually denoted as $M$, and refereed to as the moment map operators associated with the punctures, for historical reasons having to do with their manifestations in the study of the compactifications of the $6d$ $(2,0)$ theory. In our case, these are an $SU(2)\times SU(N_f)$ bifundamental, with the $SU(2)$ being the one associated with the puncture and the $SU(N_f)$ being a subgroup of the $5d$ global symmetry. 

We note that when we give the boundary conditions we have the freedom to decide to which of the two $3d$ $\mathcal{N}=2$ chiral fields to give Neumann or Dirichlet boundary conditions. This is true for all the $N_f$ doublet hypermultiplets. This gives a discrete label associated with the punctures that is refereed to as sign and color, depending on the convention. Punctures with different signs, that is whose boundary conditions differ by giving Neumann boundary conditions to different components of the hypermultiplets, are equivalent and can usually be transformed to one another by a global symmetry transformation. However, if we have two punctures of different signs on the same surface, then this difference is physical, as a global symmetry transformation will affect both punctures simultaneously. We note that the charges carried by the moment map operators under the $5d$ global symmetry depend on the sign of the puncture, and as such punctures having different signs and colors are glued in different ways.

The gluing rules involve the puncture global symmetry and its associated moment map operators, $M$. We previously noted that when gluing two punctures we need to gauge the global symmetry associated with the punctures by an $\mathcal{N}=2$ vector multiplet. Additionally, the moment map operators associated with the two punctures, $M_i$, are coupled to one another using a superpotential. The form of the coupling, however, differs depending on whether they have the same sign or opposite signs. For the most part, we dealt with the case where the punctures have the same sign, in which case, when we glue we also add $N_f$ $SU(2)$ doublets denoted as $\phi$. These then couple to the moment map operators using the superpotential $\phi(M_1-M_2)$. Note that here it is important that $M_1$ and $M_2$ carry the same charges under the $5d$ global symmetry, as they have the same sign, so the superpotential does not need to break these symmetries. This is usually refereed to as $\phi$ gluing.

We can also consider gluing punctures with opposite signs, that is the components of the bulk $4d$ $N_f$ $SU(2)$ hypermultiplets receiving Neumann boundary conditions for the first puncture receive Dirichlet boundary conditions for the second one. In this case, when gluing we do not introduce the fields $\phi$, but rather couple $M_1$ and $M_2$ directly via the superpotential $M_1 M_2$. Note that here it is important that $M_1$ and $M_2$ carry opposite charges under the $5d$ global symmetry, as they have the opposite signs, so the superpotential does not need to break these symmetries. This is usually refereed to as $S$ gluing.

The gluing rules stated so far have been observed in many cases of compactifications of $6d$ SCFTs, see for instance \cite{BTW,RVZ,KRVZ}. While this is the first case, to our knowledge, of their appearance in the study of compactifications of $5d$ SCFTs, they are still quite well established. However, there are several aspects that we have encountered in this paper that do not appear in compactifications to $4d$. We shall next wish to discuss our expectations for the gluing rules of these more novel elements, based on our experience with the $3d$ theories studied thus far. 

The first of these is the presence of Chern-Simons terms. As we have seen, when we glue punctures we need to turn on Chern-Simons terms of level $\frac{6-N_f}{2}$ of alternating signs. We can then ask how can this be incorporated in the gluing rules. The simplest option is to introduce an additional label to the puncture called $k$-sign, and then declare that when glued, punctures with positive $k$-sign are glued with a Chern-Simons term of positive level while punctures with negative $k$-sign are glued with a Chern-Simons term of negative level. We would then say that the two punctures in the basic tube of figure \ref{BTube} have opposite $k$-signs, while those in figure \ref{DTubes} have the same $k$-sign, which should be opposite to the sign of the Chern-Simons term of the gauge group in the tube. 

 However, this raises the question of what happens when we glue a puncture with positive $k$-sign to one that has a negative $k$-sign. For instance, in the basic tube of figure \ref{BTube}, we would say that the two punctures have opposite $k$-sign so gluing them together provides an examples of such gluing. We looked at this in the previous subsection, and noted that for $N_f=6$, where there is no Chern-Simons term, the results conform to our higher dimensional expectations. However, cases with the Chern-Simons term seem to behave in a somewhat different manner.

It is possible that one simply cannot glue punctures with opposite $k$-sign. However, we can perform this procedure in field theory, and so it is tempting to think that this should have some geometric interpretation. Either way, it does appear that gluing punctures of opposite $k$-sign has some less desirable effects, like breaking of part of the global symmetry. It would be interesting to better understand the origin of the Chern-Simons term, which hopefully will also shed light on these issues.

The second issue we encountered is the presence of monopole superpotentials. Specifically, we noted that when we glue theories, in addition to the superpotentials coupling the moment map operators of the punctures, we also need to turn on monopole superpotentials carrying the minimal possible charge under adjacent $SU(2)$ groups. This presents several issues when trying to pose gluing rules.

The first issue is how to describe it abstractly. Specifically, in all cases studied so far the gluing rules appear to depend only on the type of punctures and not on what happens on the rest of the surface. However, the description as a monopole superpotential with charges under adjacent $SU(2)$ groups assumes a specific description. Furthermore, the monopole superpotential cannot be represented as a product of operators associated with each of the punctures individually. We can try to get around this by defining them as operators that can be associated with the punctures in the presence of background magnetic flux to their flavor symmetry. However, the charges of monopole operators depend on all the flavors seen by the gauge group. As such, when we glue the two punctures together, the $SU(2)$ that does the gluing sees more matter than the flavor $SU(2)$ before the gluing. Thus, the monopole superpotential we turn on carries different charges than the ones that exist for the punctures in the presence of background magnetic flux\footnote{We are grateful to Shlomo Razamat for the discussion on this point.}.

It seems then that the monopole superpotential we turn on cannot be associated with any of the punctures. Rather, we need to stipulate that after the two punctures are glued, one must turn on monopole superpotentials having the property that they carry minimal magnetic charge under the gluing $SU(2)$. These do not exist in the theories we glued before the gluing, but only emerge after the gluing is done. As to how to properly identify these, the construction suggests that these must be such that they preserve the $5d$ global symmetry and R-symmetry. The examples done so far suggest that this should be enough to determine them. We also note that in almost all cases these were just the monopole operators with charges under adjacent $SU(2)$ groups. However, there are exceptions. For instance in the theories in figure \ref{3dquivers}, the monopole superpotential was actually the product of this with the flip fields. This suggests that their exact interpretation in a given $3d$ field theory may differ, but they should still exist.

This, however, does not solve all the issues. One issue that remains is how to understand the gluing of two basic tubes, the ones in figure \ref{BTube} to form the one in figure \ref{DTubes} (a). The problem here is that when we glue the two tubes we do not turn on a monopole superpotential. One interpretation is that in this case the set of monopole superpotentials with the required properties, that is that they preserve the $5d$ global symmetry and R-symmetry, is empty. However, this would appear to suggest that the gluing rules depend not only on the punctures glued but also on the surfaces, which differs from the behavior in all the other cases studied so far. Furthermore, we note that in most of the other cases when we glued we needed to turn on precisely two monopole superpotentials, regardless of other details. An alternative interpretation is that this specific tube is not actually a tube and rather the tube in figure \ref{DTubes} (a) should be interpreted as the basic tube. This might be sensible as most of the other models can be generated by gluing it rather then the tube in figure \ref{BTube}. Particularly, we can get the theories in figure \ref{3dquivers} but gluing this tube to itself. However, we noted that for $N_f=6$ we can glue the tube to itself and the results are consistent with a higher dimensional interpretation, and this theory requires the basic tube to be constructed. 

\section{Conclusions}
\label{conc}

Inspired by the more investigated compactifications of $6d$ SCFTs on Riemann surfaces to four dimensions, in this paper we initiated the study of the compactifications of $5d$ SCFTs to $3d$ $\mathcal{N}=2$ theories. More precisely, we concentrated on the $5d$ $\mathcal{N}=1$ theories known as the Seiberg rank 1 $E_{N_f+1}$ SCFTs. We first conjectured some $3d$ Lagrangians that flow in the IR to the same SCFTs obtained by compactifying the $5d$ theories on a sphere with two punctures, or a tube, with some value of flux for the $5d$ global symmetry. This was done by studying the domain walls that interpolate between two copies of the $4d$ reduction on a circle of the $5d$ theories with an holonomy and with some specific boundary conditions, where this latter data imply a specific choice for the type of the puncture. From these fundamental building blocks, we then constructed $3d$ theories that we conjecture correspond to the torus compactifications of the $5d$ rank 1 $E_{N_f+1}$ SCFTs with some value of flux.

This conjecture has been then tested in various ways, some of which are inherited from the study of $6d$ to $4d$ compactifications. Most importantly, the flux breaks the $5d$ global symmetry to a subgroup, which might not be fully manifest in the $3d$ Lagrangians. The preserved symmetry should then appear as an enhanced symmetry from the three-dimensional perspective. This enhancement can be tested by using the superconformal index and the central charges of the SCFT, where the latter can be extracted from the $\mathbb{S}^3$ partition function. Moreover, from their $5d$ origin we expect the $3d$ theories to possess some gauge invariant operators that descend from the $5d$ stress-energy tensor and conserved current multiplets. Another test that we performed was to check the presence of these states by means of the superconformal index. Finally, it is possible that different choices of flux, to which we associated different looking $3d$ Lagrangians, are actually equivalent up to an element of the Weyl group of the $5d$ $E_{N_f+1}$ global symmetry, and are thus expected to lead to the same SCFT. This implies that the two $3d$ UV Lagrangians flow to the same fixed point at low energies, that is they are dual in the IR. We encountered some examples of this phenomenon which we checked, again, using supersymmetric partition functions. In one example we have also been able to understand the duality that we can geometrically predict from the $5d$ construction as an instance of Aharony duality in three-dimensions.

The construction of torus models starting from the tubes requires some rules for how to implement the gluing at the level of the $3d$ field theory. After analyzing various examples, we tried to draw some general prescription for this. This was partly motivated by how the gluing is performed in the more understood context of $6d$ to $4d$ compactifications, but we also encountered some new features that are peculiar to three-dimensional physics. Specifically, in $3d$ we can have Chern-Simons interactions and a very special type of operators that cannot be written in terms of elementary fields, the monopole operators, which may be turned on in the superpotential. By looking at the examples that we studied, we tried to find some general pattern for these new elements that appear in the compactifications of $5d$ SCFTs on Riemann surfaces to $3d$ for the case of the rank 1 $E_{N_f+1}$ theories that has been the focus of this paper.

There are some questions that are left open and that would be worth to further investigate in the future. The first one is to better understand these new intrinsically $3d$ ingredients, the Chern-Simons terms and the monopole superpotentials. In particular, it would be interesting to understand their occurrence from a geometric perspective, since we concentrated on working them out from a purely $3d$ field theory analysis of the models at hand. Another open question is related to the torus models with half-integer fluxes, which can be constructed by self-gluing a tube corresponding to an odd number of domain walls. The half-integer flux can be accommodated provided that it is supplemented by a flux for the center of the residual non-abelian global symmetry, but at the price of breaking this further to some smaller subgroup. In field theory, this is due to the fact that the two punctures of the tube that we are trying to glue are of different types if the flux is half-integer, implying a breaking of the global symmetry of the model. Understanding precisely what should be the preserved global symmetry from the perspective of the $5d$ compactification is in general a difficult question. We presented an example for $N_f=6$ where it is possible to do so, but it would be interesting to also better understand the cases of lower $N_f$.

There are also various directions that one can follow starting from the results of our paper. For example, we focused on compactifications on tubes and tori, but one may also consider more generic Riemann surfaces. Given such a surface, one can always find at least one pair of pants decomposition for it. Hence, the additional element that we need to construct a generic surface is the sphere with three punctures, or trinion. Finding the theory corresponding to compactifications on a three-punctured sphere is a more complicated task, even in the more understood set-up of $6d$ to $4d$ compactifications. This problem has been tackled for example in \cite{Razamat:2019mdt,Razamat:2019ukg,Razamat:2020bix,Sabag:2020elc} in the context of compactifications of $6d$ SCFTs, by studying the interrelation of the flow triggered by the compactification and some flows that one can trigger by giving vacuum expectation values to some operators of the $6d$ theory and which also have a counterpart in $4d$. One possibility would then be to try to apply the same strategy to the case of compactifications of $5d$ SCFTs. Alternatively, we can try to use the knowledge of the trinion theories in $6d$ to $4d$ compactifications. We indeed observe that the tube models that we found in the present paper can also be obtained by considering the $4d$ compactification of the rank 1 E-string theory on a tube, which has been studied in \cite{KRVZ}, by compactifying it on a circle to $3d$ and turning on a suitable real mass deformation. This deformation has the effect of lifting the monopole superpotential that is dynamically generated in the $\mathbb{S}^1$ compactification \cite{Aharony:2013dha,Aharony:2013kma} without producing any Chern-Simons interaction. This is compatible with the fact that reducing the $6d$ E-string theory on a circle with a suitable holonomy we obtain a $5d$ $SU(2)$ gauge theory with 8 hypers, which can be mass deformed to an $SU(2)$ gauge theory with $N_f\le 7$ flavors. By the same token, we can try to find the theories obtained by compactifying $5d$ SCFTs on three-punctured spheres by considering the $4d$ trinion theories, reducing them on a circle and preforming a suitable real mass deformation.

It is also possible to study compactifications on spheres without punctures. As it was recently discussed in \cite{HRSS} in the context of $6d$ to $4d$ compactifications, a novel feature here is that the resulting low dimensional theories should possess an additional $SU(2)_{\text{ISO}}$ flavor symmetry that descends from the isometry of the two-sphere. One way to obtain sphere compactifications is to start from theories resulting from compactifications on a tube and close the two punctures. These is expected to be done by giving some vacuum expectation values to some operators charged under the symmetries of the punctures, with the effect of completely breaking these symmetries. On top of this, one typically also needs to introduce additional singlet fields flipping some of the gauge invariant operators to reproduce the theory that correctly corresponds to the $6d$ compactification, which can be worked out by matching the anomalies with the $6d$ predictions. The main difficulty in extending this procedure to $5d$ compactifications is that we do not have a tool like the anomalies to guide us in understanding the pattern of these singlet fields.

Another possible line of future investigation is to change the starting $5d$ SCFT rather than the surface of the compactification. The methods developed in this paper can be in principle applied to find the tube and torus compactifications of any $5d$ SCFT that admits a mass deformation to a gauge theory. In the recent years there has been a lot of progress in constructing $5d$ SCFTs using various approaches, mainly based on $(p,q)$-web diagrams in Type IIB \cite{AH,AHK,KB,Bergman:2015dpa,Zafrir:2015ftn,Hayashi:2015vhy} and on M-theory compactifications on singular Calabi--Yau threefolds \cite{Douglas:1996xp,Jefferson:2018irk,Bhardwaj:2018vuu,Bhardwaj:2018yhy,Closset:2018bjz,Apruzzi:2018nre,Apruzzi:2019opn,Apruzzi:2019enx,Bhardwaj:2019jtr,Bhardwaj:2019fzv,Saxena:2019wuy,Apruzzi:2019kgb}, and in understanding various of their properties, including gauge theory phases. One immediate generalization of the cases studied in the present paper is to consider their higher rank version. For example, we can consider $USp(2N)$ gauge theories with one antisymmetric and $N_f\le 7$ fundamental hypermultiplets. These theories have a manifest $SO(2N_f)\times SU(2)\times U(1)$ global symmetry, where the additional $SU(2)$ factor that appears for rank greater than 1 is acting on the two half-hypers contained in the antisymmetric hyper. These gauge theories have a UV completion to SCFTs with $E_{N_f+1}\times SU(2)$ global symmetry, where the enhancement pattern is the same as in the rank 1 case, with the $SU(2)$ being a spectator. We can then consider compactifications of these SCFTs on tubes and tori with fluxes for the $E_{N_f+1}$ part of the global symmetry, in analogy with the compactifications of the higher rank $6d$ E-string theory to $4d$ studied in \cite{Pasquetti:2019hxf} where the flux is turned on only for the $E_8$ part of its $E_8\times SU(2)$ global symmetry. In \cite{Pasquetti:2019hxf} it was understood that in the higher rank case the contribution of the domain wall to the tube theory is more intricate than just a bifundamental and a singlet field. Rather, it is given by a more complicated $4d$ $\mathcal{N}=1$ theory that was called $E[USp(2N)]$ theory (see also \cite{Hwang:2020wpd,Garozzo:2020pmz,HRSS}). One might thus expect that also in the compactification of the aforementioned higher rank $5d$ $E_{N_f+1}$ SCFTs the tube theory could be more complicated and may involve the dimensional reduction to $3d$ of the $E[USp(2N)]$ theory. 

Alternatively, one can consider SCFTs that admit mass deformations to $USp(2N)$ or $SU(N)$ gauge theories with fundamental hypermultiplets only and no antisymmetric, with the possibility of having a Chern-Simons level for the $SU(N)$ case already in $5d$. An interesting feature that some of these theories possess is that, for some values of the number of flavors and the Chern-Simons levels, different gauge theories, even with a different gauge group, may be UV completed by the same SCFT. In such cases, these ``UV dualities" can possibly lead to a greater wealth of tube models in $3d$ which would be interesting to analyze. Indeed, one can try to find tube models corresponding to domain walls interpolating between either two copies of the same gauge theory or a copy of one gauge theory and a copy of a dual gauge theory.

Another natural generalization of the compactifications considered in this paper can be obtained by considering $5d$ SCFTs which are the UV completions of models obtained by gauging a subgroup of the global symmetry of the $E_{N_f+1}$ SCFT while also adding hypermultiplets in the fundamental representation. This is analogous to the analysis performed in \cite{Razamat:2018gbu,Sela:2019nqa} in the context of the compactification of the $6d$ E-string theory to $4d$, which resulted in very interesting symmetry enhancement patterns and self-dualities (see also \cite{Razamat:2017wsk}), and it would be interesting to explore whether something of this nature also takes place in $5d$ compactifications to $3d$. Note also that such an analysis can be performed in the other direction, by first identifying $3d$ patterns of IR symmetry enhancement and then conjecturing the corresponding $5d$ SCFTs yielding the observed symmetries under compactification. 

Moreover, we comment on the possibility of using the study of the compactifications of $5d$ SCFTs of higher rank to find interesting applications via the AdS/CFT correspondence, where one usually considers a large $N$ limit. In particular, it would be interesting to understand the RG flow from five to three dimensions triggered by the compactification on a Riemann surface from a holographic perspective, in relation to possible dual $\text{AdS}_6$ solutions possessing an asymptotic limit to a geometry of the form $\text{AdS}_4\times\Sigma$, where $\Sigma$ is the Riemann surface (see \cite{Naka:2002jz,Bah:2018lyv,Hosseini:2018usu} for some examples of these types of solutions).

Another interesting direction would be to consider the computation of the partition functions of $5d$ SCFTs on a surface times three manifold, based on the techniques of \cite{Crichigno:2018adf,Hosseini:2018uzp}. This might then be compared against the partition functions of the proposed $3d$ theories. Alternatively, it might be used to formulate conjectures for the $3d$ theories expected to result from the compactification of these theories on the surface, as done in \cite{Razamat:2019sea}. 

Finally, we mentioned that there are no anomalies in $5d$ and $3d$, but more precisely this is true for continuous symmetries. The theories may indeed also possess discrete global symmetries, which can be both 0-form or higher form symmetries \cite{Gaiotto:2014kfa}, and these can have non-trivial mixed anomalies with continuous symmetries such as the topological symmetry, both in $5d$ and in $3d$. These have been studied for example in  \cite{Morrison:2020ool,Albertini:2020mdx,Bhardwaj:2020phs,BenettiGenolini:2020doj} in five dimensions and in \cite{Bergman:2020ifi} in three dimensions. A very interesting question would be if it is possible to use the anomalies for these discrete symmetries in $5d$ to make a prediction for the anomalies in $3d$.

\section*{Acknowledgments}
We thank Chris Beem, Seyed Morteza Hosseini and Shlomo Razamat for relevant discussions. MS and GZ are supported in part by the ERC-STG grant 637844-HBQFTNCER and by the INFN. MS is also partially supported by the University of Milano-Bicocca grant 2016-ATESP0586 and by the MIUR-PRIN contract 2017CC72MK003. OS is supported in part by Israel Science Foundation under grant no. 2289/18, by the I-CORE Program of the Planning and Budgeting Committee, by BSF grant no. 2018204, by grant No. I-1515-303./2019 from the GIF (the German-Israeli Foundation for Scientific Research and Development) and by the Clore Scholars Programme.

\appendix

\section{Some properties of the groups $E_{N_f+1}$}
\label{App:flux}

Here we collect various properties of the groups $E_{N_f+1}$, and particularly of the groups $E_6$ and $E_7$, that would be of most use to us here. We first begin by introducing the root lattice. For convenience we shall use a basis manifesting the $U(1)\times SO(2N_f)$ subgroup of $E_{N_f+1}$, for instance, $U(1)\times SO(10)$ for $E_6$ and $SU(2)\times SO(12)$ for $E_7$. We first consider the roots, where we shall use the following basis for $E_7$:

\be
(0;\pm 1, \pm 1, 0 , 0 , 0 , 0 ) + \text{permutations}, (\sqrt{2}; 0 , 0 , 0 , 0 , 0 , 0 ), (\pm \frac{\sqrt{2}}{2};\pm \frac{1}{2}, \pm \frac{1}{2}, \pm \frac{1}{2} , \pm \frac{1}{2} , \pm \frac{1}{2} , \pm \frac{1}{2} ),
\ee  
with an even number of minus signs for the six terms in the last entry. The first gives the roots of $SO(12)$,the second those of $SU(2)$ and the third gives the states in the $(\bold{2},\bold{32})$, that enhances $SU(2)\times SO(12)$ to $E_7$. Here we have chosen the normalization of the $U(1)$ Cartan of $SU(2)$ to be with the square root so that all roots have equal length with a trivial metric. This simplifies some of the expressions later on. 

For $E_6$ we use the roots:

\be
(0;\pm 1, \pm 1, 0 , 0 , 0 ) + \text{permutations}, (\pm \frac{\sqrt{3}}{2};\pm \frac{1}{2}, \pm \frac{1}{2}, \pm \frac{1}{2} , \pm \frac{1}{2} , \pm \frac{1}{2} ),
\ee  
with an even number of minus signs for the six terms in the last entry. The first gives the roots of $SO(10)$ and the second the states in the $\bold{16}^1 \oplus \overline{\bold{16}}^{-1}$, that enhances $U(1)\times SO(10)$ to $E_6$. Here we have chosen the normalization of the $U(1)$ to be with the square root so that all roots have equal length with a trivial metric. This simplifies some of the expressions later on.

Similarly, for generic $N_f<6$ we take the roots to be:

\be
(0;\underbrace{\pm 1, \pm 1, 0 , ... , 0 }_{N_f}) + \text{permutations}, (\pm\frac{\sqrt{8-N_f}}{2};\underbrace{\pm\frac{1}{2},\pm\frac{1}{2},...,\pm\frac{1}{2}}_{N_f}),
\ee
again with an even number of minus signs in the last expression. The first gives the roots of $SO(2N_f)$ and the second the states that enhances $U(1)\times SO(2N_f)$ to $E_{N_f+1}$. We have again chosen the normalization of the $U(1)$ to be with the square root so that all roots have equal length with a trivial metric.

We next want to list the possible fluxes. These are defined as a vector of numbers, $F$, obeying that $R\cdot F\in \mathbb{Z}$, for any root $R$. This in turn identify these as elements of the weight lattice, and as such are identified with representations of the groups. Here the precise choice of group is important, where if the group is $E_7$ or $E_6$ then the fluxes are associated with the weight lattice of $\frac{E_7}{\mathbb{Z}_2}$ or $\frac{E_6}{\mathbb{Z}_3}$ respectively, and similarly for other $E_{N_f+1}$ groups. Alternatively, if the group is $\frac{E_7}{\mathbb{Z}_2}$ or $\frac{E_6}{\mathbb{Z}_3}$ then the fluxes are associated with the weight lattice of $E_7$ or $E_6$ respectively. Here we take the groups to be $\frac{E_7}{\mathbb{Z}_2}$ and $\frac{E_6}{\mathbb{Z}_3}$. 

Note, that it may be possible to support fluxes for which $R\cdot F$ is not integer if one allows coupling of non-abelian symmetries, commuting with the fluxes, to non-trivial bundles, see \cite{KRVZ}.

Fluxes generically break the symmetry to a collections of $U(1)$ groups, in which the flux resides, and the non-abelian part that commutes with them. Given a flux vector, we can uncovered the preserved symmetry in the following manner. First, we can associate a flux with a representation, where that flux corresponds to its highest weight. That representation can be written by its Dynkin labels $[i_1,i_2,...]$, where to each Dynkin label is associated a node in the Dynkin diagram. The preserved non-abelian symmetry is then given by the group whose Dynkin diagram is built from the nodes whose Dynkin label is zero. This can be used to find the flux preserving a specific symmetry.

Alternatively, the roots of the preserved non-abelian symmetry are given by the subset of roots that are orthogonal to the flux. This is useful to find the symmetry preserved by a specific flux. For instance consider the flux vector $(\frac{\sqrt{2}}{2};\frac{1}{2},\frac{1}{2},\frac{1}{2},\frac{1}{2},\frac{1}{2},\frac{1}{2})$ of $E_7$. We note that it is orthogonal to the roots: $(0;1,-1,0,0,0,0) +$ permutations, $(\frac{\sqrt{2}}{2};\frac{1}{2},\frac{1}{2},-\frac{1}{2},-\frac{1}{2},-\frac{1}{2},-\frac{1}{2}) +$ permutations and $(-\frac{\sqrt{2}}{2};-\frac{1}{2},-\frac{1}{2},\frac{1}{2},\frac{1}{2},\frac{1}{2},\frac{1}{2}) +$ permutations. Here the orthogonality is done with respect to the trivial metric, which is a result of our specific choice of normalization of the first $U(1)$. We note that the first term are the roots of $SU(6)$, and the second and third terms form the $SU(6)$ representations of ${\bf \overline{15}}$ and ${\bf 15}$. Together, these give the roots of $SO(12)$. Thus, the flux vector $(\frac{\sqrt{2}}{2};\frac{1}{2},\frac{1}{2},\frac{1}{2},\frac{1}{2},\frac{1}{2},\frac{1}{2})$ preserves the $U(1)\times SO(12)$ subgroup of $E_7$

Finally, we note that different values of flux are associated with different theories, unless the flux vectors differ by the action of the Weyl group. We next give some additional details on the cases of $E_7$ and $E_6$.

\subsection*{$E_7$}

Here we introduce the fluxes for $N_f=6$ case. These are given by the possible weights. The smallest possible flux is given by the smallest representation which is the $\bold{56}$. This is only possible if the group is $\frac{E_7}{\mathbb{Z}_2}$, as we shall take the group to be. In the case of $E_7$, the smallest possible flux is the one associated with the representation $\bold{133}$. Next we present a list illustrating examples of various fluxes and the symmetries they preserve, which appears in the tables below.

\begin{table}[!h]
\centering
\begin{tabular}{|c|c|c|c|}
\hline
 Node & Associated & Dynkin & Commutant in \\
 Number & representation & index & $E_7$ \\
\hline 
1 & $\bold{133}$ & $[1,0,0,0,0,0,0]$ & $U(1)\times SO(12)$ \\
\hline
2 & $\bold{912}$ & $[0,1,0,0,0,0,0]$ & $U(1)\times SU(7)$ \\
\hline
3 & $\bold{8645}$ & $[0,0,1,0,0,0,0]$ & $U(1)\times SU(2) \times SU(6)$ \\
\hline
4 & $\bold{365750}$ & $[0,0,0,1,0,0,0]$ & $U(1)\times SU(2) \times SU(3)\times SU(4)$ \\
\hline
5 & $\bold{365750}$ & $[0,0,0,0,1,0,0]$ & $U(1)\times SU(3) \times SU(5)$ \\
\hline
6 & $\bold{27664}$ & $[0,0,0,0,0,1,0]$ & $U(1)\times SU(2) \times SO(10)$ \\
\hline
7 & $\bold{56}$ & $[0,0,0,0,0,0,1]$ & $U(1)\times E_6$ \\
\hline
\end{tabular}
\label{FluxE71}
\end{table}

\begin{table}[!h]
\centering
\begin{tabular}{|c|c|c|c|}
\hline
 Node  & Associated & Commutant in & Associated \\
Number & Representation & $E_7$ & flux vectors \\
\hline 
 &  &  & $(0;1,1,0,0,0,0)$, \\
1 & $\bold{133}$ & $U(1)\times SO(12)$ & $(\frac{\sqrt{2}}{2};\frac{1}{2},\frac{1}{2},\frac{1}{2},\frac{1}{2},\frac{1}{2},\frac{1}{2})$, \\
 &  &  & $(\sqrt{2};0,0,0,0,0,0)$ \\
\hline
&  &  & $(\frac{\sqrt{2}}{2};1,1,1,0,0,0)$, \\
2 & $\bold{912}$ & $U(1)\times SU(7)$ & $(\sqrt{2};\frac{1}{2},\frac{1}{2},\frac{1}{2},\frac{1}{2},\frac{1}{2},-\frac{1}{2})$, \\
&  &  & $(0;\frac{3}{2},\frac{1}{2},\frac{1}{2},\frac{1}{2},\frac{1}{2},\frac{1}{2})$ \\
\hline
&  &  & $(\frac{3\sqrt{2}}{2};\frac{1}{2},\frac{1}{2},\frac{1}{2},\frac{1}{2},\frac{1}{2},\frac{1}{2})$, \\
3 & $\bold{8645}$ & $U(1)\times SU(2) \times SU(6)$ &  $(\sqrt{2};1,1,1,1,0,0)$, \\
&  &  & $(\frac{\sqrt{2}}{2};\frac{3}{2},\frac{3}{2},\frac{1}{2},\frac{1}{2},\frac{1}{2},\frac{1}{2})$,  \\
&  &  & (0;2,1,1,0,0,0) \\
\hline
&  &  & $(2\sqrt{2};1,1,1,1,0,0)$, \\
4 & $\bold{365750}$ & $U(1)\times SU(2)$ &  $(\frac{3\sqrt{2}}{2};\frac{3}{2},\frac{3}{2},\frac{3}{2},\frac{1}{2},\frac{1}{2},-\frac{1}{2})$, \\
 & & $\times SU(3)\times SU(4)$ & $(\sqrt{2};2,2,1,1,0,0)$, \\
&  &  &  $(\frac{\sqrt{2}}{2};\frac{5}{2},\frac{3}{2},\frac{3}{2},\frac{1}{2},\frac{1}{2},\frac{1}{2})$ \\
\hline
&  &   & $(\frac{3\sqrt{2}}{2};1,1,1,0,0,0)$,  \\
5 & $\bold{27664}$ & $U(1)\times SU(3) \times SU(5)$ &  $(\sqrt{2};\frac{3}{2},\frac{3}{2},\frac{1}{2},\frac{1}{2},\frac{1}{2},-\frac{1}{2})$, \\
&  &  & $(\frac{\sqrt{2}}{2};2,1,1,1,0,0)$,  \\
&  &  &  $(0;\frac{5}{2},\frac{1}{2},\frac{1}{2},\frac{1}{2},\frac{1}{2},-\frac{1}{2})$ \\
\hline
&  &  & $(\sqrt{2};1,1,0,0,0,0)$, \\
6 & $\bold{1539}$ & $U(1)\times SU(2) \times SO(10)$ &  $(0;1,1,1,1,0,0)$, \\
&  &  & $(\frac{\sqrt{2}}{2};\frac{3}{2},\frac{1}{2},\frac{1}{2},\frac{1}{2},\frac{1}{2},-\frac{1}{2})$ \\
\hline
7 & $\bold{56}$ & $U(1)\times E_6$ & $(\frac{\sqrt{2}}{2};1,0,0,0,0,0)$,  \\
&  &  &  $(0;\frac{1}{2},\frac{1}{2},\frac{1}{2},\frac{1}{2},\frac{1}{2},-\frac{1}{2})$ \\
\hline
\end{tabular}
\label{FluxE72}
\end{table}

\subsection*{$E_6$}

Here we introduce the fluxes for $N_f=5$ case. These are given by the possible weights. The smallest possible flux is given by the smallest representation which is the $\bold{27}$. This is only possible if the group is $\frac{E_6}{\mathbb{Z}_3}$, as we shall take the group to be. In the case of $E_6$, the smallest possible flux is the one associated with the representation $\bold{78}$. Next we present a list illustrating examples of various fluxes and the symmetries they preserve, which appears in the tables below.

\begin{table}[!h]
\centering
\begin{tabular}{|c|c|c|c|}
\hline
 Node & Associated  & Dynkin & Commutant in  \\
Number & representation &  index &  $E_6$ \\
\hline 
1, 6 & $\bold{27}$, $\overline{\bold{27}}$ & $[1,0,0,0,0,0]$, $[0,0,0,0,0,1]$ & $U(1)\times SO(10)$ \\
\hline
2 & $\bold{78}$ & $[0,1,0,0,0,0,0]$ & $U(1)\times SU(6)$ \\
\hline
3, 5 & $\bold{351}$, $\overline{\bold{351}}$ & $[0,0,1,0,0,0]$, $[0,0,0,0,1,0]$ & $U(1)\times SU(2) \times SU(5)$ \\
\hline
4 & $\bold{2925}$ & $[0,0,0,1,0,0]$ & $U(1)\times SU(2) \times SU(3)^2$ \\
\hline
\end{tabular}
\label{FluxE61}
\end{table}

\begin{table}[!h]
\centering
\begin{tabular}{|c|c|c|c|}
\hline
 Node  & Associated & Commutant in & Associated \\
Number & Representation & $E_6$ & flux vectors \\
\hline 
 & & & $(\frac{2\sqrt{3}}{3};0,0,0,0,0)$, \\
1, 6 & $\bold{27}$, $\overline{\bold{27}}$ & $U(1)\times SO(10)$ &  $(\frac{\sqrt{3}}{6};\frac{1}{2},\frac{1}{2},\frac{1}{2},\frac{1}{2},-\frac{1}{2})$, \\
 & & & $(\frac{\sqrt{3}}{3};1,0,0,0,0)$ \\
\hline
2 & $\bold{78}$ & $U(1)\times SU(6)$ & $(0;1,1,0,0,0)$, $(\frac{\sqrt{3}}{2};\frac{1}{2},\frac{1}{2},\frac{1}{2},\frac{1}{2},\frac{1}{2})$ \\
\hline
 & & & $(\frac{5\sqrt{3}}{6};\frac{1}{2},\frac{1}{2},\frac{1}{2},\frac{1}{2},-\frac{1}{2})$, \\
3, 5 & $\bold{351}$, $\overline{\bold{351}}$ & $U(1)\times SU(2) \times SU(5)$ &  $(\frac{2\sqrt{3}}{3};1,1,0,0,0)$, \\
 & & & $(\frac{\sqrt{3}}{3};1,1,1,0,0)$, $(\frac{\sqrt{3}}{6};\frac{3}{2},\frac{1}{2},\frac{1}{2},\frac{1}{2},\frac{1}{2})$\\
\hline
4 & $\bold{2925}$ & $U(1)\times SU(2) \times SU(3)^2$ & $(\sqrt{3};1,1,1,0,0)$, $(\frac{\sqrt{3}}{2};\frac{3}{2},\frac{3}{2},\frac{1}{2},\frac{1}{2},\frac{1}{2})$, \\
& & & $(0;2,1,1,0,0)$ \\
\hline
\end{tabular}
\label{FluxE62}
\end{table}

\subsection{Weyl group}

Here we discuss the Weyl groups of $E_6$ and $E_7$. The Weyl group is the isometry group of the root system. Recall that the Weyl group can be generated by reflections in hyperplanes orthogonal to each of the simple roots. Additionally, each simple root is associated with a node in the Dynkin diagram. As such given the Dynkin diagram of a Lie group, we can associate with each node a generator of the Weyl group. Denoting these as $s_i$, these obey:

\begin{enumerate}

\item $s^2_i = 1$ for all $i$. 

\item $\underbrace{s_i s_j s_i s_j ...}_{p} = \underbrace{s_j s_i s_j s_i ...}_{p}$, for $i\neq j$, where $p=2$ for unconnected nodes, $p=3$ for nodes connected by a single edge and $p=2n$ for nodes connected by an edge of multiplicity $n>1$.

\end{enumerate}

The first condition is a consequence of these being reflections. The second is a result of the fact that the edges between nodes reflect the angle between the roots. For instance, unconnected nodes obey that the respected roots are orthogonal to one another, and so the reflections commute. These conditions then can be used to define the Weyl group as an abstract group.

We next discuss the Weyl group for the cases of $E_6$ and $E_7$, that will be of interest to us here.

\subsection*{$E_7$}

First we consider the Weyl group of $E_7$. Its Coxeter diagram is:

\begin{center}
\includegraphics[width=0.37\textwidth]{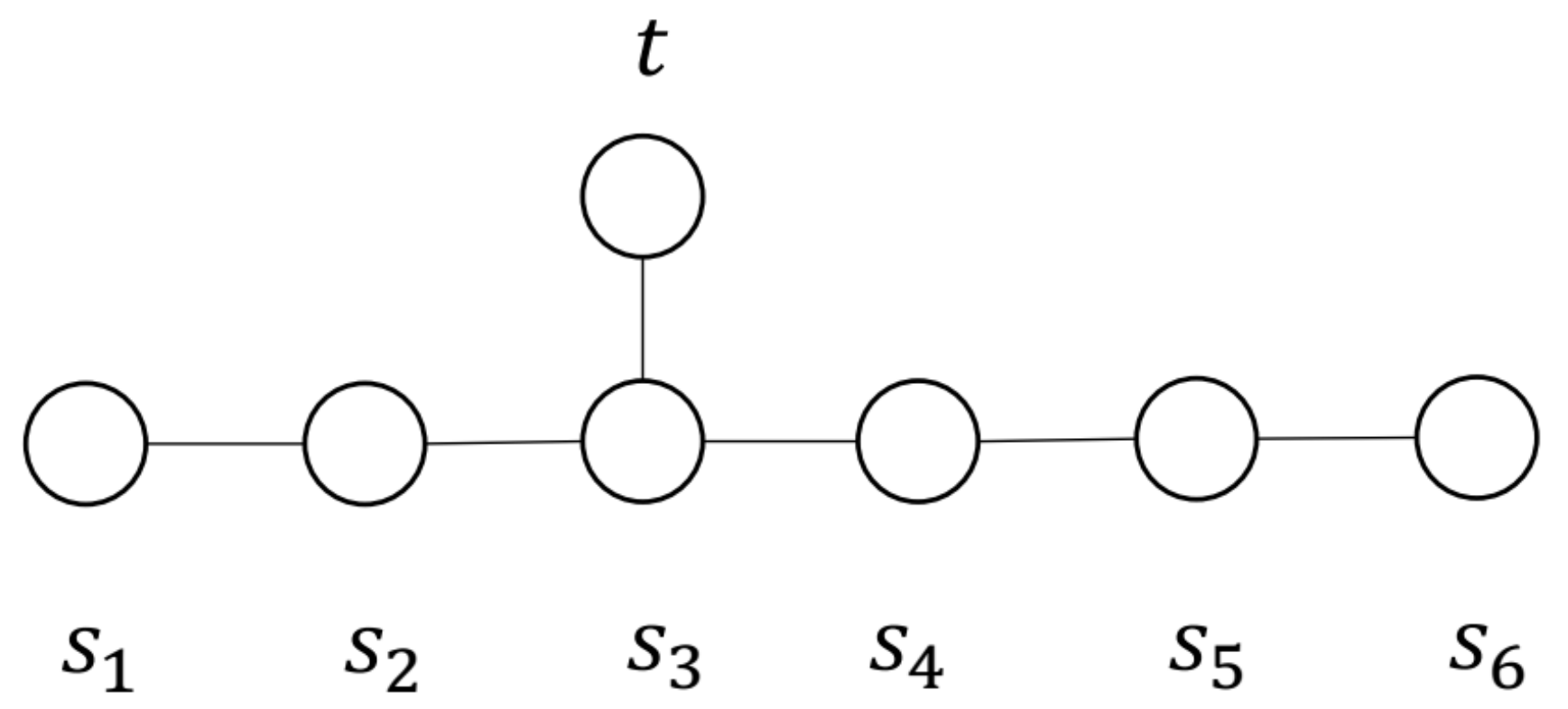}
\end{center}

All the generators are of order $2$. The seven generators obey the following braiding relations: $s_i s_{i+1} s_i = s_{i+1} s_i s_{i+1}$, $t s_3 t = s_3 t s_3$, for $i=1,2,3,4,5$ with all other elements commuting. Spanning the space $\mathbb{C}^{7}$ with the coordinates $z_i$, this structure can be represented by the transformations of the $z_i$ coordinates, represented in terms of the matrices:

\begin{equation}
\makebox[\linewidth][c]{\scalebox{1}{$
\begin{split} \nonumber
& s_6 = \begin{pmatrix}
  1 & 0 & 0 & 0 & 0 & 0 & 0\\
  0 & 0 & 1 & 0 & 0 & 0 & 0\\
  0 & 1 & 0 & 0 & 0 & 0 & 0\\
	0 & 0 & 0 & 1 & 0 & 0 & 0\\
	0 & 0 & 0 & 0 & 1 & 0 & 0\\
	0 & 0 & 0 & 0 & 0 & 1 & 0\\
	0 & 0 & 0 & 0 & 0 & 0 & 1
\end{pmatrix},
s_5 = \begin{pmatrix}
  1 & 0 & 0 & 0 & 0 & 0 & 0\\
  0 & 1 & 0 & 0 & 0 & 0 & 0\\
  0 & 0 & 0 & 1 & 0 & 0 & 0\\
	0 & 0 & 1 & 0 & 0 & 0 & 0\\
	0 & 0 & 0 & 0 & 1 & 0 & 0\\
	0 & 0 & 0 & 0 & 0 & 1 & 0\\
	0 & 0 & 0 & 0 & 0 & 0 & 1
\end{pmatrix}, 
s_4 = \begin{pmatrix}
  1 & 0 & 0 & 0 & 0 & 0 & 0\\
  0 & 1 & 0 & 0 & 0 & 0 & 0\\
  0 & 0 & 1 & 0 & 0 & 0 & 0\\
	0 & 0 & 0 & 0 & 1 & 0 & 0\\
	0 & 0 & 0 & 1 & 0 & 0 & 0\\
	0 & 0 & 0 & 0 & 0 & 1 & 0\\
	0 & 0 & 0 & 0 & 0 & 0 & 1
\end{pmatrix}, \\ \nonumber&
s_3 = \begin{pmatrix}
  1 & 0 & 0 & 0 & 0 & 0 & 0\\
  0 & 1 & 0 & 0 & 0 & 0 & 0\\
  0 & 0 & 1 & 0 & 0 & 0 & 0\\
	0 & 0 & 0 & 1 & 0 & 0 & 0\\
	0 & 0 & 0 & 0 & 0 & 1 & 0\\
	0 & 0 & 0 & 0 & 1 & 0 & 0\\
	0 & 0 & 0 & 0 & 0 & 0 & 1
\end{pmatrix},
t = \begin{pmatrix}
  1 & 0 & 0 & 0 & 0 & 0 & 0\\
  0 & 1 & 0 & 0 & 0 & 0 & 0\\
  0 & 0 & 1 & 0 & 0 & 0 & 0\\
	0 & 0 & 0 & 1 & 0 & 0 & 0\\
	0 & 0 & 0 & 0 & 1 & 0 & 0\\
	0 & 0 & 0 & 0 & 0 & 0 & 1\\
	0 & 0 & 0 & 0 & 0 & 1 & 0
\end{pmatrix},
s_2 = \begin{pmatrix}
  1 & 0 & 0 & 0 & 0 & 0 & 0\\
  0 & 1 & 0 & 0 & 0 & 0 & 0\\
  0 & 0 & 1 & 0 & 0 & 0 & 0\\
	0 & 0 & 0 & 1 & 0 & 0 & 0\\
	0 & 0 & 0 & 0 & 1 & 0 & 0\\
	0 & 0 & 0 & 0 & 0 & 0 & -1\\
	0 & 0 & 0 & 0 & 0 & -1 & 0
\end{pmatrix}, \\ \nonumber &
s_1 = \frac{1}{4}\begin{pmatrix}
  2 & \sqrt{2} & \sqrt{2} & \sqrt{2} & \sqrt{2} & \sqrt{2} & \sqrt{2}\\
  \sqrt{2} & 3 & -1 & -1 & -1 & -1 & -1\\
  \sqrt{2} & -1 & 3 & -1 & -1 & -1 & -1\\
	\sqrt{2} & -1 & -1 & 3 & -1 & -1 & -1\\
	\sqrt{2} & -1 & -1 & -1 & 3 & -1 & -1\\
	\sqrt{2} & -1 & -1 & -1 & -1 & 3 & -1\\
	\sqrt{2} & -1 & -1 & -1 & -1 & -1 & 3
\end{pmatrix}.
\end{split}$}}
\end{equation}

As we are interested in a description manifesting the $SO(12)$ subgroup, we have used a presentation of the Weyl group manifesting the $SO(12)$ Weyl group, here generated by $s_2$, $s_3$, $s_4$, $s_5$, $s_6$ and $t$.

\subsection*{$E_6$}

For the case of the Weyl group of $E_6$, the Coxeter diagram is:

\begin{center}
\includegraphics[width=0.33\textwidth]{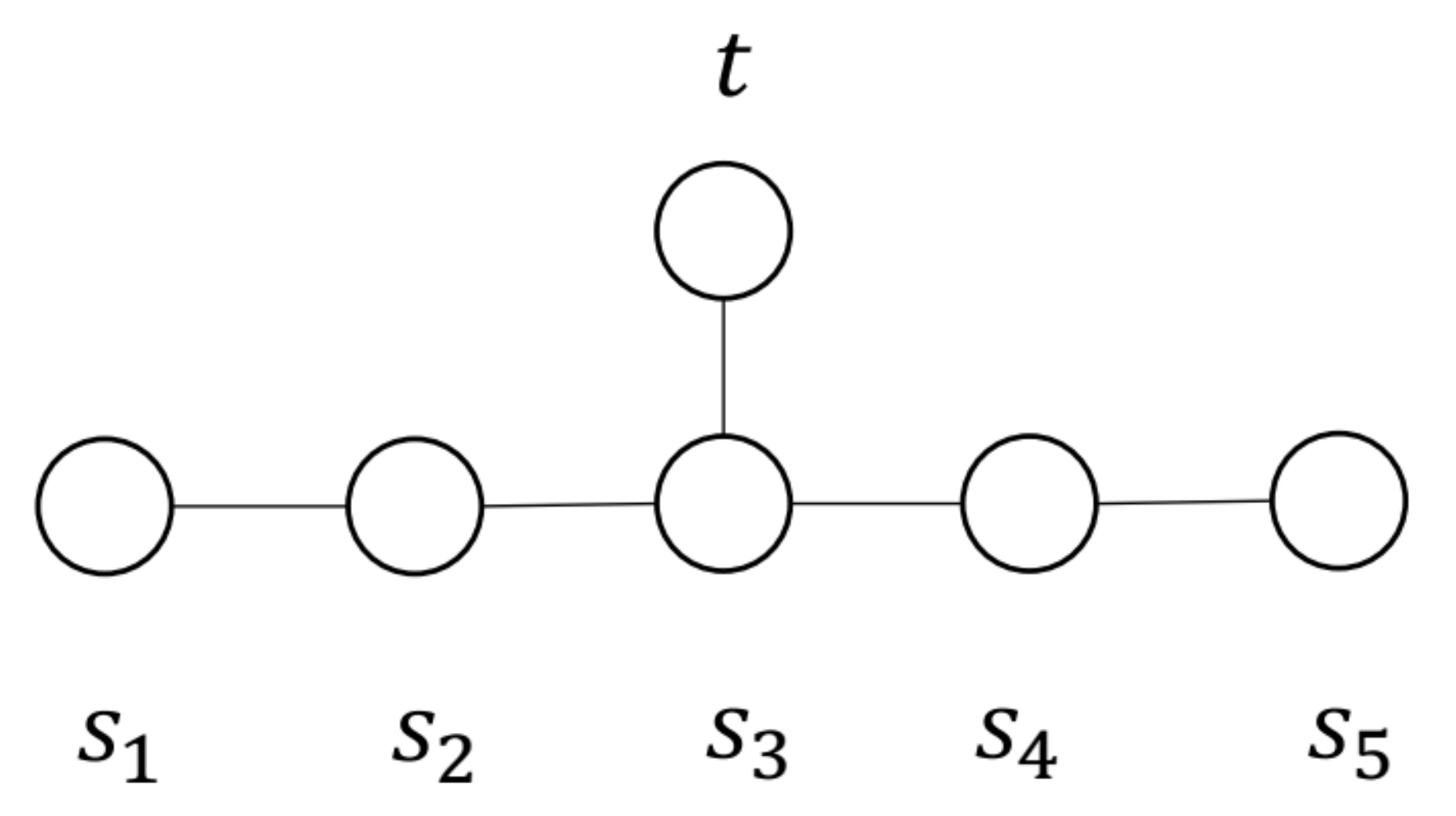}
\end{center}

All the generators are of order $2$. The six generators obey the following braiding relations: $s_i s_{i+1} s_i = s_{i+1} s_i s_{i+1}$, $t s_3 t = s_3 t s_3$, for $i=1,2,3,4$ with all other elements commuting. These can be represented in terms of the matrices:

\bea \nonumber
& & s_1 = \begin{pmatrix}
  1 & 0 & 0 & 0 & 0 & 0\\
  0 & 0 & 1 & 0 & 0 & 0\\
  0 & 1 & 0 & 0 & 0 & 0\\
	0 & 0 & 0 & 1 & 0 & 0\\
	0 & 0 & 0 & 0 & 1 & 0\\
	0 & 0 & 0 & 0 & 0 & 1
\end{pmatrix},   
s_2 = \begin{pmatrix}
  1 & 0 & 0 & 0 & 0 & 0\\
  0 & 1 & 0 & 0 & 0 & 0\\
  0 & 0 & 0 & 1 & 0 & 0\\
	0 & 0 & 1 & 0 & 0 & 0\\
	0 & 0 & 0 & 0 & 1 & 0\\
	0 & 0 & 0 & 0 & 0 & 1
\end{pmatrix},
s_3 = \begin{pmatrix}
  1 & 0 & 0 & 0 & 0 & 0\\
  0 & 1 & 0 & 0 & 0 & 0\\
  0 & 0 & 1 & 0 & 0 & 0\\
	0 & 0 & 0 & 0 & 1 & 0\\
	0 & 0 & 0 & 1 & 0 & 0\\
	0 & 0 & 0 & 0 & 0 & 1
\end{pmatrix}, \\ \nonumber & &
t = \begin{pmatrix}
  1 & 0 & 0 & 0 & 0 & 0\\
  0 & 1 & 0 & 0 & 0 & 0\\
  0 & 0 & 1 & 0 & 0 & 0\\
	0 & 0 & 0 & 1 & 0 & 0\\
	0 & 0 & 0 & 0 & 0 & 1\\
	0 & 0 & 0 & 0 & 1 & 0
\end{pmatrix},
s_4 = \begin{pmatrix}
  1 & 0 & 0 & 0 & 0 & 0\\
  0 & 1 & 0 & 0 & 0 & 0\\
  0 & 0 & 1 & 0 & 0 & 0\\
	0 & 0 & 0 & 1 & 0 & 0\\
	0 & 0 & 0 & 0 & 0 & -1\\
	0 & 0 & 0 & 0 & -1 & 0
\end{pmatrix}, \\ \nonumber & &
s_5 = \frac{1}{4}\begin{pmatrix}
  1 & -\sqrt{3} & -\sqrt{3} & -\sqrt{3} & -\sqrt{3} & -\sqrt{3}\\
  -\sqrt{3} & 3 & -1 & -1 & -1 & -1\\
  -\sqrt{3} & -1 & 3 & -1 & -1 & -1\\
	-\sqrt{3} & -1 & -1 & 3 & -1 & -1\\
	-\sqrt{3} & -1 & -1 & -1 &  3 & -1\\
	-\sqrt{3} & -1 & -1 & -1 & -1 &  3
\end{pmatrix}.
\eea 

As we are interested in a description manifesting the $SO(10)$ subgroup, we have used a presentation of the Weyl group manifesting the $SO(10)$ Weyl group, here generated by $s_1$, $s_2$, $s_3$, $s_4$ and $t$.

\section{3d partition functions and central charges}
\label{App:PF}

In this appendix we review the definition and computation of the three dimensional partition functions used throughout this paper, and the way global symmetries and central charges are encoded in them. We begin with a discussion of the 3d superconformal index which is used extensively in the sections above, followed by a discussion of the $\mathbb{S}^{3}$ partition function and the way the superconformal R-charge and global symmetry central charges can be extracted from it. 

\subsection*{Superconformal index}

The first quantity we consider is the superconformal index of a three dimensional theory with $\mathcal{N}=2$ supersymmetry, given by the following weighted trace over the states of the theory quantized on ${\mathbb S}^2\times {\mathbb R}$ \cite{Bhattacharya:2008zy,Kim:2009wb,Imamura:2011su,Krattenthaler:2011da,Kapustin:2011jm,Willett:2016adv},
\begin{equation}
\label{ind3dTr}
{\cal I}\left(q;{u_a}\right)=\mathrm{Tr}_{{\mathbb S}^2}\left[(-1)^{2J_3} e^{-\beta \delta} x^{2J_3+R} \prod_{a}u_{a}^{e_{a}} \right].
\end{equation}
Here $x$ is the superconformal fugacity, $J_3$ is the Cartan generator of the Lorenz $SO(3)$ isometry group of ${\mathbb S}^2$, $R$ is the $U(1)_R$ charge and $u_{a}$ and $e_{a}$ are the fugacities and charges of the global symmetries, respectively. Moreover, $\delta$ is defined as an anti-commutator of a Poincare supercharge as follows, 
\begin{equation}
\label{delta}
\delta\equiv\left\{ \mathcal{Q},\mathcal{Q}^{\dagger}\right\} =E-J_3-R\,,
\end{equation}
where $E$ is the conformal dimension. Note that even though the chemical potential $\beta$ appears in the definition \eqref{ind3dTr}, the index is in fact independent of it since the states with $\delta>0$ come in boson/fermion pairs and therefore cancel. Only the states with $\delta=0$, corresponding to short multiplets of the superconformal algebra, contribute.

Note also that the signs in \eqref{ind3dTr} are determined using the operator $2J_3$ instead of the fermion number $F$ (as in the definition of the 4d index) since in the presence of monopole operators there is a magnetic field that shifts the spins of states according to their electric charges (see the discussion in \cite{Aharony:2013dha} for more details).

The most salient feature of the index, which also applies to other partition functions, is its invariance under the RG flow \cite{Festuccia:2011ws,Dumitrescu:2016ltq}. That is, by calculating the index of an asymptotically free theory at the UV fixed point (using however the IR R-charge), one automatically obtains the index of the theory in the IR where the interactions are strong. This, in turn, allows us to study certain properties of the IR SCFT using a simple calculation in the UV. An example for this is given by the expansion of the index in powers of $x$, where only relevant operators contribute at orders $x^{\#<2}$, while at order $x^2$ marginal operators contribute with a positive sign and conserved currents with a negative one \cite{Razamat:2016gzx}. As a result, by investigating the expansion of the index in powers of $x$ one can extract nontrivial information about the operator spectrum and symmetries of the IR theory. 

In addition to the trace formula given in \eqref{ind3dTr}, the superconformal index is also simply related to the supersymmetric partition function of the theory on the space $\mathbb{S}^{2}\times\mathbb{S}^{1}$. Then, using localization one can compute the contributions of the different ingredients to the index and find the following results. First, the contribution of a chiral field with R-charge $R$, coupled with a unit charge to a $U(1)$ gauge field with flux $m$, is given by\footnote{We follow the notations of \cite{Aharony:2013kma}.}
\begin{equation}
{\cal I}_{\chi}(z;m;R)=\left(x^{1-R}z^{-1}\right)^{\frac{|m|}{2}}\prod_{l=0}^{\infty}\frac{1-(-1)^{m}z^{-1}x^{|m|+2-R+2l}}{1-(-1)^{m}zx^{|m|+R+2l}}
\end{equation}
where $z$ is the $U(1)$ fugacity. For general ({\it e.g.} nonabelian) gauge groups, one should simply introduce the corresponding fugacities $z_i$ and fluxes $m_i$ (for each generator in the Cartan subalgebra) and multiply several copies of ${\cal I}_{\chi}$ according to the representation of the chiral field under the group. For such general gauge groups, the contribution of a vector multiplet (including the Haar measure) is 
\begin{equation}
{\cal I}_{V}=\prod_{\alpha\in\mathcal{R}}x^{-\left|\alpha(m)\right|/2}\left(1-\left(-1\right)^{\alpha(m)}z^{\alpha}x^{\left|\alpha(m)\right|}\right)
\end{equation}
where $\mathcal{R}$ denotes the set of roots of the gauge group. Furthermore, in addition to the standard terms in the Lagrangian involving vector and chiral multiplets, we can also include Chern-Simons interactions in three dimensional theories. For a level $k$ term corresponding to a $U(1)$ gauge multiplet with fugacity $z$ and flux $m$, the contribution to the index is given by $z^{km}$ and the generalization to more general groups is straightforward. 

Finally, after all the different contributions are added, the index is obtained by integrating over the gauge fugacities and summing over the gauge fluxes.

\subsection*{$\mathbb{S}^{3}$ partition function}

We next turn to discuss the supersymmetric partition function defined on the round sphere $\mathbb{S}^{3}$. It depends only on the real masses for the symmetries and as in the previous case of the ${\mathbb S}^2\times {\mathbb S}^1$ index, the different contributions are calculated using localization \cite{Jafferis:2010un}. A chiral field with a unit charge under a $U(1)$ group with real mass $u$ contributes to this function a factor of 
\begin{equation}
\label{chis3}
\exp\left[l\left(1-R+iu\right)\right],
\end{equation}
where $R$ is its R charge and the function $l\left(z\right)$ is given by
\begin{equation}
l\left(z\right)=-z\log\left(1-e^{2\pi iz}\right)+\frac{i}{2}\left[\pi z^{2}+\frac{1}{\pi}\mathrm{Li}_{2}\left(e^{2\pi iz}\right)\right]-\frac{i\pi}{12}\,.
\end{equation}
In the case of a general group, the contribution is obtained in a straightforward way by multiplying factors of the form \eqref{chis3} according to the appropriate representation of the chiral field under the group. A vector multiplet corresponding to a general gauge group with real masses $\sigma_i$, on the other hand, contributes a factor of 
\begin{equation}
\frac{1}{\left|W\right|}\prod_{\alpha\in\mathcal{R}}\left|2\sinh\left(\pi\alpha(\sigma)\right)\right|,
\end{equation}
where $\left|W\right|$ is the order of the Weyl group and $\mathcal{R}$ denotes as before the set of roots. As for Chern-Simons terms, a pure $U(1)_k$ term with real mass $u$ contributes a factor of $\exp (\pi iku^2)$ while a mixed term a factor of $\exp (2\pi iku_1u_2)$, and the generalization to more general groups is straightforward. 

Once all the different contributions are included, we are only left with integrating over the real masses of the gauge group.

\subsection*{Superconformal R-charge and central charges}

The $\mathbb{S}^{3}$ partition function discussed above is very useful for finding the superconformal R-symmetry of an IR SCFT with a weakly-coupled UV description, as well as the central charges associated with its global symmetry currents. These latter quantities appear in the correlation functions of such currents, and serve as nontrivial CFT data of the theory. 

More concretely, let us consider the flat space two-point function (at separated points) of a $U(1)$ global symmetry current $J^{\mu}$. Conformal invariance then restricts it to take the following form,
\begin{equation}
\label{JJ}
\langle J^{\mu}\left(x\right)J^{\nu}\left(0\right)\rangle=\frac{C}{16\pi^{2}}\left(\delta^{\mu\nu}\partial^{2}-\partial^{\mu}\partial^{\nu}\right)\frac{1}{x^{2}}\,,
\end{equation}
and the number $C$, which is positive in a unitary theory, is defined as the corresponding central charge. 

To see how we can use the $\mathbb{S}^{3}$ partition function in order to find the IR R symmetry and the central charges, let us consider the space of R-symmetries parameterized by the mixing coefficients with all Abelian flavor symmetries $U(1)_I$. That is, we consider the trial R-symmetry 
\begin{equation}
R\left(t\right)=R_{0}+\sum_{I}t^{I}Q_{I}\,,
\end{equation}
where $R_{0}$ is some reference R-symmetry and $t^I$ and $Q_I$ are the mixing coefficients and charges of $U(1)_I$, respectively. The $\mathbb{S}^{3}$ partition function, denoted by $Z$, is then a function of $t$ when evaluated with respect to the R-symmetry $R(t)$. As shown in \cite{Jafferis:2010un,Closset:2012vg}, the value of $t$ that minimizes $\left|Z(t)\right|$ is the one corresponding to the R-symmetry which appears in the $\mathcal{N}=2$ superconformal algebra. As a result, using this $Z$-minimization principle one is able to find the IR R symmetry if the partition function $Z(t)$ is known. 

This principle can also be formulated in terms of the real part of the free energy, 
\begin{equation}
\textrm{Re}\,F(t)=-\textrm{Re}\log Z\,.
\end{equation}
In this case, the superconformal R symmetry locally maximizes $\textrm{Re}\,F(t)$ over the space of trial R symmetries $R(t)$. Denoting the corresponding value of $t$ by $t_{SC}$, we therefore have 
\begin{equation}
\left.\frac{\partial}{\partial t^{I}}\textrm{Re}\,F\right|_{t=t_{SC}}=0\,.
\end{equation}
The second derivative of $\textrm{Re}\,F(t)$ at $t_{SC}$ also turns out to have an interesting meaning, and in fact encodes the central charge $C$ defined in \eqref{JJ}. More explicitly, it is given by \cite{Closset:2012vg} (see also \cite{Gang:2019jut})
\begin{equation}
\label{d2F}
\left.\left(\frac{\partial}{\partial t^{I}}\right)^{2}\textrm{Re}\,F\right|_{t=t_{SC}}=-\frac{\pi^{2}}{2}C_{I}\,,
\end{equation}
where $C_I$ is the central charge of $U(1)_I$. We see that in addition to the superconformal R symmetry, the $\mathbb{S}^{3}$ partition function can also be used to compute the global symmetry central charges of the IR SCFT. 

\section{Confining Aharony duality for $SU(2)$ quivers}
\label{App:AharonyDuality}

In this appendix we revisit the problem of mapping monopole operators when applying Aharony duality \cite{Aharony:1997gp} locally on a node of a quiver gauge theory. This issue was first investigated in \cite{Pasquetti:2019uop,Pasquetti:2019tix} in the case of unitary gauge groups and later in \cite{Benvenuti:2020wpc,Benvenuti:2020gvy} in the case of $SU$, $USp$ and $SO$ groups (see also appendix A of \cite{Giacomelli:2020ryy}). Nevertheless, in all of these papers the case in which the dualized gauge node is confined wasn't considered, and this is exactly the case we are interested in. In particular, we will focus on quivers with $SU(2)$ groups only in which one of the nodes confines after the application of Aharony duality. The results of this appendix are used to derive the duality between quivers that we discussed in subsection 4.1.3.\footnote{The technique of deriving dualities between quiver gauge theories by applying more fundamental dualities locally on a node has recently been exploited a lot, see \cite{Benvenuti:2017kud,Giacomelli:2017vgk,Pasquetti:2019uop,Pasquetti:2019tix,Garozzo:2019xzi,Hwang:2020wpd,Benvenuti:2020gvy,Giacomelli:2020ryy} for some examples in three dimensions where one can apply either Aharony duality \cite{Aharony:1997gp} or variants with monopole superpotentials \cite{Benini:2017dud}.}

Let us consider a $3d$ $\mathcal{N}=2$ quiver where we have three adjacent $SU(2)$ gauge nodes
\bea\label{quiver}
\begin{tikzpicture}[baseline=0, font=\scriptsize]
\node[draw=none] (a1) at (0,0) {\large$\cdots$};
\node[draw, circle] (a2) at (1.5,0) {\large 2};
\node[draw, circle] (a3) at (3.5,0) {\large 2};
\node[draw, circle] (a4) at (5.5,0) {\large 2};
\node[draw=none] (a5) at (7,0) {\large $\cdots$};
\draw[draw, solid] (a1)--(a2);
\draw[draw, solid] (a2)--(a3);
\draw[draw, solid] (a3)--(a4);
\draw[draw, solid] (a4)--(a5);
\node[below] at (0.8,0) {$r_1$};
\node[below] at (2.5,0) {$r_2$};
\node[below] at (4.5,0) {$r_3$};
\node[above] at (2.5,0) {$P_1$};
\node[above] at (4.5,0) {$P_2$};
\node[below] at (6.25,0) {$r_4$};
\end{tikzpicture}
\eea
In the drawing we are also representing the R-charges of the various fields that are relevant for our discussion and we are denoting by $P_1$ and $P_2$ the two chiral fields attached to the middle node.

The right and left $SU(2)$ nodes can then be attached to any kind of matter. For definiteness, let us assume that the left node has $N_L$ additional fundamental chirals, while the right node has $N_R$ additional fundamental chirals, since this is the kind of situation we always have in the quivers in the main text.
The crucial thing is that the central $SU(2)$ node is only attached to the left and right $SU(2)$ nodes so that it effectively sees 4 chirals only. In this way we can dualize this middle node using Aharony duality. This is a duality between $SU(2)$ with 4 chirals and a WZ model of $6+1$ chiral fields that we call $M_{ij}$ and $S$ respectively, with $i<j=1,\cdots,4$. The singlets $M$ are mapped to the mesons in the antisymmetric representation of $SU(4)$ on the original side, while the singlet $S$ is mapped to the fundamental monopole of $SU(2)$ on the original side. The superpotential of the dual theory is
\bea
\mathcal{W}=S\,\mathrm{Pf}\,M\,.
\eea

When we apply such a duality to the middle node of the quiver \eqref{quiver}, the chirals are separated into $2+2$ and so the dual singlets are naturally decomposed into an $SU(2)\times SU(2)$ bifundamental field $B$ and two singlets $\alpha$, $\beta$. The resulting theory is thus
\bea\label{quiverdual}
\begin{tikzpicture}[baseline=0, font=\scriptsize]
\node[draw=none] (a1) at (0,0) {\large$\cdots$};
\node[draw, circle] (a2) at (1.5,0) {\large 2};
\node[draw, circle] (a3) at (3.5,0) {\large 2};
\node[draw=none] (a4) at (5,0) {\large$\cdots$};
\draw[draw, solid] (a1)--(a2);
\draw[draw, solid] (a2)--(a3);
\draw[draw, solid] (a3)--(a4);
\node[below] at (0.8,0) {$r_1$};
\node[below] at (2.5,-0.2) {$r_2+r_3$};
\node[above] at (2.5,0.2) {$B$};
\node[below] at (4.3,0) {$r_4$};
\node[thick] at (2.5,0) {\large $\times$};
\end{tikzpicture}
\eea
with the superpotential
\bea
\mathcal{W}=S\left(B^2+\alpha\,\beta\right)\,.
\eea
In the drawing we are representing the singlet field $S$ with a cross, but we are not representing the singlets $\alpha$ and $\beta$.

We are interested in how the monopole operators in the original quiver are mapped across the duality. Since we acted locally on the central node, the only monopole operators that map non-trivially are those that have non-vanishing magnetic flux w.r.t.~the only three nodes we are drawing in \eqref{quiver}. The basic monopole operators, that is those that can't be written as the composition of more fundamental monopole operators, are those that have magnetic flux only under adjacent nodes. In the following we list them as well as the corresponding R-charges, with the notation that a $\bullet$ corresponds to a unit of flux under the associated node and a zero corresponds to no flux:
\bea
&&R[\mathfrak{M}^{0,\bullet,0}]=2-2r_2-2r_3\nn\\
&&R[\mathfrak{M}^{\bullet,\bullet,0}]=N_L-N_Lr_1-2r_2-2r_3\nn\\
&&R[\mathfrak{M}^{0,\bullet,\bullet}]=N_R-2r_2-2r_3-N_Rr_4\nn\\
&&R[\mathfrak{M}^{\bullet,0,0}]=N_L-N_Lr_1-2r_2\nn\\
&&R[\mathfrak{M}^{0,0,\bullet}]=N_R-2r_3-N_Rr_4\nn\\
&&R[\mathfrak{M}^{\bullet,\bullet,\bullet}]=-2+N_L+N_R-N_Lr_1-2r_2-2r_3-N_Rr_4
\eea
In the dual theory \eqref{quiverdual} we only have 3 basic monopole operators, whose R-charges are
\bea
&&R[\mathfrak{M}^{\bullet,\bullet}]=-2+N_L+N_R-N_Lr_1-2r_2-2r_3-N_Rr_4\nn\\
&&R[\mathfrak{M}^{\bullet,0}]=N_L-N_Lr_1-2r_2-2r_3\nn\\
&&R[\mathfrak{M}^{0,\bullet}]=N_R-2r_2-2r_3-N_Rr_4
\eea
Moreover, in the dual frame we also have the singlets
\bea
&&R[S]=2-2r_2-2r_3\nn\\
&&R[\alpha]=2r_2\nn\\
&&R[\beta]=2r_3
\eea
With this information, we can try to understand how the monopoles map. We already know that
\bea
\mathfrak{M}^{0,\bullet,0}\leftrightarrow S
\eea
But what about the other monopole operators? We propose that they are mapped as follows:
\bea\label{monopolemap}
\mathfrak{M}^{\bullet,\bullet,0}\leftrightarrow\mathfrak{M}^{\bullet,0},\quad \mathfrak{M}^{0,\bullet,\bullet}\leftrightarrow \mathfrak{M}^{0,\bullet},\quad \mathfrak{M}^{\bullet,0,0}\leftrightarrow\beta\,\mathfrak{M}^{\bullet,0},\quad \mathfrak{M}^{0,0,\bullet}\leftrightarrow\alpha\,\mathfrak{M}^{0,\bullet},\quad \mathfrak{M}^{\bullet,\bullet,\bullet}\leftrightarrow \mathfrak{M}^{\bullet,\bullet}\nn\\
\eea
This mapping is indeed compatible with the assignment of R-charges written above.


\begin{thebibliography}{40}

\bibitem{Dimofte:2011ju}
  T.~Dimofte, D.~Gaiotto and S.~Gukov,
  Commun. Math. Phys. \textbf{325} (2014), 367-419
  [arXiv:1108.4389 [hep-th]].

\bibitem{Gang:2018wek}
  D.~Gang and K.~Yonekura,
  JHEP \textbf{07} (2018), 145
  [arXiv:1803.04009 [hep-th]].

\bibitem{SEI}
  N.~Seiberg, 
  Phys.\ Lett.\  B388:753-760 (1996)
  [arXiv:9608111 [hep-th]].

\bibitem{SM}
  N.~Seiberg, D.~R.~Morrison,
  Nucl.\ Phys.\  B483:229-247 (1997)
  [arXiv:9609070 [hep-th]].

\bibitem{SMI}
  N.~Seiberg, D.~R.~Morrison and K.~Intriligator,
  Nucl.\ Phys.\  B497:56-100 (1997)
  [arXiv:9702198 [hep-th]].

\bibitem{AH}
  O.~Aharony, A.~Hanany,
   Nucl.\ Phys.\  B504:239-271 (1997)
  [arXiv:9704170 [hep-th]].

\bibitem{AHK}
  O.~Aharony, A.~Hanany, and B.~Kol,
   JHEP {\bf 9801}, 002 (1998)
  [arXiv:9710116 [hep-th]].

\bibitem{KB}
  I.~Brunner, A.~Karch,
   Phys.\ Lett.\  B409:109-116 (1997)
  [arXiv:9705022 [hep-th]].

\bibitem{Bergman:2015dpa}
  O.~Bergman and G.~Zafrir,
  JHEP \textbf{12} (2015), 163
  [arXiv:1507.03860 [hep-th]].

\bibitem{Zafrir:2015ftn}
  G.~Zafrir,
  JHEP \textbf{03} (2016), 109
  [arXiv:1512.08114 [hep-th]].

\bibitem{Hayashi:2015vhy}
  H.~Hayashi, S.~S.~Kim, K.~Lee, M.~Taki and F.~Yagi,
  JHEP \textbf{10} (2016), 126
  [arXiv:1512.08239 [hep-th]].

\bibitem{Douglas:1996xp}
  M.~R.~Douglas, S.~H.~Katz and C.~Vafa,
  Nucl. Phys. B \textbf{497} (1997), 155-172
  [arXiv:hep-th/9609071 [hep-th]].

\bibitem{Jefferson:2018irk}
  P.~Jefferson, S.~Katz, H.~C.~Kim and C.~Vafa,
  JHEP \textbf{04} (2018), 103
  [arXiv:1801.04036 [hep-th]].

\bibitem{Bhardwaj:2018yhy}
  L.~Bhardwaj and P.~Jefferson,
  JHEP \textbf{07} (2019), 178
  [arXiv:1809.01650 [hep-th]].

\bibitem{Bhardwaj:2018vuu}
  L.~Bhardwaj and P.~Jefferson,
  JHEP \textbf{10} (2019), 282
  [arXiv:1811.10616 [hep-th]].

\bibitem{Closset:2018bjz}
  C.~Closset, M.~Del Zotto and V.~Saxena,
  SciPost Phys. \textbf{6} (2019) no.5, 052
  [arXiv:1812.10451 [hep-th]].
	
\bibitem{Apruzzi:2018nre}
  F.~Apruzzi, L.~Lin and C.~Mayrhofer,
  JHEP \textbf{05} (2019), 187
  [arXiv:1811.12400 [hep-th]].

\bibitem{Apruzzi:2019opn}
  F.~Apruzzi, C.~Lawrie, L.~Lin, S.~Sch\"afer-Nameki and Y.~N.~Wang,
  JHEP \textbf{11} (2019), 068
  [arXiv:1907.05404 [hep-th]].

\bibitem{Apruzzi:2019enx}
  F.~Apruzzi, C.~Lawrie, L.~Lin, S.~Sch\"afer-Nameki and Y.~N.~Wang,
  JHEP \textbf{03} (2020), 052
  [arXiv:1909.09128 [hep-th]].

\bibitem{Bhardwaj:2019jtr}
  L.~Bhardwaj,
  JHEP \textbf{09} (2020), 007
  [arXiv:1909.09635 [hep-th]].

\bibitem{Bhardwaj:2019fzv}
  L.~Bhardwaj, P.~Jefferson, H.~C.~Kim, H.~C.~Tarazi and C.~Vafa,
  JHEP \textbf{12} (2020), 151
  [arXiv:1909.11666 [hep-th]].

\bibitem{Saxena:2019wuy}
  V.~Saxena,
  JHEP \textbf{04} (2020), 198
  [arXiv:1911.09574 [hep-th]].

\bibitem{Apruzzi:2019kgb}
  F.~Apruzzi, S.~Schafer-Nameki and Y.~N.~Wang,
  JHEP \textbf{08} (2020), 153
  [arXiv:1912.04264 [hep-th]].

\bibitem{Chan:2000qc}
  C.~S.~Chan, O.~J.~Ganor and M.~Krogh,
  Nucl. Phys. B \textbf{597} (2001), 228-244
  [arXiv:hep-th/0002097 [hep-th]].

\bibitem{RVZ}
  S.~S.~Razamat, C.~Vafa, and G.~Zafrir,
  JHEP {\bf 1704}, 064 (2017) 
  [arXiv:1610.09178 [hep-th]].

\bibitem{Gai} 
  D.~Gaiotto,
  JHEP {\bf 1208}, 034 (2012)
  [arXiv:0904.2715 [hep-th]].

\bibitem{BTW} 
  F.~Benini, Y.~Tachikawa, and B.~Wecht,
	JHEP {\bf 1001}, 088 (2010)
  [arXiv:0909.1327 [hep-th]].

\bibitem{BBBW}
  I.~Bah, C.~Beem, N.~Bobev, and B.~Wecht,
	JHEP {\bf 1206}, 005 (2012)
  [arXiv:1203.0303 [hep-th]].

\bibitem{Gaiotto:2015usa}
  D.~Gaiotto and S.~S.~Razamat,
  JHEP \textbf{07} (2015), 073
  [arXiv:1503.05159 [hep-th]].

\bibitem{Bah:2017gph}
  I.~Bah, A.~Hanany, K.~Maruyoshi, S.~S.~Razamat, Y.~Tachikawa and G.~Zafrir,
  JHEP \textbf{06} (2017), 022
  [arXiv:1702.04740 [hep-th]].

\bibitem{KRVZ} 
  H.~-C.~Kim, S.~S.~Razamat, C.~Vafa, and G.~Zafrir,
	Fortsch.Phys. 66 (2018) no.1, 1700074
  [arXiv:1709.02496 [hep-th]].

\bibitem{KRVZ1} 
  H.~-C.~Kim, S.~S.~Razamat, C.~Vafa, and G.~Zafrir,
	JHEP {\bf 1806}, 058 (2018)
  [arXiv:1802.00620 [hep-th]].

\bibitem{KRVZ2} 
  H.~-C.~Kim, S.~S.~Razamat, C.~Vafa, and G.~Zafrir,
	JHEP {\bf 1809}, 110 (2018)
  [arXiv:1806.07620 [hep-th]].

\bibitem{Razamat:2018gro}
  S.~S.~Razamat and G.~Zafrir,
  Phys. Rev. D \textbf{98} (2018) no.6, 066006
  [arXiv:1806.09196 [hep-th]].

\bibitem{Zafrir:2018hkr}
  G.~Zafrir,
  JHEP \textbf{10} (2019), 040
  [arXiv:1809.04260 [hep-th]].

\bibitem{Chen:2019njf}
  J.~Chen, B.~Haghighat, S.~Liu and M.~Sperling,
  JHEP \textbf{01} (2020), 152
  [arXiv:1907.00536 [hep-th]].

\bibitem{Razamat:2019mdt}
  S.~S.~Razamat, E.~Sabag and G.~Zafrir,
  JHEP \textbf{12} (2019), 108
  [arXiv:1907.04870 [hep-th]].

\bibitem{Pasquetti:2019hxf}
  S.~Pasquetti, S.~S.~Razamat, M.~Sacchi and G.~Zafrir,
  SciPost Phys. \textbf{8} (2020) no.1, 014
  [arXiv:1908.03278 [hep-th]].
	
\bibitem{Razamat:2019ukg}
  S.~S.~Razamat and E.~Sabag,
  JHEP \textbf{01} (2020), 086
  [arXiv:1910.03603 [hep-th]].

\bibitem{Razamat:2020bix}
  S.~S.~Razamat and E.~Sabag,
  JHEP \textbf{09} (2020), 028
  [arXiv:2006.03480 [hep-th]].

\bibitem{Sabag:2020elc}
  E.~Sabag,
  JHEP \textbf{10} (2020), 139
  [arXiv:2007.13567 [hep-th]].

\bibitem{Mitev:2014jza}
  V.~Mitev, E.~Pomoni, M.~Taki and F.~Yagi,
  JHEP \textbf{04} (2015), 052
  [arXiv:1411.2450 [hep-th]].

\bibitem{Zafrir:2016wkk}
  G.~Zafrir,
  JHEP \textbf{01} (2017), 097
  [arXiv:1605.08337 [hep-th]].

\bibitem{Morrison:2020ool}
  D.~R.~Morrison, S.~Schafer-Nameki and B.~Willett,
  JHEP \textbf{09} (2020), 024
  [arXiv:2005.12296 [hep-th]].

\bibitem{Albertini:2020mdx}
  F.~Albertini, M.~Del Zotto, I.~Garc\'\i{}a Etxebarria and S.~S.~Hosseini,
  JHEP \textbf{12} (2020), 203
  [arXiv:2005.12831 [hep-th]].

\bibitem{Bhardwaj:2020phs}
  L.~Bhardwaj and S.~Sch\"afer-Nameki,
	JHEP \textbf{02} (2021), 159
  [arXiv:2008.09600 [hep-th]].

\bibitem{BenettiGenolini:2020doj}
  P.~Benetti Genolini and L.~Tizzano,
  [arXiv:2009.07873 [hep-th]].

\bibitem{Chang:2017cdx}
  C.~M.~Chang, M.~Fluder, Y.~H.~Lin and Y.~Wang,
  JHEP \textbf{03} (2018), 123
  [arXiv:1710.08418 [hep-th]].

\bibitem{Gang:2018huc}
  D.~Gang and M.~Yamazaki,
  Phys. Rev. D \textbf{98} (2018) no.12, 121701
  [arXiv:1806.07714 [hep-th]].

\bibitem{Ganor:1996pc}
  O.~J.~Ganor, D.~R.~Morrison and N.~Seiberg,
  Nucl. Phys. B \textbf{487} (1997), 93-127
  [arXiv:hep-th/9610251 [hep-th]].

\bibitem{Cordova:2016xhm}
  C.~Cordova, T.~T.~Dumitrescu and K.~Intriligator,
  JHEP \textbf{11} (2016), 135
  [arXiv:1602.01217 [hep-th]].

\bibitem{GKSTW}
  D.~Green, Z.~Komargodski, N.~Seiberg, Y.~Tachikawa and B.~Wecht,
  JHEP {\bf 1006}, 106 (2010)
  [arXiv:1005.3546 [hep-th]].

\bibitem{BRZtoapp}
  C.~Beem, S.~S.~Razamat and G.~Zafrir,
  to appear, (See S. S. Razamat, Geometrization of
  relevance", talk at ‘Avant-garde methods for quantum field theory and gravity, Nazareth 2/2019
  (https://phsites.technion.ac.il/the-fifth-israeli-indian-conference-on-string-theory/program/).

\bibitem{Razamat:2016gzx}
  S.~S.~Razamat and G.~Zafrir,
  JHEP \textbf{11} (2016), 061
  doi:10.1007/JHEP11(2016)061
  [arXiv:1609.02089 [hep-th]].

\bibitem{GB} 
  C.~Beem, A.~Gadde,
	JHEP {\bf 1404}, 036 (2014)
  [arXiv:1212.1467 [hep-th]].
  
  \bibitem{Aharony:1997bx}
O.~Aharony, A.~Hanany, K.~A.~Intriligator, N.~Seiberg and M.~J.~Strassler,
Nucl. Phys. B \textbf{499} (1997), 67-99
doi:10.1016/S0550-3213(97)00323-4
[arXiv:hep-th/9703110 [hep-th]].

\bibitem{Intriligator:2013lca}
K.~Intriligator and N.~Seiberg,
JHEP \textbf{07} (2013), 079
doi:10.1007/JHEP07(2013)079
[arXiv:1305.1633 [hep-th]].

\bibitem{Aharony:1997gp}
  O.~Aharony,
  Phys. Lett. B \textbf{404} (1997), 71-76
  [arXiv:hep-th/9703215 [hep-th]].

\bibitem{Ohmori:2018ona}
  K.~Ohmori, Y.~Tachikawa and G.~Zafrir,
  JHEP \textbf{04} (2019), 006
  [arXiv:1812.04637 [hep-th]].

\bibitem{Kim:2012gu}
  H.~C.~Kim, S.~S.~Kim and K.~Lee,
  JHEP \textbf{10} (2012), 142
  [arXiv:1206.6781 [hep-th]].

\bibitem{Aharony:2013dha}
  O.~Aharony, S.~S.~Razamat, N.~Seiberg and B.~Willett,
  JHEP \textbf{07} (2013), 149
  [arXiv:1305.3924 [hep-th]].

\bibitem{Aharony:2013kma}
  O.~Aharony, S.~S.~Razamat, N.~Seiberg and B.~Willett,
  JHEP \textbf{08} (2013), 099
  [arXiv:1307.0511 [hep-th]].

\bibitem{HRSS}
  C.~Hwang, S.~S.~Razamat, E.~Sabag and M.~Sacchi,
  [arXiv:2103.09149 [hep-th]].

\bibitem{Hwang:2020wpd}
  C.~Hwang, S.~Pasquetti and M.~Sacchi,
  JHEP \textbf{09} (2020), 047
  [arXiv:2002.12897 [hep-th]].

\bibitem{Garozzo:2020pmz}
  I.~Garozzo, N.~Mekareeya, M.~Sacchi and G.~Zafrir,
  JHEP \textbf{06} (2020), 159
  [arXiv:2003.07373 [hep-th]].

\bibitem{Razamat:2018gbu}
  S.~S.~Razamat, O.~Sela and G.~Zafrir,
  JHEP \textbf{10}, 163 (2018)
  [arXiv:1809.00541 [hep-th]].

\bibitem{Sela:2019nqa}
  O.~Sela and G.~Zafrir,
  JHEP \textbf{12}, 052 (2019)
  [arXiv:1910.03629 [hep-th]].

\bibitem{Razamat:2017wsk}
  S.~S.~Razamat, O.~Sela and G.~Zafrir,
  Phys. Rev. Lett. \textbf{120}, no.7, 071604 (2018)
  [arXiv:1711.02789 [hep-th]].

\bibitem{Naka:2002jz}
  M.~Naka,
  [arXiv:hep-th/0206141 [hep-th]].
	
\bibitem{Bah:2018lyv}
  I.~Bah, A.~Passias and P.~Weck,
  JHEP \textbf{01} (2019), 058
  [arXiv:1807.06031 [hep-th]].

\bibitem{Hosseini:2018usu}
  S.~M.~Hosseini, K.~Hristov, A.~Passias and A.~Zaffaroni,
  JHEP \textbf{12} (2018), 001
  [arXiv:1809.10685 [hep-th]].

\bibitem{Crichigno:2018adf}
  P.~M.~Crichigno, D.~Jain and B.~Willett,
  JHEP \textbf{11} (2018), 058
  [arXiv:1808.06744 [hep-th]].

\bibitem{Hosseini:2018uzp}
  S.~M.~Hosseini, I.~Yaakov and A.~Zaffaroni,
  JHEP \textbf{11} (2018), 119
  [arXiv:1808.06626 [hep-th]].

\bibitem{Razamat:2019sea}
  S.~S.~Razamat and B.~Willett,
  Phys. Rev. D \textbf{101} (2020) no.6, 065004
  [arXiv:1911.00956 [hep-th]].

\bibitem{Gaiotto:2014kfa}
  D.~Gaiotto, A.~Kapustin, N.~Seiberg and B.~Willett,
  JHEP \textbf{02} (2015), 172
  [arXiv:1412.5148 [hep-th]].

\bibitem{Bergman:2020ifi}
  O.~Bergman, Y.~Tachikawa and G.~Zafrir,
  JHEP \textbf{07} (2020), 077
  [arXiv:2004.05350 [hep-th]].

\bibitem{Bhattacharya:2008zy}
  J.~Bhattacharya, S.~Bhattacharyya, S.~Minwalla and S.~Raju,
  JHEP \textbf{02}, 064 (2008)
  [arXiv:0801.1435 [hep-th]].

\bibitem{Kim:2009wb}
  S.~Kim,
  Nucl. Phys. B \textbf{821}, 241-284 (2009)
  [erratum: Nucl. Phys. B \textbf{864}, 884 (2012)]
  [arXiv:0903.4172 [hep-th]].

\bibitem{Imamura:2011su}
  Y.~Imamura and S.~Yokoyama,
  JHEP \textbf{04}, 007 (2011)
  [arXiv:1101.0557 [hep-th]].

\bibitem{Krattenthaler:2011da}
  C.~Krattenthaler, V.~P.~Spiridonov and G.~S.~Vartanov,
  JHEP \textbf{06}, 008 (2011)
  [arXiv:1103.4075 [hep-th]].

\bibitem{Kapustin:2011jm}
  A.~Kapustin and B.~Willett,
  [arXiv:1106.2484 [hep-th]].

\bibitem{Willett:2016adv}
  B.~Willett,
  J. Phys. A \textbf{50}, no.44, 443006 (2017)
  [arXiv:1608.02958 [hep-th]].

\bibitem{Festuccia:2011ws}
  G.~Festuccia and N.~Seiberg,
  JHEP \textbf{06}, 114 (2011)
  [arXiv:1105.0689 [hep-th]].

\bibitem{Dumitrescu:2016ltq}
  T.~T.~Dumitrescu,
  J. Phys. A \textbf{50}, no.44, 443005 (2017)
  [arXiv:1608.02957 [hep-th]].

\bibitem{Jafferis:2010un}
  D.~L.~Jafferis,
  JHEP \textbf{05}, 159 (2012)
  [arXiv:1012.3210 [hep-th]].

\bibitem{Closset:2012vg}
  C.~Closset, T.~T.~Dumitrescu, G.~Festuccia, Z.~Komargodski and N.~Seiberg,
  JHEP \textbf{10}, 053 (2012)
  [arXiv:1205.4142 [hep-th]].

\bibitem{Gang:2019jut}
  D.~Gang and M.~Yamazaki,
  JHEP \textbf{02}, 102 (2020)
  [arXiv:1912.09617 [hep-th]].

\bibitem{Pasquetti:2019uop}
  S.~Pasquetti and M.~Sacchi,
  JHEP \textbf{11} (2019), 081
  [arXiv:1903.10817 [hep-th]].

\bibitem{Pasquetti:2019tix}
  S.~Pasquetti and M.~Sacchi,
  JHEP \textbf{01} (2020), 061
  [arXiv:1905.05807 [hep-th]].

\bibitem{Benvenuti:2020wpc}
  S.~Benvenuti, I.~Garozzo and G.~Lo Monaco,
  [arXiv:2012.08556 [hep-th]].

\bibitem{Benvenuti:2020gvy}
  S.~Benvenuti, I.~Garozzo and G.~Lo Monaco,
  [arXiv:2012.09773 [hep-th]].

\bibitem{Giacomelli:2020ryy}
  S.~Giacomelli, N.~Mekareeya and M.~Sacchi,
  [arXiv:2012.12852 [hep-th]].

\bibitem{Benvenuti:2017kud}
  S.~Benvenuti and S.~Giacomelli,
  JHEP \textbf{10} (2017), 173
  [arXiv:1706.04949 [hep-th]].

\bibitem{Giacomelli:2017vgk}
  S.~Giacomelli and N.~Mekareeya,
  JHEP \textbf{03} (2018), 126
  [arXiv:1711.11525 [hep-th]].

\bibitem{Garozzo:2019xzi}
  I.~Garozzo, N.~Mekareeya and M.~Sacchi,
  JHEP \textbf{11} (2019), 053
  [arXiv:1909.02832 [hep-th]].

\bibitem{Benini:2017dud}
  F.~Benini, S.~Benvenuti and S.~Pasquetti,
  JHEP \textbf{08} (2017), 086
  [arXiv:1703.08460 [hep-th]].

\bibitem{Cordova:2016emh}
  C.~Cordova, T.~T.~Dumitrescu and K.~Intriligator,
  JHEP \textbf{03} (2019), 163
  [arXiv:1612.00809 [hep-th]].

\end{thebibliography}
\end{document}